\documentclass[aps,prx,longbibliography,twocolumn,amsmath,amssymb,floatfix,eqsecnum,superscriptaddress]{revtex4-1}

\usepackage{amsmath}
\usepackage{amssymb}
\usepackage{amsthm}
\usepackage{mathtools}
\usepackage{mathdots}
\usepackage{amsfonts}
\usepackage{bbm}

\usepackage[pdftex]{color} 
\usepackage{graphicx}
\usepackage{dcolumn} 
\usepackage{bm} 
\usepackage{float} 
\usepackage[driverfallback=dvipdfm]{hyperref} 
\usepackage{longtable}
\usepackage{ulem}   
\newcommand{\bs}[1]{{\boldsymbol{#1}}}

\newcommand{\sgn}{\mathcal{\text{sgn}}}
\newcommand{\Pf}{\mathcal{\text{Pf}}}

\renewcommand\[{\begin{equation}}
\renewcommand\]{\end{equation}}

\begin{document}

\title{ Theory of Hofstadter Superconductors}

\author{Daniel Shaffer}
\thanks{
These authors contributed equally to the work.}
\affiliation
{
Department  of  Physics,  Emory  University,  400 Dowman Drive, Atlanta,  GA  30322,  USA
}

\author{Jian Wang}
\thanks{
These authors contributed equally to the work.}
\affiliation
{
Department  of  Physics,  Emory  University,  400 Dowman Drive, Atlanta,  GA  30322,  USA
}

\author{Luiz H. Santos}
\affiliation
{
Department  of  Physics,  Emory  University,  400 Dowman Drive, Atlanta,  GA  30322,  USA
}

\begin{abstract}
We study mean-field states resulting from the pairing of electrons in time-reversal broken fractal Hofstadter bands, which arise in two-dimensional lattices where the unit cell traps magnetic flux $\Phi = (p/q)\Phi_0$ comparable to the flux quantum $\Phi_0 = h/e$. It is established that the dimension and degeneracy of the irreducible representations of the magnetic translation group (MTG) furnished by the charge 2e pairing fields have different properties from those furnished by single particle Bloch states, and in particular are shown to depend on the parity of the denominator $q$. We explore this symmetry analysis to formulate a Ginzburg-Landau theory describing the thermodynamic properties of Hofstadter superconductors at arbitrary rational flux $\Phi = (p/q)\Phi_0$  in terms of a multicomponent order parameter that describes the finite momentum pairing of electrons across different Fermi surface patches. This phenomenological theory leads to a rich phase diagram characterized by different symmetry breaking patterns of the MTG, which can be interpreted as distinct classes of vortex lattices. A class of $\mathbb{Z}_q$-symmetric Hofstadter SCs is identified, in which the MTG breaks down to a $\mathbb{Z}_q$ subgroup. We study the topological properties of such $\mathbb{Z}_q$-symmetric Hofstadter SCs and show that the parity of the Chern numbers is fixed by the parity of $q$. We identify the conditions for the realization of Bogoliubov Fermi surfaces in the presence of parity and MTG symmetries, establishing a novel topological invariant capturing the existence of such charge-neutral gapless excitations. Our findings, which could bear relevance to the description of re-entrant superconductivity in moir\'e systems in the Hofstadter regime, establish Hofstadter SC as a fertile setting to explore symmetry broken and topological orders.
\end{abstract}

\date{\today}

\maketitle

\section{Introduction}

The goal of this paper is to provide a framework for describing charge 2e condensates formed by pairing of electrons in Hofstadter bands,
which arise from the interplay of an external magnetic field and a two-dimensional (2D) lattice periodic potential \cite{Harper55,Azbel64,Zak64_1,Brown64,Hofstadter76},
producing a magnetic flux per unit cell $\Phi = (p/q)\Phi_0$, where $\Phi_{0} = h/e$ is the magnetic flux quantum.
Hofstadter systems are characterized by a rich fractal spectrum of topological Chern bands \cite{TKNN, Kohmoto85}, which have long been recognized as an arena to realize integer quantum Hall topological states through complete \cite{TKNN, Kohmoto85, Haldane1988} filling of Chern bands. In recent years, great focus has been given towards identifying fractional topological insulating orders in partially filled Chern bands \cite{Neupert-2011,Sheng-2011,Tang-2011,Sun-2011,Regnault2011,parameswaran2013fractional,Kol1993,Moller2015,murthyshankar2012,Sohal-2018,AndrewsNeupert21}.
However, the properties of superconducting condensates formed by paired electrons in partially filled Hofstadter bands have received considerably less attention. 

Superconductivity in high magnetic fields, specifically above the upper critical field \(H_{c2}\), known as re-entrant superconductivity \cite{Tesanovic89, AkeraMacDonald91, Tesanovic91, Rajagopal91, MacDonald92, Norman92, RasoltTesanovich92,MacDonald93, RyanRajagopal93,DukanTesanovic97,Maska02, Ueta13, ScherpelzRajagopal13}, has been predicted theoretically \cite{Gruenberg68} as a result of an increased density of states of the Landau levels in the absence of a lattice, which can be understood as the \(q\rightarrow\infty\) limit of the Hofstadter systems. Phenomenon similar to re-entrant superconductivity has also been studied in the context of superconducting networks \cite{Alexander83,Pannetier84} and Josephson arrays \cite{Niu89,Kato13}.
However, recent advances in 2D ``twisted" and nanopatterned quantum materials exhibiting large unit cells have enabled the experimental realization of fermionic Hofstadter bands \cite{Dean13,Ponomarenko13,Hunt13,Forsythe18,Wang15,Spanton18,Saito21} by trapping large orbital magnetic flux, $\Phi/\Phi_{0} \sim 1$, in the super unit cell.
These highly tunable 2D quantum materials -- which offer an alternative
to realizing fractal bands in optical lattices \cite{Mueller04,Gerbier10, Aidelsburger11, Hauke12, Celi14} --  provide a motivation for seeking a deeper understanding of the properties of paired electrons in a regime characterized by breaking of time-reversal symmetry and strong lattice effects.
We note that though re-entrant superconductivity has not been conclusively observed, it may have been seen in some recent experiments in UTe\(_2\) and twisted trilayer graphene \cite{Ran19, Mineev20, Lebed20, CaoTTG21}, and a similar phenomenon has been observed in superconducting networks \cite{Pannetier84}. Interestingly, as the re-entrant phase in UTe\(_2\) has been observed at magnetic fields not aligned with any symmetry direction of the lattice, it has been argued that the system may in fact be in the Hofstadter regime with a large effective flux per unit cell \cite{Park20}.

Hofstadter lattices possess rich single particle properties. In contrast to Landau levels obtained when $\Phi/\Phi_0 \ll 1$, Hofstadter bands have finite bandwidths and are characterized by a wider range of Chern numbers, which gives way to unconventional quantum critical phenomena controlled by lattice effects \cite{wang_classification_2020,Lee-PRX-2018}.
Furthermore, the band structure  non-trivially depends on the magnetic flux per unit cell, which can be controlled by the strength of the external field. All these properties combined make the microscopic analysis of the pairing problem rich, yet involved. 
However, we show in this work that it is possible to extract universal properties of Hofstadter superconductors despite the complexity of the single particle states.
To that purpose, we pursue a symmetry based approach that captures some of the essential features of Hofstadter systems and, thus, leads to a general framework describing the thermodynamic properties of time-reversal broken superconductivity in a regime dominated by strong lattice effects.

In this work we focus on describing the symmetry properties of the pairing order parameter $\hat{\Delta}_{a b} \propto \langle \psi_{a} \psi_{b} \rangle$ under U(1) and magnetic translation symmetries for a general Hofstadter lattice with magnetic flux $\Phi = (p/q)\Phi_0$ per unit cell, with $p$ and $q$ coprime integers.
The magnetic translation group (MTG) is generated by non-commuting lattice translation operators in the presence of a magnetic field \cite{Zak64_1, Zak64_2, Brown64}, and plays an essential role in our symmetry analysis of the order parameter.
A key result of this work is establishing the properties of the irreducible representations (irreps) of the MTG furnished by the charge 2e pairing fields $\hat{\Delta}$, and showing that they are different from the well-known \(q\)-dimensional irreps of the MTG furnished by single particle Bloch states \cite{Zak64_1, Zak64_2, Brown64}.
Specifically, while the irreps furnished by the pairing order parameter have the same dimension \(q\) for odd \(q\), for even \(q\) there are four distinct irreps of dimension \(q/2\). In fact, as pointed out in \cite{Florek97}, the possibility of such charge 2e irreps has already been identified mathematically in \cite{Zak64_1} but were originally rejected as unphysical.
We show that this property stems from the interplay between the generators of the U(1) group and the MTG, which are \textit{simultaneously} broken in the Hofstadter charge 2e condensate.
We emphasize that having a charge 2e order parameter is key to our findings. In particular, the paired states in fermionic Hofstadter systems analyzed here have different symmetry properties under the action of the MTG from those occurring in the \emph{bosonic} Hofstadter model \cite{Balents05, PowellDasSarma10, PowellDasSarma11, NatuDasSarma16, Song19} for which the order parameter $\langle b^{\dagger} \rangle$ is a charge 1e operator that furnishes irreps analogous to single particle Bloch states.

From the study of group representations stems a comprehensive Ginzburg-Landau phenomenological theory of Hofstadter superconductors for generic magnetic flux.
The analysis shows that Hofstadter condensates are characterized by multi-component order parameters, uncovering an ideal scenario to explore a number of unconventional superconducting orders \cite{SigristUeda91} such as pair-density waves \cite{ZhaiOktel10,Iskin15_1,Iskin15_2,UmucalilarIskin16,SohalFradkin20}, multi-band chiral topological condensates beyond conventional chiral p-wave superconductors \cite{ReadGreen00},
as well as condensates with gapless fermionic excitations such as symmetry-protected critical points and Bogoliubov Fermi surfaces
\cite{Agterberg17,BrydonAgterberg18,santos_pdw_2019,YuanFu18,SumitaYanase19,MenkeBrydon19,Link20,Lapp20,Shaffer20,ZhuFu20,KobayashiSato14,ZhaoSchnyder16}. 
As such, this work provides a symmetry-based framework to study thermodynamic phases --  and their corresponding phase transitions -- realized by charge 2e condensates in the presence of large magnetic flux and lattice effects,  establishing superconductivity in Hofstadter systems as a fertile setting to explore symmetry broken and topological orders.

Moreover, the analysis of irreps of the MTG shows that the Hofstadter superconducting phase behaves as a vortex lattice, in the sense that the phase of the components of the order parameter winds as a parallel transport takes place around the unit cell.
Therefore, these findings provide a general framework that justifies the numerical observation of vortex lattices in fermionic \cite{ZhaiOktel10,SohalFradkin20} and bosonic \cite{PowellDasSarma11} Hofstadter systems.
However, we note that the order parameter \(\hat{\Delta}\) need not vanish anywhere in space, unlike in heterostructures combing quantum Hall systems and regular Abrikosov vortex lattices \cite{Zocher16, SahuJain18, SchirmerJain20, ChaudharyMacDonald20, ChaudharyMacDonald21, WeeksFranz07, JeonJain19}.
In the limit \(q\rightarrow\infty\), we expect the vortex lattice to approach the Abrikosov vortex lattice at the upper critical field \(H_{c2}\).

The group theory classification of Hofstadter pairing discussed in this work sheds light on earlier studies of pairing in fermionic Hofstadter systems done mostly in the context of synthetic gauge fields in optical lattices of cold atoms \cite{Wang14, Peotta15, Iskin15_2, UmucalilarIskin17, Iskin18, Maska02, ZhaiOktel10, Iskin15_1, Iskin15_2, UmucalilarIskin16,  JeonJain19, Iskin19, SohalFradkin20}. All of these works carried out numerical analyses restricted to small values of \(q\), with $\Phi = (p/q)\Phi_0$ being the flux per unit cell.
Refs. \cite{Maska02, ZhaiOktel10, SohalFradkin20} considered the role of MTG symmetries, noting that it implies the presence of order parameters with multiple finite pairing momenta similar to Fulde-Ferrell-Larkin-Ovchinnikov and pair-density wave phases \cite{ZhaiOktel10, SohalFradkin20}, and indicates translational symmetry breaking \cite{FF,LO,Hu06,AgterbergKaur07,Radzihovsky09,Radzihovsky11,ChoMoore12,Zheng13, Agterberg20}.
Furthermore, earlier numerical observation \cite{SohalFradkin20} that some magnetic translation symmetries may also be broken is corroborated by our group theory analysis that shows that at least one magnetic translation symmetry is \emph{necessarily} broken in the paired state, resulting in \(q\)- or \(q/2\)-fold degenerate ground states for odd and even \(q\) respectively \footnote{A similar result was found in \cite{PowellDasSarma11} for charge 1e condensates in synthetic magnetic fields}.
We also show that the phase relations between order parameters with different pairing momenta found in \cite{ZhaiOktel10} for \(q=3\) and \(q=4\) result from an unbroken cyclic \(\mathbb{Z}_q\) subgroup of the MTG, which also implies the coexistence of these order parameters as discussed in \cite{ZhaiOktel10, Iskin15_1, Iskin15_2, UmucalilarIskin16, UmucalilarIskin17, SohalFradkin20}. 
We establish exact results for the phase relations for all values of \(q\) and note that at least a part of the MTG is broken in the paired state. The latter is reflected in the fact that the irreducible representations (irreps) of the MTG furnished by the order parameters of the paired state are multi-dimensional.

This work is organized as follows. In Sec. \ref{section:HofstadterSystems} we set our notation, review the action of the MTG on single particle electronic states, emphasising -- see Fig. \ref{fig:qfoldFS} -- that the $q$ dimensional irrep of the single particle states gives rise to $q$ Fermi surface patches, which forms the low energy space for electronic pairing. We then discuss the constraints imposed by the MTG on interactions projected onto Hofstadter bands, thus providing an effective microscopic description for the pairing instability of Hofstadter electrons.
In Sec. \ref{section:sym}, we present a comprehensive classification of the MTG irreps furnished by the pairing matrix $\hat{\Delta}$. We show that the dimension of the irreps depends upon the parity of $q$: while the irreps furnished by the pairing order parameter have dimension \(q\) for odd \(q\), for even \(q\) there are four distinct irreps of dimension \(q/2\).
We emphasize that while we use an effective microscopic model projected to a single band for illustration purposes, our symmetry-based analysis generalizes when multiple bands are present, including the strong-coupling regime.
In particular, Sec. \ref{GL} presents a phenomenological Ginzburg-Landau theory of Hofstadter superconductors described by a multi-component order parameter, which reflects the multi-dimensionality of the irreps discussed in Sec. \ref{section:sym}. We analytically and numerically analyse the phase diagram of some representative cases, discussing their symmetry breaking patterns. In particular, we discuss a class of $\mathbb{Z}_q$-symmetric Hofstadter superconductors formed when the MTG breaks down to a discrete $\mathbb{Z}_{q}$ subgroup resulting in ground state degeneracy equal to the dimension of the irrep. The remaining subgroup may also be broken in the ground state, as we find numerically for \(q\geq 5\), further enlarging the degeneracy of the ground state.

Finally, in Sec. \ref{Chiral} we analyse the spectrum of fermionic excitations of such $\mathbb{Z}_{q}$-symmetric Hofstadter superconductors, deriving some general results about the bulk topology and nature of quantum critical points. First, we show that chiral Hofstadter superconductors provide a natural setting for realizing topological superconductivity with tunable Chern numbers. We demonstrate, under general conditions, that particle-hole and parity symmetries constraint the change of the Chern number across quantum critical points to even values $\Delta C \in 2 \mathbb{Z}$, implying the conservation of Chern number parity and therefore the non-Abelian character of bulk fermionic states.
Furthermore, we show that when Hofstadter superconductors with \(\mathbb{Z}_q\) MTG symmetry also possess parity symmetry, their spectrum  \textit{necessarily} supports Bogoliubov Fermi surfaces (BFS). For odd $q$, there is only one BFS, which is stable even if the $\mathbb{Z}_q$ symmetry is broken as long as particle-hole and parity symmetries are preserved and protected by the same topological invariant as constructed in \cite{Agterberg17}; for even $q$, there are two degenerate BFS when the gap function belongs to two out of the four irreps, but that are unstable if the $\mathbb{Z}_q$ symmetry is broken. The stability of the doubly degenerate BFS for even \(q\) is therefore protected by a new topological invariant that only exists in the presence of the $\mathbb{Z}_q$ symmetry, as we establish.

\section{Hofstadter Systems and the Magnetic Translation Group}
\label{section:HofstadterSystems}

\begin{figure*}
\centering
\includegraphics[width=0.95\textwidth]{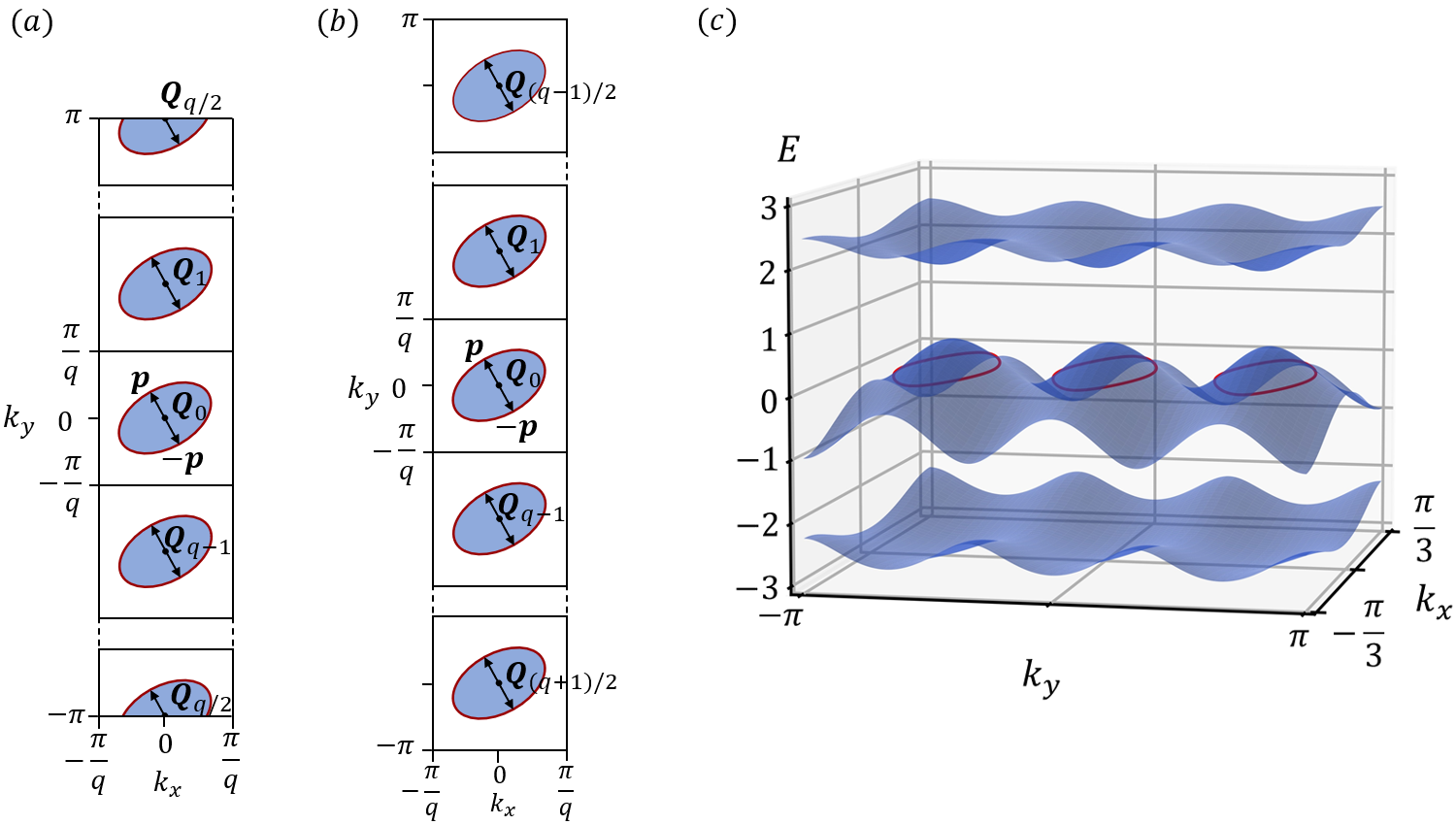}
\caption{Brillouin zone and Fermi surfaces for a square lattice for (a) even and  (b) odd \(q\), and (c) the Hofstadter bands for the original Hofstadter tight binding model on a square lattice \cite{Harper55,Hofstadter76,HasegawaWiegmann89} for \(q=3\). A single band in the absence of the magnetic field splits into \(q\) bands, \(q\)-fold symmetric under translations by \(\mathbf{Q}\). The red contours indicate the Fermi level that we take to be at \(E=0\). The Brillouin zone is folded along the \(k_x\) direction by a factor of \(q\) relative to the Brillouin zone in the absence of the magnetic field. Due to the \(\hat{T}_1\) magnetic translation symmetry, the band structure repeats \(q\) times along the \(k_y\) direction. As a result, there are \(q\) copies of a Fermi surface centered at momenta \(\mathbf{Q}_\ell=\ell\mathbf{Q}\). The interactions are projected onto the single band that crosses the Fermi level.}
\label{fig:qfoldFS}
\end{figure*}

We consider a 2D electronic system in a perpendicular magnetic field \(\mathbf{B}\) on a lattice with basis lattice vectors \(\mathbf{a}_1\) and \(\mathbf{a}_2\). We assume \(\mathbf{a}_1\) points along the \(x\)-direction. As spin will play no role in the analysis of the MTG, we will consider spin polarized fermions for simplicity, but note that the analysis applies to systems with both spins included. The primitive vectors of the reciprocal lattice are then \(\mathbf{b}_1\) and \(\mathbf{b}_2=b_2\hat{\mathbf{y}}\) and we work with the vector potential \(\mathbf{A}=x B \mathbf{a}_2/a_{2y}\) in the Landau gauge, where \(a_{2y}\) is the \(y\) component of \(\mathbf{a}_2\).
We further assume that the magnetic flux per unit cell is a rational multiple of the flux quantum: \(\Phi=\frac{p}{q}\Phi_0\), with \(\Phi_0=h/e\). The unit cell is therefore extended by a factor of \(q\) along the \(x\) axis, and the Brillouin zone is folded along the \(\hat{\mathbf{b}}_1\) direction by the same factor (see Fig.\ref{fig:qfoldFS}). After the folding, in general each energy band in the absence of the magnetic field is split into \(q\) Hofstadter sub-bands \cite{Harper55, Azbel64, Hofstadter76, Zak64_1, Zak64_2, Brown64} carrying non-trivial Chern numbers that depend on the particular lattice \cite{TKNN, Kohmoto85}.

The \(q\)-fold splitting of the bands can be understood from a symmetry perspective. The non-trivial vector potential breaks the translation symmetry \(T_1\) of the lattice along the \(x\) direction, but the system remains symmetric under \emph{magnetic} translation \(\hat{T}_1\) that is a composition of \(T_1\) and the gauge transformation \(\mathbf{A}(x)\rightarrow \mathbf{A}(x)-Ba_1\mathbf{a}_2/(2a_{2y})\). Importantly, the two magnetic translation symmetries \(\hat{T}_1\) and \(\hat{T}_2=T_2\) do not commute but satisfy \(\hat{T}_1\hat{T}_2=\omega^p_q\hat{T}_2\hat{T}_1\) where \(\omega^p_q=e^{2\pi i p/q}\) is a \(q^{th}\) root of unity. Together \(\hat{T}_1\) and \(\hat{T}_2\) generate the magnetic translation group (MTG), which includes a subgroup of \(U(1)\)  transformations generated by the commutator \(\hat{T}_1\hat{T}_2\hat{T}_1^{-1}\hat{T}_2^{-1}=\omega^p_q\). As a result of the non-commutativity of the MTG, its irreducible representations (irreps) formed by the electron states are \(q\) dimensional \cite{Zak64_1, Zak64_2, Brown64}, and as a result the bands in the absence of the magnetic field split into \(q\) sub-bands with dispersions \(\varepsilon_\alpha(\mathbf{k})\), the index \(\alpha=0,\dots,q-1\) labeling the sub-bands being defined modulo \(q\) \footnote{The irreducible representations of the MTG can also be considered as projective irreducible representations of the regular translation group.}. 
We assume that the dispersion is symmetric under inversion, \(\varepsilon_\alpha(\mathbf{k})=\varepsilon_\alpha(-\mathbf{k})\), which is necessary for the pairing instability.

We define \(d_{\alpha\mathbf{k}}\) to be the second-quantized annihilation operators corresponding to the \(\alpha\) sub-band. With our gauge choice, the quasimomentum component \(k_x\) can be restricted to \(\left(-\frac{\pi}{q},\frac{\pi}{q}\right]\) due to the folding of the Brillouin zone. Under the magnetic translations,
\begin{align} \label{dT1T2}
\hat{T}_1d_{\alpha,\mathbf{k}}\hat{T}_1^{\dagger}&=e^{-ik_x}d_{\alpha,\mathbf{k}+\mathbf{Q}}\,,\nonumber\\
\hat{T}_2d_{\alpha,\mathbf{k}}\hat{T}_2^{\dagger}&=e^{-i\mathbf{k}\cdot\mathbf{b}_2}d_{\alpha,\mathbf{k}}\,,
\end{align}
where \(\mathbf{Q}=\frac{2\pi p}{q}\hat{\mathbf{b}}_2\). Notice that \(\hat{T}_1\) therefore acts as a translation operator also on the reciprocal lattice, but along a perpendicular direction. As a result, each sub-band is \(q\)-fold degenerate within the Brillouin zone, \(\varepsilon_\alpha(\mathbf{k})=\varepsilon_\alpha(\mathbf{k}+\mathbf{Q})\), which reflects the fact that with a different gauge the unit cell could be extended along \(\mathbf{a}_2\) instead of \(\mathbf{a}_1\), with the Brillouin zone folded in the \(\hat{\mathbf{b}}_2\) direction. Since the dispersion is gauge invariant it must remain the same with both choices, and is therefore periodic in both directions (see Fig \ref{fig:qfoldFS}).
Consequently, the Brillouin zone can be further folded along \(\hat{\mathbf{b}}_2\), producing the reduced Brillouin zone in which each band is \(q\)-fold degenerate at each momentum (the unit cell in real space is correspondingly extended by a factor of \(q\times q\)). We therefore define \(d_{\alpha,\mathbf{p},\ell}=d_{\alpha,\mathbf{p}+\ell\mathbf{Q}}\) with \(\mathbf{p}\) restricted to the reduced Brillouin zone with the patch index \(\ell=0,\dots,q-1\) (defined modulo \(q\)) labeling the degeneracy. With this relabeling and using Eq.\eqref{dT1T2}, the magnetic translation symmetries act as matrices in patch indices
\begin{align}
\label{eq: dT1dT2 patch}
\hat{T}_1d_{\alpha,\mathbf{p},\ell}\hat{T}_1^{\dagger}&=(\mathcal{T}_1)_{\ell,\ell'}d_{\alpha,\mathbf{p},\ell'}\,,\nonumber\\
\hat{T}_2d_{\alpha,\mathbf{p},\ell}\hat{T}_2^{\dagger}&=(\mathcal{T}_2)_{\ell,\ell'}d_{\alpha,\mathbf{p},\ell'}
\end{align}
(with implicit summation over \(\ell'\) on the RHS), with \(\mathcal{T}_1=e^{-ip_x}\hat{\tau}\) and \(\mathcal{T}_2=e^{-i\mathbf{p}\cdot\mathbf{b}_2}\hat{\sigma}^*\), where
\begin{align}\label{tausigma}
    \hat{\tau}_{\ell,\ell'}&=\delta_{\ell,\ell'-1}\,,\nonumber\\
    \hat{\sigma}_{\ell,\ell'}&=\omega_q^{\ell p}\delta_{\ell,\ell'}
\end{align}
are the \(q\times q\) shift and clock matrices, respectively.
The transformation properties of electron operators described in Eq. \eqref{eq: dT1dT2 patch} will play a central role in understanding the properties of the superconducting state; in particular, they will lead to a general understanding of the irreps realized by the superconducting gap function to be discussed in Sec. \ref{section:sym} that we will see are distinct from the single electron irreps relevant for the Hofstadter bands constructed in  \cite{Brown64,Zak64_1,Zak64_2}.

\subsection*{Projected Interactions}

Although our symmetry analysis of superconductivity presented below is quite general and can be easily extended to include pairing between multiple bands \(\alpha\) relevant in the strong-coupling regime, since the MTG symmetries act trivially on the band index as seen in Eqs. (\ref{dT1T2}-\ref{eq: dT1dT2 patch}), it is sufficient to consider the pairing instability of the Fermi surfaces restricted to a single band \(\alpha\) in the weak-coupling regime. We therefore consider a scenario where the chemical potential lies within a single band \(\alpha\) and project general momentum-conserving pairing interaction Hamiltonian onto this band (with the $\alpha$ band index henceforth omitted)
\[\label{Hint}
H_{int}=\sum g^{(\ell)}_{n,m}(\mathbf{p;p}')d^\dagger_{\mathbf{p},\ell+n}d^\dagger_{-\mathbf{p},-n}d_{\mathbf{p}',\ell+m}d_{-\mathbf{p}',-m}\,,\]
where \(\ell, n, m=0,\dots,q-1\), with \(\ell\) labeling the total momentum of the interacting pair. The sum is over all momenta and indices. Momentum conservation means that the interactions respect the \(\hat{T}_2 =T_2 \) symmetry. However, according to Eq. \eqref{eq: dT1dT2 patch}, \(\hat{T}_1\) places the additional constraint
\begin{equation}
\label{eq: g relation under T1}
g^{(\ell)}_{n,m}=g^{(\ell+2)}_{n-1,m-1}  
\end{equation}
on the couplings \footnote{There are additional constraints from hermiticity: \(g^{(\ell)}_{n,m}(\mathbf{p;k})=g^{(\ell)*}_{m,n}(\mathbf{k;p})\). Moreover, anti-commutation relations imply that we can further take \(g^{(\ell)}_{n,m}(\mathbf{p;k})=-g^{(\ell)}_{-\ell-n,m}(\mathbf{-p;k})=-g^{(\ell)}_{n,-\ell-m}(\mathbf{p;-k})=g^{(\ell)}_{-\ell-n,-\ell-m}(\mathbf{-p;-k})\).}.
With this, the normal state is invariant under the MTG and the global \(U(1)\) transformation associated with charge conservation. As we will see in Sec. \ref{section:sym}, the fact that the MTG contains \(U(1)\) transformations implies that paired states that break the \(U(1)\) symmetry necessarily break the MTG down to a smaller subgroup.

\section{Symmetry Analysis of Pairing in Hofstadter Systems}
\label{section:sym}

We describe the state obtained by pairing of spin polarized electrons in a single Hofstadter band using the standard mean-field pairing Hamiltonian
\begin{subequations}
\begin{equation}
\label{HSC}
\begin{split}
H &=
\sum_{\ell,\mathbf{p}}\varepsilon(\mathbf{p})d^\dagger_{\mathbf{p},\ell}d_{\mathbf{p},\ell}+\frac{1}{2}\sum_{\ell,\ell',\mathbf{p}}\left[\hat{\Delta}_{\ell,\ell'}(\mathbf{p})d^\dagger_{\mathbf{p},\ell}d^\dagger_{-\mathbf{p},\ell'}+h.c.\right]\\
&=\frac{1}{2}\sum_{\ell,\ell',\mathbf{p}}\Psi_{\mathbf{p},\ell}^\dagger\left[\mathcal{H}_{BdG}(\mathbf{p})\right]_{\ell,\ell'}\Psi_{\mathbf{p},\ell'}\,, 
\end{split}    
\end{equation}
\begin{equation}
\label{BdG}  
\mathcal{H}_{BdG}(\mathbf{p})=\left(\begin{array}{cc}
     \varepsilon(\mathbf{p}) \openone_{q \times q} &  \hat{\Delta}(\mathbf{p})\\
     \hat{\Delta}^\dagger(\mathbf{p}) & -\varepsilon(-\mathbf{p}) \openone_{q \times q}
\end{array}\right)\,,
\end{equation}
\end{subequations}
where \(\varepsilon(\mathbf{p})\) is the electron dispersion and 
\(\hat{\Delta}(\mathbf{p})\) are gap functions that are \(q\times q\) matrices. \(\mathcal{H}_{BdG}\) is the Bogoliubov-de Gennes (BdG) Hamiltonian in the basis of Nambu spinors \(\Psi_{\mathbf{p},\ell}=(d_{\mathbf{p},\ell},d_{-\mathbf{p},\ell}^\dagger)\). The unphysical redundancy of the BdG formalism is encoded in the anti-unitary particle hole symmetry (PHS) of the BdG Hamiltonian, which acts as \(\mathcal{C}=\tau^x\mathcal{K}\) on the Nambu spinors where \(\tau^j\) are Pauli matrices acting on the particle/hole sectors and \(\mathcal{K}\) is complex conjugation. Under the PHS, \(\mathcal{C}^{-1}\mathcal{H}_{BdG}(\mathbf{p})\mathcal{C} = -\mathcal{H}_{BdG}(-\mathbf{p})\), which implies that \(\hat{\Delta}(\mathbf{p}) = -\hat{\Delta}^T(-\mathbf{p})\) consistent with anti-commutation relations.

In the ground state, the gap functions further satisfy the self-consistent gap equation obtained by minimizing the free energy. Close below the transition temperature the gap function is small and approximately satisfies the linearized gap equation
\[\hat{\Delta}_{\ell+n,-n}(\mathbf{p})=-\nu\log\frac{1.13\Lambda}{T}\sum_{\mathbf{p}'m}g^{(\ell)}_{n,m}(\mathbf{p;p}')\hat{\Delta}_{\ell+m,-m}(\mathbf{p}')\,,\label{LinGapEq}\]
where \(\nu\) is the density of states per patch at the Fermi level and \(\Lambda\) is the high energy cutoff (see Appendix \ref{A} for derivation). Notice there are $q$ patches, as shown Fig. \ref{fig:qfoldFS}, so that the total density of states is $q\,\nu$.

The solutions of the linearized gap equation can be classified by the irreducible representations (irreps) of the symmetry groups of the system according to which they transform \cite{SigristUeda91}. Our goal in what follows is therefore to determine the transformation properties of the gap functions under the MTG symmetries, and thus the irrep of the MTG the gap functions belong to. Although the irreps of the MTG formed by \textit{single} electron Bloch states are well-known to be \(q\) dimensional \cite{Brown64,Zak64_1,Zak64_2}, we show that the \textit{paired} states transform according to different irreps depending on the parity of \(q\) \cite{Florek97}: a single irrep of dimension \(q\) for odd \(q\), or four irreps of dimension \(q/2\) for even \(q\). As a result, symmetry dictates that the linearized gap equation falls apart into \(q\) or \(q/2\) independent equations for odd and even \(q\) respectively, with solutions that have the same \(T_c\) \cite{SigristUeda91}. Any linear combination of these independent solutions remains a solution of the linearized gap equation, but this degeneracy is lifted by non-linear terms in the full gap equation or, equivalently, by higher order terms in the free energy that we study in Sec. \ref{GL}.

In order to address these questions, we first review the general symmetry action on the BdG Hamiltonian in Sec. \ref{BdGSym}. We will then utilize this formalism in Sec. \ref{section:irreps} to address the role of MTG on the pairing Hamiltonian and construct the corresponding irreps furnished by the gap functions.

\subsection{Symmetry Action on the Gap Function and the Linearized Gap Equation}\label{BdGSym}

We now review how symmetries other than the PHS act on the BdG Hamiltonian and in particular how they act on the gap functions \(\hat{\Delta}\) \cite{SigristUeda91,SatoAndo17,Ono19}.
First, recall a \(U(1)\) symmetry \(U(\theta)\) acts on the annihilation operators as \(d_{\mathbf{p},\ell}\rightarrow e^{i\theta}d_{\mathbf{p},\ell}\). By requiring that the Hamiltonian in Eq. (\ref{BdG}) is invariant under this transformation, we conclude that the gap function transforms as
\[\hat{\Delta}(\mathbf{p})\xrightarrow[]{U(\theta)} e^{2i\theta}\hat{\Delta}(\mathbf{p})\,,\label{U1}\]
which implies that the \(U(1)\) symmetry is broken down to a \(\mathbb{Z}_2\) symmetry in the SC state.

More generally, suppose that \(S(\mathbf{p})\) is some (possibly momentum dependent) unitary matrix representing a symmetry acting on the normal state Hamiltonian. In the \textit{normal state}, we then have a family of symmetries \(S(\mathbf{p},\theta_0)=U(\theta_0/2)S(\mathbf{p})\) parametrized by \(\theta_0\). Since the gap function breaks the \(U(1)\) symmetry, at most one member of this family may remain unbroken, and so we have to consider each possibility in our analysis. By the same argument as for the \(U(1)\) symmetry itself, we find that \(S(\mathbf{p},\theta_0)\) acts on the gap function as
\[\hat{\Delta}(\mathbf{p})\xrightarrow[]{S(\theta_0)} e^{i\theta_0}S(\mathbf{p})\hat{\Delta}(\mathbf{p})S^T(-\mathbf{p})\label{DeltaTrans}\,.\]
Requiring the gap function to be invariant under this symmetry, we note that the phase \(\theta_0\) is not arbitrary but instead determined by the possible solutions of 
\begin{equation}
\label{eq: S transformation of delta}
S(\mathbf{p})\hat{\Delta}(\mathbf{p})S^T(-\mathbf{p})=e^{-i\theta_0}\hat{\Delta}(\mathbf{p}) \,.
\end{equation}
Typically there is only a finite set of allowed \(\theta_0\), which determine the irrep to which \(\hat{\Delta}\) belongs. In particular, \(e^{i\theta_0}\) determine the characters of the irrep \cite{Ono19}, and for finite irreps there is only a finite set of characters. For example, if \(S^q(\mathbf{p})=1\), we must have \(\theta_0=\frac{2\pi}{q}N\) with \(N\) being some integer. This is the case for the MTG symmetries: \(\hat{T}_1^q=\hat{T}_2^q=1\). Using Eqs. (\ref{eq: dT1dT2 patch}) and (\ref{DeltaTrans}), we find that under \(\hat{T}_1(\theta_1)=U(\theta_1/2)\hat{T}_1\) and \(\hat{T}_2(\theta_2)=U(\theta_2/2)\hat{T}_2\) the gap function transforms as
\begin{subequations}
\begin{align}
    \hat{\Delta}&\xrightarrow[]{\hat{T}_1(\theta_1)}e^{i\theta_1}\hat{\tau}\hat{\Delta}\hat{\tau}^T\label{DeltaT1mat}\,, \\
    \hat{\Delta}&\xrightarrow[]{\hat{T}_2(\theta_2)}e^{i\theta_2}\hat{\sigma}\hat{\Delta}\hat{\sigma}\,,
    \label{DeltaT2mat}
\end{align}
\end{subequations}
where \(\hat{\tau}\) and \(\hat{\sigma}\) are the shift and clock matrices defined in Eq. (\ref{tausigma}). More explicitly, the elements of \(\hat{\Delta}\) transform as
\begin{subequations}
\begin{align}
    \hat{\Delta}_{\ell,\ell'}&\xrightarrow[]{\hat{T}_1(\theta_1)}e^{i\theta_1}\hat{\Delta}_{\ell-1,\ell'-1}\,,\label{DeltaT1}\\
    \hat{\Delta}_{\ell,\ell'}&\xrightarrow[]{\hat{T}_2(\theta_2)}e^{i\theta_2}\omega^{p(\ell+\ell')}_q\hat{\Delta}_{\ell,\ell'}\label{DeltaT2}\,.
\end{align}
\end{subequations}
As mentioned above, \(\theta_1\) and \(\theta_2\) in Eqs. (\ref{DeltaT1}-\ref{DeltaT2}) are restricted to the values:
\begin{subequations}
\begin{align}
\theta_1&=-\frac{2\pi p}{q}M\label{theta1}\,,\\
\theta_2&=-\frac{2\pi p}{q}L\label{theta2}\,,
\end{align}
\end{subequations}
where \(M, L\) are integers defined modulo \(q\). In particular, note that \(e^{i\theta_j}\) (\(j=1,2\)) are actually elements of the MTG since
\(\hat{T}_2\hat{T}_1\hat{T}_2^{-1}\hat{T}_1^{-1}=U(-2\pi p/q)\), a \(U(1)\) transformation acting as Eq. (\ref{U1}) on the gap function.

The first important conclusion is that there is no non-zero \(\hat{\Delta}\) that satisfies both (\ref{DeltaT1}) and (\ref{DeltaT2}) for \(q>2\) (\(q=2\) is an exception, as we will discuss below), for any choice of \(\theta_1\) and \(\theta_2\). This means that any pairing order \emph{necessarily} breaks at least some MTG symmetries. A similar result was shown for the superfluid bosonic condensate in a strong magnetic field, essentially the bosonic version of the Hofstadter model, in Ref.~\cite{PowellDasSarma11}. We note that the transformation properties of the bosonic fields considered in Ref.~\cite{PowellDasSarma11} under the MTG differ from Eqs. (\ref{DeltaT1}-\ref{DeltaT2}), i.e. they belong to a different irreducible representation (see Sec. \ref{section:irreps} below). The essential reason for the symmetry breaking is however the same: in both cases, either the superfluid or superconducting condensates break the \(U(1)\) symmetry, and since the MTG contains a subgroup of \(U(1)\), it too must be broken. The fact that the MTG contains a subgroup of \(U(1)\), in turn, is a consequence of the non-trivial commutation relations of the magnetic translations.

Before moving on to the analysis of the irreps realized by the gap function in the next subsection, let us consider the implications of the MTG symmetries for the linearized gap equation Eq. (\ref{LinGapEq}), which reads:
\[\hat{\Delta}_{\ell+n,-n}(\mathbf{p})=-\nu\log\frac{1.13\Lambda}{T}\sum_{\mathbf{p}'m}g^{(\ell)}_{n,m}(\mathbf{p;p}')\hat{\Delta}_{\ell+m,-m}(\mathbf{p}')\,.\]
The first observation is that only gap function elements \(\hat{\Delta}_{\ell,\ell'}\) with the same value of \(L=\ell+\ell'\) appear on both sides of the equation, and, as a result, the linearized gap equation splits into \(q\) independent equations for each value of \(L\), which corresponds to the \(L^{th}\) anti-diagonal of the matrix \(\hat{\Delta}\) and labels the total momentum of the Cooper pairs along the \(\hat{\mathbf{b}}_2\) direction. 
This decoupling is a consequence of momentum conservation along that direction due to the \(\hat{T}_2=T_2\) symmetry. 
As such, for a particular \(L\), the decoupled solution is a matrix with non-zero elements only along the \(L^{th}\) anti-diagonal:
\[\hat{\Delta}^{(L)}=\left(\begin{array}{cccccccc}
    & & & &\hat{\Delta}_{0L} &  \\
     & & & \hat{\Delta}_{1,L-1} & & \\
    &  \iddots & &  & & \\
    \hat{\Delta}_{L0}  \\
    & & & & & &&\hat{\Delta}_{L+1,q}\\
     & & & &&&  \iddots &\\
      & & &&& \hat{\Delta}_{q,L+1}\\
\end{array}\right)\,.\label{DeltaL}\]
We refer to such a matrix as an \(L^{th}\) anti-diagonal matrix. 
From Eq. (\ref{DeltaT2}),
\[\hat{\Delta}^{(L)}\xrightarrow[]{\hat{T}_2(\theta_2)}e^{i\theta_2}\omega^{pL}_q\hat{\Delta}^{(L)}\]
thus enforcing the condition \(e^{i\theta_2}=\omega^{-pL}_q\), 
which identifies the \(L\) index in Eq. (\ref{theta2}) with the momentum of the Cooper pair. As such,
\(\hat{\Delta}^{(L)}\) are precisely the gap functions symmetric under \(\hat{T}_2(\theta_2)\) symmetry with \(\theta_2=-\frac{2\pi p}{q}L\). Conversely, \(\hat{\Delta}^{(L)}\) break \(\hat{T}_2(\theta_2)\) for any other value of \(\theta_2\) and they also break \(\hat{T}_1(\theta_1)\) for any value of \(\theta_1\) (with the exception of the case \(q=2\) discussed below).

We will consider gap functions symmetric under \(\hat{T}_1(\theta_1)\) symmetries in Sec. \ref{GL} when we study the effect of non-linear terms in the gap equation, but for the purposes of the linearized gap equation it is sufficient to look at \(\hat{\Delta}^{(L)}\). The effect of the \(\hat{T}_1\) symmetry on \(\hat{\Delta}^{(L)}\) is to shift it to \(\hat{\Delta}^{(L-2)}\).
Notice that the matrices \(\hat{\Delta}^{(L)}\) defined in Eq.\eqref{DeltaL} have well defined transformation under $\hat{T}_2$. However, the phases between  \(\hat{\Delta}^{(L)}\) and \(\hat{\Delta}^{(L')}\) with \(L'\neq L\) are arbitrary. We then fix these phases by defining \(\hat{\Delta}^{(L)}\) such that
\begin{subequations}
\begin{align}
    \hat{\Delta}^{(L)}&\xrightarrow[]{\hat{T}_1(0)}\hat{\Delta}^{(L-2)}\label{DeltaLT1}\,,\\
    \hat{\Delta}^{(L)}&\xrightarrow[]{\hat{T}_2(0)}\omega^{pL}_q\hat{\Delta}^{(L)}\label{DeltaLT2}\,.
\end{align}
\end{subequations}
Eqs. (\ref{DeltaT1},\ref{DeltaLT1}) in particular imply that \(\hat{\Delta}^{(L)}_{\ell,\ell'}=\hat{\Delta}^{(L-2)}_{\ell-1,\ell'-1}\).

At this stage, it becomes necessary to distinguish between the cases of even and odd \(q\). In the latter case, note that applying \(\hat{T}_1\) to \(\hat{\Delta}^{(0)}\) generates all of \(\hat{\Delta}^{(L)}\) since \(L+1\equiv L+2\frac{q+1}{2}\) mod \(q\); the parity of \(L\), in other words, is not well-defined 
\footnote{For example, for $q=3$, the $L = 0, 1, 2$ values, under the action of $T_{1}$, are cycled as $0 \rightarrow 2 \rightarrow 1 \rightarrow 0 \rightarrow ...$; for $q=5$, $0 \rightarrow 2 \rightarrow 4 \rightarrow 1 \rightarrow 3
\rightarrow 0
\rightarrow ..., etc.$}.
For odd \(q\), we therefore define
\[\Delta_{\ell}\equiv\hat{\Delta}^{(L)}_{\left[\frac{L+\ell}2\right]_q,\left[\frac{L-\ell}2\right]_q}\label{DeltaEllOdd}\]
that are independent of \(L\). Here \([\ell/2]_q=\ell/2\) if \(\ell\) is even but \([\ell/2]_q=(\ell+q)/2\) if \(\ell\) is odd. As a concrete example, for \(q=3\) there is a single irrep with three irrep components \(\hat{\Delta}^{(L)}\) given by
\begin{widetext}
\[\begin{array}{ccc}
    \hat{\Delta}^{(0)}=\left(\begin{array}{ccc}
    \Delta_0 & 0 & 0 \\
    0 & 0 & \Delta_2\\
    0 & \Delta_1 & 0
\end{array}\right), &\hat{\Delta}^{(1)}=\left(\begin{array}{cccc}
    0 & 0  & \Delta_1\\
    0 & \Delta_0 & 0\\
     \Delta_2 & 0 & 0
\end{array}\right), &\hat{\Delta}^{(2)}=\left(\begin{array}{cccc}
    0 & \Delta_2  & 0\\
    \Delta_1 & 0 & 0\\
    0 & 0 & \Delta_0
\end{array}\right)
\end{array}\,.
\label{q3irrep}\]
\end{widetext}

For even \(q\), however, \(\hat{\Delta}^{(L)}\) are not all generated by applying \(\hat{T}_1\) to \(\hat{\Delta}^{(0)}\); rather, \(\hat{\Delta}^{(L)}\) split into two groups for even and odd \(L\),
\begin{equation}
\label{eq: even/odd Delta sequences}
\begin{split}
&\,
\hat{\Delta}^{(0)} \xrightarrow[]{\hat{T}_1(0)}\hat{\Delta}^{(q-2)}
\xrightarrow[]{\hat{T}_1(0)}\hat{\Delta}^{(q-4)}
\xrightarrow[]{\hat{T}_1(0)}...
\\
&\,
\hat{\Delta}^{(1)} \xrightarrow[]{\hat{T}_1(0)}\hat{\Delta}^{(q-1)}
\xrightarrow[]{\hat{T}_1(0)}\hat{\Delta}^{(q-3)}
\xrightarrow[]{\hat{T}_1(0)}...
\end{split}    
\end{equation}
that are not mapped to each other by \(\hat{T}_1\) or any other MTG symmetry (in Sec.\ref{section:irreps} we show that each of these two groups actually splits into two more, resulting in 4 irreps for even q).
For even \(q\) we therefore define
\[\Delta_{\ell}\equiv\hat{\Delta}^{(L)}_{\frac{L+\ell}2,\frac{L-\ell}2}\label{DeltaEllEven}\]
but with \(\ell\) defined modulo \(2q\) and restricted to be of the same parity as \(L\), i.e. \(\ell=0,2,\dots,2(q-1)\) for even \(L\) and \(\ell=1,3,\dots,2q-1\) for odd \(L\). With the indices \(\frac{L\pm\ell}2\) defined modulo \(q\) this can be seen to properly index all the elements of \(\hat{\Delta}^{(L)}\). As another concrete example, for \(q=2\) we have
\[\hat{\Delta}^{(0)}=\left(\begin{array}{cc}
    \Delta_0 & 0 \\
    0 & \Delta_2
\end{array}\right)\label{q2irrep0}\]
and
\[\hat{\Delta}^{(1)}=\left(\begin{array}{cc}
    0 & \Delta_3 \\
   \Delta_1 & 0
\end{array}\right)\,.\label{q2irrep1}\]

In terms of the functions \(\Delta_{\ell}\), the linearized gap equation for each value of \(L\) becomes
\[\Delta_{\ell}(\mathbf{p})=-\nu\log\frac{1.13\Lambda}{T}\sum_{\mathbf{p}'\ell'}g^{(L)}_{\left[\frac{\ell-L}2\right]_q,\left[\frac{\ell'-L}2\right]_q}(\mathbf{p;p}')\Delta_{\ell'}(\mathbf{p}')\,.\label{ReducedLinGapEq}\]
Because the interactions satisfy Eq. (\ref{eq: g relation under T1}) due to the \(\hat{T}_1\) symmetry, for odd \(q\) we can take \(g^{(L)}_{[(\ell-L)/2]_q,[(\ell'-L)/2]_q}=g^{(0)}_{[\ell/2]_q,[\ell'/2]_q}\). Recall that \([\ell/2]_q=\ell/2\) if \(\ell\) is even but \([\ell/2]_q=(\ell+q)/2\) if \(\ell\) is odd; note that for even \(L\), \(\ell\) is also even. For even \(q\) and even \(L\) we can take \(g^{(L)}_{(\ell-L)/2,(\ell'-L)/2}=g^{(0)}_{\ell/2,\ell'/2}\); for odd \(L\), we instead take \(g^{(L)}_{(\ell-L)/2,(\ell'-L)/2}=g^{(1)}_{(\ell-1)/2,(\ell'-1)/2}\). Recall that for even \(q\), \(\ell\) and \(\ell'\) are both defined modulo \(2q\) and restricted to have the same parity as \(L\).

The second important conclusion that follows from these considerations of \(\hat{T}_1\) symmetry is that, for odd \(q\), each of the \(L\) equations Eq. (\ref{ReducedLinGapEq}) that the linearized gap equation Eq. (\ref{LinGapEq}) decouples into are the \emph{same} equation; for even \(q\), the equations are the same for a given parity of \(L\) but may be different for even and odd \(L\). This means in particular that the \(T_c\) determined by these equations is also the same, and moreover any linear combination
\[\hat{\Delta}=\sum_L\eta_L\hat{\Delta}^{(L)}
\label{eq: irrep expansion}\]
with some complex coefficients \(\eta_L\) is another solution of the linearized gap equation. As we will see in Sec. \ref{GL}, this includes gap functions symmetric under \(\hat{T}_1(\theta_1)\) and indeed any other element of the MTG symmetries. This large degeneracy of solutions of the linearized gap equation is lifted when non-linear terms are included, as we also show in Sec. \ref{GL}. The higher order terms thus determine which MTG symmetries, if any, remain unbroken in the ground state.

\subsection{Gap Functions as Irreducible Representations of the MTG}
\label{section:irreps}

The symmetry analysis of the previous section shows that pairing matrix \(\hat{\Delta}\) belongs to a multidimensional irrep, with \(\hat{\Delta}^{(L)}\) forming the components of the irrep, as described in Eq.~\eqref{eq: irrep expansion}. The dimension of the irrep, i.e. the number of its components, is thus \(q\) for odd \(q\) and \(q/2\) for even \(q\). The fact that the linearized gap equation Eq. (\ref{LinGapEq}) decouples into \(q\) equations, all or half of which are degenerate (i.e. have the same \(T_c\)) for odd and even \(q\) respectively, is a general consequence of this fact. The existence of non-trivial irreps implies that \(\hat{\Delta}\) must break at least part of the MTG symmetries,
as only gap functions belonging to a trivial irrep, which is one dimensional, respect all the symmetries. 
While the analysis of irreps of the MTG acting on single particle Bloch states is well-known \cite{Brown64, Zak64_1, Zak64_2}, a comprehensive study of these irreps in the context of charge $2$e condensates is lacking. The goal of this section is therefore to classify pairing functions that furnish the irreps of the MTG. In the course of this analysis, we will encounter an interesting dependence on dimension of the irreps as a function of parity of $q$, which fundamentally contrasts with the well-known irreps furnished by one particle states.

A representation is a group homomorphism \(\Gamma:G\rightarrow GL(V)\) between the symmetry group \(G\) -- the MTG in our case -- and the group of linear transformations \(GL(V)\) that act on a particular vector space \(V\). For irreps realized by \(\hat{\Delta}\), \(\hat{\Delta}^{(L)}\) form the basis of this vector space. The linear transformations in question, \(\Gamma(S)\) for an element \(S\) of the MTG, act on the gap function via Eq. (\ref{DeltaTrans}). 
The representation is reducible if there is a non-trivial subspace of \(V\) that is mapped to itself under the action of any symmetry operation. To clarify the terminology, we say that \(\hat{\Delta}\) belongs to, realizes, or transforms as the irrep, and that \(\hat{\Delta}^{(L)}\) are the components of the irreps. The homomorphism is always understood to be the one determined by Eq. (\ref{DeltaTrans}). 
For the components $\hat{\Delta}^{(L)}$ of the irrep, the transformations are explicitly given by Eqs. \eqref{DeltaLT1} and \eqref{DeltaLT2}.

In order to explicitly construct an irrep, the general procedure \cite{Brown64} is to start with some fixed gap function \(\hat{\Delta}^{(0)}\) symmetric under a particular symmetry, \(\hat{T}_2(0)\) in our case, and then apply all other symmetries to obtain the vector space and the rest of the irrep components \(\hat{\Delta}^{(L)}\). In our case it was sufficient to apply \(\hat{T}_1(0)\), and we thus proved in Sec. \ref{BdGSym} that the representation realized by \(\hat{\Delta}^{(L)}\) with \(L=0,1,\dots,q-1\) and satisfying the defining relations Eqs. (\ref{DeltaLT1}-\ref{DeltaLT2}) is irreducible of dimension \(q\) for odd \(q\).
However, in the even \(q\) case, the representation reduces to irreps of dimension \(q/2\) with a basis still given by \(\hat{\Delta}^{(L)}\), but with \(L\) restricted to be even or odd. In what follows, we provide additional information to fully specify the irreps for even \(q\).

Before discussing the irreps for even \(q\) in more detail, let us note that the fact that they are \(q/2\) dimensional irreps is in contrast with the earlier result found by Brown and Zak in Refs. \cite{Brown64, Zak64_1, Zak64_2}. The reason for this difference is that the MTG contains a subgroup of the \(U(1)\) gauge symmetry that gives rise to charge conservation; therefore, in order to specify the irrep of the MTG, it is necessary to specify the irrep of the \(U(1)\) subgroup, which is equivalent to specifying the charge of the particle modulo \(q\). In Refs. \cite{Brown64, Zak64_1, Zak64_2}, the authors were interested in Bloch states, i.e. single particle states with elementary charge. They therefore required that the MTG element
\[\hat{L}=\hat{T}_2\hat{T}_1\hat{T}_2^{-1}\hat{T}_1^{-1}=U(-2\pi p/q)\]
is represented by \(\Gamma(U(-2\pi p/q))=e^{\frac{2\pi i p}{q}}\). This is simply the Aharonov-Bohm phase picked up by an electron moving around a loop encircling the original, non-magnetic, unit cell described by the operation \(\hat{L}\). Applying \(\hat{L}\) on the gap function via Eq. (\ref{DeltaTrans}), however, yields a different result, since from Eq. (\ref{U1}) we have
\[\hat{\Delta}^{(L)}\xrightarrow[]{\hat{T}_2\hat{T}_1\hat{T}_2^{-1}\hat{T}_1^{-1}}e^{-2\frac{2\pi p}{q}i}\hat{\Delta}^{(L)}\label{vorticity}\,,\]
consistent also with Eqs. (\ref{DeltaT1}-\ref{DeltaT2}), and corresponding to an Aharonov-Bohm phase picked up by a particle of charge \(2e\). This is of course as one would expect for a Cooper pair.

Although Refs. \cite{Brown64, Zak64_1, Zak64_2} rejected irreps with \(\Gamma(U(\theta))=e^{iQ\theta}\) with \(Q\neq1\) as unphysical, here we find that they naturally correspond to irreps realized by condensates of charge \(Q\). Indeed, there always exists the trivial irrep under which the MTG elements are mapped to the identity, and it corresponds to \(Q=0\). This general observation has been made earlier in Ref. \cite{Florek97} in the context of states of pairs of electrons, though it did not explicitly discuss superconductivity. The main conclusion is that the irreps realized by the gap functions \(\hat{\Delta}\) cannot be classified by the same irreps as considered in Refs. \cite{Brown64, Zak64_1, Zak64_2} in the context of single particle states. This is also the main difference between the fermionic Hofstadter SC problem and the bosonic Hofstadter superfluid considered in Refs. \cite{Balents05,PowellDasSarma10,PowellDasSarma11, NatuDasSarma16, Song19}.

We also note that Eq. (\ref{vorticity}) implies that the phase of \(\hat{\Delta}^{(L)}\) winds as one goes around the non-magnetic unit cell, indicating that the Hofstadter SC phase is a vortex lattice. 
Thus, our irrep analysis provides a general framework that justifies the numerical observation of vortex lattices in fermionic \cite{ZhaiOktel10,SohalFradkin20} and bosonic \cite{PowellDasSarma11} Hofstadter systems.
Unlike regular Abrikosov vortices, however, note that \(\hat{\Delta}^{(L)}\) need not vanish anywhere in space. A similar phenomenon occurs in Josephson vortices that also have a non-vanishing gap in their cores \cite{BlatterLarkin94}. In the limit \(q\rightarrow\infty\), we expect the vortex lattice to approach the Abrikosov vortex lattice at the upper critical field \(H_{c2}\).

\subsubsection*{Irreps for Even \(q\)}

Although the MTG irreps realized by \(\hat{\Delta}^{(L)}\) are distinct from the single particle irreps of the MTG, for odd \(q\) they are qualitatively similar as they are of the same dimension. For even \(q\), on the other hand, we saw that the parity of \(L\) is well defined, and \(\hat{\Delta}^{(L)}\) split into two \(q/2\) dimensional irreps for each parity of \(L\), as in Eq. \eqref{eq: even/odd Delta sequences}. The irreps for even $q$ are thus qualitatively different. As we will now show, there are in fact four such irreps. One can anticipate that there must be four \(q/2\) irreps from a version of Schur's orthogonality relations proven in Refs. \cite{Brown64,Florek97} that state that the sum of the squares of the dimensions of all distinct irreps for a fixed value of the \(U(1)\) charge must equal the order of the MTG, i.e. the number of its elements, that is \(q^2\). There is thus only one irrep for odd \(q\), but there must be four \(q/2\)-dimensional irreps for even \(q\), as shown also in Ref. \cite{Florek97}.

The reason there are additional irreps is that there is an additional symmetry that may remain unbroken by the paired state. To see this, note that applying \(\hat{T}_1(\theta_1)\) \(q/2\) times on \(\hat{\Delta}\) brings an element on the \(L^{th}\) anti-diagonal to the same anti-diagonal but not to the same diagonal, as can be seen from Eq. (\ref{DeltaT1}):
\[\hat{\Delta}_{\ell,\ell'}\xrightarrow[]{\hat{T}_1^{q/2}(\theta_1)}e^{i\theta_1q/2}\hat{\Delta}_{\ell-q/2,\ell'-q/2}
\,.\]
The element on the RHS is on the same anti-diagonal since \(\ell-q/2+\ell'-q/2\equiv \ell+\ell'\) mod \(q\), but it is not the same element since \(q/2\neq 0\) mod \(q\). Note that from Eq. (\ref{theta1}) we have \(e^{i\theta_1q/2}=(-1)^{pM}=(-1)^{M}\) (using the fact that \(p\) cannot be even if \(q\) is even), consistent with the fact that \(\hat{T}_1^{q/2}\) squares to the identity. The parity of \(M\) provides the additional character, in addition to the parity of \(L\), that yields the four \(q/2\)-dimensional irreps as claimed.

We label the irrep components 
\(\hat{\Delta}^{(L,\pm)}\), and these satisfy
\[\hat{\Delta}^{(L,\pm)}_{\ell-q/2,\ell'-q/2}=\pm \hat{\Delta}^{(L,\pm)}_{\ell,\ell'}\,.\label{DeltaPM}\]
This implies that the functions \(\Delta_\ell\) in Eq. (\ref{DeltaEllEven}) additionally satisfy \(\Delta_{\ell+q}=\pm\Delta_\ell\) (recall that \(\ell\) in this case is defined modulo \(2q\)). In addition to having only a single anti-diagonal, \(\hat{\Delta}^{(L,\pm)}\) has a \(\frac{q}{2}\times\frac{q}{2}\) block structure:
\[\hat{\Delta}^{(L,\pm)}=\left(\begin{array}{cc}
   \hat{A}  & \hat{B} \\
   \pm \hat{B}  & \pm\hat{A}
\end{array}\right)\,,\label{ABblocks}\]
where \(\hat{A}\) and \(\hat{B}\) are \(\frac{q}{2}\times\frac{q}{2}\) matrices. Notice that \(\hat{\Delta}^{(L,+)\dagger}\hat{\Delta}^{(L,-)}\) is odd under \(\hat{T}_1^{q/2}\) and so such terms are not allowed in the free energy (also their trace vanishes), as expected for products of elements from different irreps. This implies in particular that \(\hat{\Delta}^{(L,+)}\) and \(\hat{\Delta}^{(L,-)}\) decouple in the linearized gap equation and in general have different critical temperatures.

To illustrate the irreps and the additional symmetry, it is helpful to consider again the special case of \(q=2\). The relation Eq. (\ref{DeltaPM}) places an additional constraint on the gap functions in Eqs. (\ref{q2irrep0}-\ref{q2irrep1}), so
the components of the two irreps symmetric under \(\hat{T}_2(0)\) are
\[\hat{\Delta}^{(0,\pm)}=\left(\begin{array}{cc}
    \Delta_0 & 0 \\
    0 & \pm\Delta_0
\end{array}\right)\label{q2irrep0pm}\,,\]
while the  components of the two irreps anti-symmetric under \(\hat{T}_2(0)\) (symmetric under \(\hat{T}_2(\pi)\)) are
\[\hat{\Delta}^{(1,\pm)}=\left(\begin{array}{cc}
    0 & \pm\Delta_1 \\
   \Delta_1 & 0
\end{array}\right)\label{q2irrep1pm}\,.\]
The \(\pm\) corresponds to gaps symmetric or anti-symmetric under \(\hat{T}_1(0)\) respectively.

A more generic example is provided by \(q=4\), for which the irrep components are given by
\begin{widetext}
\begin{align}
    &\hat{\Delta}^{(0,\pm)}=\left(\begin{array}{cccc}
    \Delta_0 & 0 & 0 & 0\\
    0 & 0& 0 & \Delta_2\\
    0 & 0 & \pm \Delta_0 & 0\\
    0 & \pm \Delta_2 & 0 & 0
\end{array}\right), &\hat{\Delta}^{(2,\pm)}=\left(\begin{array}{cccc}
    0 & 0 &  \pm\Delta_2 & 0\\
    0 & \Delta_0& 0 & 0\\
    \Delta_2 & 0 & 0 & 0\\
    0 & 0 & 0 & \pm \Delta_0
\end{array}\right), \nonumber\\
&\hat{\Delta}^{(1,\pm)}=\left(\begin{array}{cccc}
    0 & \Delta_3 & 0 & 0\\
    \Delta_1 & 0& 0 & 0\\
    0 & 0 & 0 & \pm \Delta_3\\
    0 & 0 & \pm \Delta_1 & 0
\end{array}\right), &\hat{\Delta}^{(3,\pm)}=\left(\begin{array}{cccc}
    0 & 0 &  0 & \pm\Delta_1\\
    0 & 0 &  \Delta_3 & 0\\
    0 & \Delta_1 & 0 & 0\\
    \pm \Delta_3 & 0 & 0 & 0
\end{array}\right)\,,
\label{q4irrep}
\end{align}
\end{widetext}
which correspond to states symmetric under \(\hat{T}_2(\theta_2)\) with \(\theta_2=\frac{\pi p}{2} L\) and symmetric/anti-symmetric under \(\hat{T}_1^2(0)\) for \(\pm\) respectively (\(p\) can only be \(1\) or \(3\) in this case). Note the block structure that is in agreement with Eq. (\ref{ABblocks}) that also holds for the \(q=2\) case.

We stress that the irrep construction presented here for both even and odd \(q\) can be applied without change if \(\hat{\Delta}\) carries additional indices (e.g. spin or band indices), as long as the MTG symmetries do not act on these indices. The construction is therefore quite general, and additional lattice symmetries can be included in a straightforward way (see Appendix \ref{B}). It is however not unique since the irrep is multi-dimensional: instead of irrep components symmetric with respect to \(\hat{T}_2\), we could have worked with irrep components that are symmetric with respect to \(\hat{T}_1\), or, in general, any \(q\) linearly independent combinations of \(\hat{\Delta}^{(L)}\). We will consider such combinations in Sec. \ref{GL} when studying the minima of the Ginzburg-Landau free energy.

\section{Effective Ginzburg-Landau Theory}\label{GL}

As discussed in Sec. \ref{section:sym}, the gap functions that solve the linearized gap equation are arbitrary linear combinations of the irrep components
\[\hat{\Delta}(\mathbf{p})=\sum_{L}\eta_L \hat{\Delta}^{(L)}(\mathbf{p})\label{DeltaEtaL}\,.\]
The irreps are \(q\) dimensional for odd \(q\), with \(L=0,\dots,q-1\) and \(\hat{\Delta}^{(L)}\) satisfying the defining properties given in Eq. (\ref{DeltaLT1}-\ref{DeltaLT2}). For even \(q\), we assume that one of the four \(q/2\)-dimensional irreps has the highest \(T_c\) so that the rest can be neglected, and we therefore drop the \(\pm\) in \(\hat{\Delta}^{(L,\pm)}\) and restrict \(L\) to be even or odd.

The degeneracy of the linearized gap equation is, however, lifted by even infinitesimal non-linear terms, resulting in spontaneous symmetry breaking of the MTG symmetries. In order to study this symmetry breaking, here we use the irreps to construct a phenomenological effective Ginzburg-Landau free energy, following the same procedure used for unconventional superconductors \cite{SigristUeda91}. In the GL theory, we ignore the microscopic details, encoded in the functional form of \(\hat{\Delta}^{(L)}(\mathbf{p})\) determined ultimately by the interactions through the gap equation, and write the most general form of the free energy for the complex fields \(\eta_L\) constrained by the \(U(1)\) and MTG symmetries. The vector of complex numbers
\[\boldsymbol{\eta}=(\eta_0,\dots,\eta_{q-1})\]
constitutes the order parameter of the paired state (for even \(q\), we take \(\boldsymbol{\eta}\) to be a \(q/2\) component vector instead but use the same notation below). We emphasize that while we originally obtained the microscopic gap functions \(\hat{\Delta}^{(L)}\) in the context of a weak-coupling pairing within a single Hofstadter band, the effective GL theory presented here is insensitive to the details of the microscopic theory and therefore remains valid both in the presence of additional degrees of freedom (including additional Hofstadter bands) and in the strong-coupling limit.

Note that the action of \(\hat{T}_1\) and \(\hat{T}_2\) on \(\Delta^{(L)}\) as given in Eq. (\ref{DeltaLT1}-\ref{DeltaLT2}) is equivalent to an action on the components of \(\boldsymbol{\eta}\):
\begin{subequations}
\begin{align}
    \eta_L&\xrightarrow[]{\hat{T}_1(0)}\eta_{L+2}\,,
    \label{etaLT1}\\
    \eta_L&\xrightarrow[]{\hat{T}_2(0)}\omega^{- p L}_q\eta_L\,,
    \label{etaLT2}\\
    \eta_L&\xrightarrow[]{U(\theta)}e^{-2i\theta}\eta_L\,.
    \label{etaU1}
\end{align}
\end{subequations}
We can see that \(\hat{T}_1\) and \(\hat{T}_2\) act on it as \(\hat{\tau}^2\) and \(\hat{\sigma}^*\) respectively, while \(U(1)\) transformations \(U(\theta)\) act as \(e^{-2i\theta}\). This determines the irrep of the MTG realized by \(\boldsymbol{\eta}\) itself in place of the set of \(\hat{\Delta}^{(L)}\) \footnote{Eq. (\ref{etaLT1}) needs to be modified to \(\eta_L\xrightarrow[]{\hat{T}_1(2\pi p/q)}\eta_{L+2}\) for \(\hat{\Delta}^{(L,-)}\) irrep components for even \(q\) since \(\hat{T}^{q/2}_1(0)\) acts as \(-1\) on that irrep.}.

Using these transformation properties, we determine the most general form of the GL free energy to fourth order in \(\boldsymbol{\eta}\), consistent with MTG and U(1) symmetries:
\begin{widetext}
\[\mathcal{F}=\alpha\left|\boldsymbol{\eta}\right|^2+\sum_{MN}\beta_{MN}\sum_L\eta_{L+M}^*\eta_{L-M}^*\eta_{L+N}\eta_{L-N}+\sum_{jj'}\kappa_{jj'}\left(\tilde{D}_j\boldsymbol{\eta}\right)^*\cdot\left(\tilde{D}_{j'}\boldsymbol{\eta}\right)\,.
\label{F}\]
\end{widetext}
Note that terms of the form \(\eta_L^*\eta_{L'}\) are ruled out for \(L'\neq L\) by \(\hat{T}_2\) symmetry (including gradient terms), while \(\hat{T}_1\) implies that \(|\eta_L|^2\) terms have equal coefficients and that \(\beta_{MN}\) does not depend on \(L\). \(\tilde{D}_j=\partial_j+\frac{2ie}{c}\tilde{A}_j\) (with \(j=x,y\)) is the covariant derivative where \(\tilde{\mathbf{A}}=\mathbf{A}-x B \mathbf{a}_2/a_{2y}\) is the gauge field associated with the spatial variations of the overall phase of \(\boldsymbol{\eta}\). This choice imposes gauge invariance while ensuring that in the ground state \(\eta_L\) are spatially uniform. We emphasize that the relation between the spatial variations of the  overall phase of \(\boldsymbol{\eta}\) and the vector potential is only valid for very small spacial variations of \(\eta_L\), i.e. on a scale much larger than the magnetic unit cell, as the transformation properties of the gap functions \(\hat{\Delta}\) under gauge transformations, inherited from those of the \(d_{\mathbf{p},\ell}\) operators, are highly non-trivial within the magnetic unit cell. A free energy with very similar symmetries was constructed as a dual theory in a different context in Ref. \cite{Balents05}; a similar energy density was also obtained in the context of the bosonic Hofstadter model in Refs. \cite{PowellDasSarma10,PowellDasSarma11}. As discussed in Sec. \ref{section:irreps}, the difference from our work is that the order parameter considered in those works belonged to a charge \(1e\) irrep of the MTG, rather than \(2e\) irreps.

We note that there is some redundancy in the parameters, and in particular we take \(\beta_{MN}=\beta_{M,-N}=\beta_{-MN}=\beta_{NM}^*\) (the last equality to make the free energy real; the rest are not strictly necessary but account for redundancy) and \(\kappa_{jj'}=\kappa_{j'j}^*\). For odd \(q\) only \(\frac{(q+1)(q+3)}{4}\) of the \(\beta_{MN}\) parameters are thus distinct, a total of \(\frac{(q+1)^2}{4}\) parameters counting real and imaginary parts separately. For even \(q\), \(M\) and \(N\) must be both even or both odd with only even or only odd \(L\), depending on the irrep. 
The number of independent parameters, again counting the real and imaginary parts separately, is \(\frac{(q+2)^2}{8}\) if \(q/2\neq1\) is odd, with the exception of \(q=2\), for which there is an additional identification between \(\beta_{00}\) and \(\beta_{11}\). If \(q/2\) is even we additionally have that \(\beta_{00}=\beta_{q/2,q/2}\) and \(\beta_{0,q/2}=\beta_{q/2,0}\) and are therefore real, so the number of independent real parameters is \(\frac{q^2-q-6}{2}\). Terms with \(M\) and \(N\) of different parity are allowed and couple the even and odd irreps, but this can be ignored if the critical temperatures for the two irreps are sufficiently different at second order.

In addition, we observe that the free energy Eq. (\ref{F}) has an accidental symmetry at fourth order. Namely, it is symmetric under \(\mathcal{I}:\eta_L\rightarrow\eta_{-L}\) (and combinations of \(\mathcal{I}\) with MTG symmetries). This accidental symmetry is broken by sixth order terms, however:
\begin{widetext}
\[\mathcal{F}^{(6)}=\sum_{MM'NN'}\gamma_{MM'NN'}\sum_L\eta_{L+M}^*\eta_{L+M'}^*\eta_{L-M-M'}^*\eta_{L+N}\eta_{L+N'}\eta_{L-N-N'}\,.\]
\end{widetext}
We will not consider these higher order terms below but note that the accidental symmetry is in general explicitly broken. \(\mathcal{I}\) may be an actual symmetry if crystalline symmetries are present, for example a rotation by \(\pi\) as considered in Refs. \cite{Balents05,PowellDasSarma10,PowellDasSarma11} and as we show in Appendix \ref{B}.

Minimizing the free energy Eq. (\ref{F}) with respect to \(\eta_L^*\) (and integrating by parts), we obtain the Ginzburg-Landau equations:
\[\kappa_{jj'}\tilde{D}_j\tilde{D}_{j'}\eta_L=\alpha\eta_L+2\sum_{MN}\beta_{MN}\eta_{L+2M}^*\eta_{L+M+N}\eta_{L+M-N}\,.\label{GLeq}\]
Below we will only discuss the uniform phases, in which the left hand side vanishes, and leave the non-uniform solutions for a future study. As already noted in Sec. \ref{section:irreps}, there is no non-zero gap function \(\hat{\Delta}\), and hence no non-zero configuration of the field \(\boldsymbol{\eta}\), that respects all the MTG symmetry. The non-trivial solutions of the GL equations therefore necessarily break some but not necessarily all of the MTG symmetries, as we will show below. Note that the solutions correspond to local extrema of the free energy, while the ground state is determined by the global minimum.

With our choice of irrep elements, the simplest type of symmetric solutions are the ones that respect the \(\hat{T}_2(\theta_2)\) symmetry with \(\theta_2=\frac{2\pi p}{q}L\) (breaking it for any other choice of \(\theta_2\)), in which case \(\eta_{L'}=0\) unless \(L'=L\), and the only non-zero component is \(\eta_L=\sqrt{\frac{-\alpha}{2\beta_{00}}}\). There are \(q\) or \(q/2\) solutions corresponding to each choice of \(L\) (for \(q\) odd and even respectively). The value of the free energy at these extrema is
\[\mathcal{F}_{0,1}=-\frac{\alpha^2}{4\beta_{00}}\label{F01}\]
(the indexing will be explained below). The \(\hat{T}_1(\theta_1)\) symmetry is broken for any choice of \(\theta_1\) and maps the degenerate extrema to each other. If this solution is a global minimum, we refer to the corresponding ground state as a \(\mathbb{Z}_q\) symmetric Hofstadter SC phase. The order of the symmetry is \(q\) for either even or odd \(q\).

There is no fundamental difference between the \(\hat{T}_2(\theta_2)\) and \(\hat{T}_1(\theta_1)\) symmetries of course, and so we naturally expect solutions of the GL equation that respect \(\hat{T}_1(\theta_1)\) as well as other symmetries of the MTG while breaking \(\hat{T}_2(\theta_2)\) for any choice of \(\theta_2\). Indeed, since the irreps are multidimensional, the definition of the irrep components \(\hat{\Delta}^{(L)}\) is not unique, and we could always take the basis of the irrep to be any linearly independent orthogonal combinations of \(\hat{\Delta}^{(L)}\), including those that respect other MTG symmetries. In particular, there are solutions that respect the symmetry \(\hat{T}(\theta_0)=\hat{T}_2^{N_2}(\theta_2)\hat{T}_1^{N_1}(\theta_1)\) for any choice of \(\theta_0=N_1\theta_1+N_2\theta_2\) with \(N_1, N_2=0,\dots,q-1\). We index the corresponding solutions as \(\boldsymbol{\eta}^{(M)_{N_1,N_2}}\). For \(N_1\neq 0\), \(\hat{T}_2(\theta_2)\) is broken and generates degenerate solutions by shifting the index \(M=0,\dots,q-1\). The solutions found above that do respect \(\hat{T}_2(\theta_2)\) can be considered as a special case with \(N_1=0\), \(N_2=1\), and can be labeled \(\boldsymbol{\eta}^{(L)_{0,1}}\) (we reserve \(L\) to label the \(\hat{\mathbf{b}}_2\) component of the total momentum of the Cooper pairs).

Before explicitly constructing such linear combinations and showing that they do indeed yield additional solutions of the GL equations in Sec. \ref{otherZqSymmetries}, we illustrate this fact for the cases of \(q\leq4\) that can be partially analyzed analytically. Some of these cases have earlier been considered in Refs. \cite{ZhaiOktel10,SohalFradkin20}, and similar free energies with some additional symmetries were also analyzed in Ref. \cite{Balents05} in a different context. In contrast to those earlier works, here we explicitly determine the symmetries of the ground states, and identify additional possible symmetric phases that were not previously considered.

\subsection{Solutions of GL Equations for \(q\leq4\)}\label{GLsmallq}

We start with the simplest non-trivial case, \(q=2\) (only \(p=1\) is allowed), which has four 1D irreps. The free energy of each irrep is trivial and of the form (omitting gradient terms)
\[\mathcal{F}=\alpha|\eta|^2+\beta|\eta|^4\]
for the single order parameter \(\eta=\eta_0\) or \(\eta_1\) and a unique solution for \(|\eta|\), implying that the ground state always has some \(\mathbb{Z}_2\) symmetry. The corresponding gap functions \(\hat{\Delta}^{(L,\pm)}\) with \(L=0,1\) are simply the irrep components given in Eqs. (\ref{q2irrep0pm}-\ref{q2irrep1pm}):
\[\hat{\Delta}^{(0,\pm)}=\left(\begin{array}{cc}
    \Delta_0 & 0 \\
    0 & \pm\Delta_0
\end{array}\right)\]
and
\[\hat{\Delta}^{(1,\pm)}=\left(\begin{array}{cc}
    0 & \pm\Delta_1 \\
   \Delta_1 & 0
\end{array}\right)\label{q2Delta}\]
with \(\Delta_\ell\) as defined in Eq. (\ref{DeltaEllEven}). This is the only case in which case the MTG may be unbroken, in particular it is unbroken by \(\hat{\Delta}^{(0,+)}\). Note that \(L=0,1\) corresponds to gap functions symmetric/anti-symmetric under \(\hat{T}_2(0)\), while \(\pm\) corresponds to gap functions symmetric/anti-symmetric under \(\hat{T}_1(0)\). As these are all the symmetries of the MTG, in this case there are no other symmetric linear combinations.

\subsubsection{\(q=4\)}

The second-simplest non-trivial case and the only other case that can be completely solved analytically is \(q=4\) (with \(p=1\) or \(3\)), in which case there are four 2D irreps corresponding to even or odd \(L\) and gaps symmetric or anti-symmetric under \(\hat{T}_2^2(0)\). The free energy is
\begin{widetext}
\begin{align}
    \mathcal{F}_{+}&=\alpha_0(|\eta_0|^2+|\eta_2|^2)+\beta_{00}(|\eta_0|^4+|\eta_2|^4)+4\beta_{11}|\eta_0|^2|\eta_2|^2+4\beta_{02}|\eta_0|^2|\eta_2|^2\cos\varphi_{02}\,,\nonumber\\
    \mathcal{F}_-&=\alpha_1(|\eta_1|^2+|\eta_3|^2)+\beta_{00}(|\eta_1|^4+|\eta_3|^4)+4\beta_{11}|\eta_1|^2|\eta_3|^2+4\beta_{02}|\eta_1|^2|\eta_3|^2\cos\varphi_{13}
\end{align}
\end{widetext}
for even and odd \(L\) respectively, where \(\varphi_{02}=2(\phi_0-\phi_2)\) and \(\varphi_{13}=2(\phi_1-\phi_3)\), with \(\eta_L=|\eta_L|e^{i\phi_L}\). Note that \(\beta_{02}\) can actually be taken to be real in this case. Terms with \(\beta_{01}\) and \(\beta_{12}\) are allowed by symmetry but couple different irreps, so we ignore them assuming that \(\alpha_0\) and \(\alpha_1\) are sufficiently far apart. The free energies have the same mathematical form, and are also similar to the free energy for PDW order parameters \cite{Agterberg20}. For \(\beta_{02}>0\), the phases can always be minimized by setting \(\phi_L=\phi_{L+2}\pm \pi/2\), i.e. the two orders are out of phase by a factor of \(\pm i\); the two minima for \(\beta_{02}<0\) are \(\phi_L=\phi_{L+2}\) and \(\phi_L=\phi_{L+2}+\pi\), i.e. the order parameters are both real and either equal or opposite. Whether the two order coexist at all, however, depends on the ratio \((\beta_{02}-\beta_{11})/\beta_{00}\) with a phase transition at \(\beta_{02}-\beta_{11}=\beta_{00}/2\) (note that \(\beta_{00}\) and \(\beta_{11}\) are also real).

The non-coexisting solutions are precisely the \(\hat{T}_2(\theta_2)\) symmetric solutions, with the corresponding gap functions being the irrep components as given in Eq. (\ref{q4irrep}):
\begin{widetext}
\begin{align}
    &\hat{\Delta}^{(0,\pm)}=\left(\begin{array}{cccc}
    \Delta_0 & 0 & 0 & 0\\
    0 & 0& 0 & \Delta_2\\
    0 & 0 & \pm \Delta_0 & 0\\
    0 & \pm \Delta_2 & 0 & 0
\end{array}\right), &\hat{\Delta}^{(2,\pm)}=\left(\begin{array}{cccc}
    0 & 0 &  \pm\Delta_2 & 0\\
    0 & \Delta_0& 0 & 0\\
    \Delta_2 & 0 & 0 & 0\\
    0 & 0 & 0 & \pm \Delta_0
\end{array}\right), \nonumber\\
&\hat{\Delta}^{(1,\pm)}=\left(\begin{array}{cccc}
    0 & \Delta_3 & 0 & 0\\
    \Delta_1 & 0& 0 & 0\\
    0 & 0 & 0 & \pm \Delta_3\\
    0 & 0 & \pm \Delta_1 & 0
\end{array}\right), &\hat{\Delta}^{(3,\pm)}=\left(\begin{array}{cccc}
    0 & 0 &  0 & \pm\Delta_1\\
    0 & 0 &  \Delta_3 & 0\\
    0 & \Delta_1 & 0 & 0\\
    \pm \Delta_3 & 0 & 0 & 0
\end{array}\right)\,,
\end{align}
which correspond to states symmetric under \(\hat{T}_2(\theta_2)\) with \(\theta_2=\frac{\pi p}{2} L\) and symmetric/anti-symmetric under \(\hat{T}_1^2(0)\) for \(\pm\) respectively. As for \(q=2\), \(\Delta_\ell\) are defined in Eq. (\ref{DeltaEllEven}).

For the \(\hat{\Delta}^{(L,-)}\) irreps, there is however a phase transition into a state with gap functions forming linear combinations (assuming \(\beta_{02}>0\))
\begin{align}
    \hat{\Delta}^{(\pm 1,+)_{1,0}}&=\hat{\Delta}^{(0,-)}\pm i\hat{\Delta}^{(2,-)}\,,\nonumber\\
    \hat{\Delta}^{(\pm 1,-)_{1,0}}&=\hat{\Delta}^{(1,-)}\pm i\hat{\Delta}^{(3,-)}\,.
\end{align}
As can be checked directly, \(\hat{\Delta}^{(M,\pm)_{1,0}}\) are symmetric under \(\hat{T}_1(\theta_1)\) with \(\theta_1=\frac{\pi p}{2} M\) with odd \(M=\pm1\), consistent with the notation. In addition, as will generalize to all even \(q\), \(\hat{\Delta}^{(M,\pm)_{1,0}}\) are symmetric/anti-symmetric under \(\hat{T}_2^2\) respectively, requiring the additional \(\pm\) index.
In a pattern that will also generalize to other even \(q\) (and trivially holds also for the previous example of \(q=2\)), the corresponding gap functions 
are even/odd checkerboard matrices \cite{Jones18} symmetric/anti-symmetric under \(\hat{T}_2^2(0)\) respectively:
\[\hat{\Delta}^{(\pm 1,+)_{1,0}}=\left(\begin{array}{cccc}
    \Delta_0 & 0 & \mp i\Delta_2 & 0\\
    0 &  \pm i\Delta_0 & 0 & \Delta_2\\
    \pm i\Delta_2 & 0 & -\Delta_0 & 0\\
    0 & -\Delta_2 & 0 & \mp i\Delta_0
\end{array}\right),\qquad \hat{\Delta}^{(\pm1,-)_{1,0}}=\left(\begin{array}{cccc}
    0 & \Delta_3 & 0 & \mp i\Delta_1 \\
    \Delta_1 & 0 & \pm i\Delta_3 & 0 \\
    0 & \pm i\Delta_1 & 0 & -\Delta_3\\
    \mp i\Delta_3 & 0 & -\Delta_1 & 0 
\end{array}\right)\,.\label{q4DeltaM10}\]
These happen to be the phases found in Ref. \cite{ZhaiOktel10} for the special case of Hubbard interactions.

For the \(\hat{\Delta}^{(L,+)}\) irreps, again assuming \(\beta_{02}>0\), the phase transition is instead into
\begin{align}
    \hat{\Delta}^{(\pm 1,+)_{1,1}}&=\hat{\Delta}^{(0,+)}\pm i\hat{\Delta}^{(2,+)}\,,\nonumber\\
    \hat{\Delta}^{(\pm 1,-)_{1,1}}&=\hat{\Delta}^{(1,+)}\pm i\hat{\Delta}^{(3,+)}
\end{align}
that as the notation indicates are symmetric under \(\hat{T}_2(\theta_1)\hat{T}_1(\theta_2)\) with \(\theta_0=\theta_1+\theta_2=\frac{\pi p}{2} M\) with odd \(M=\pm1\). Explicitly they are given by the even/odd checkerboard matrices symmetric/anti-symmetric under \(\hat{T}_2^2(0)\) respectively:
\[\hat{\Delta}^{(\pm 1,+)_{1,1}}=\left(\begin{array}{cccc}
    \Delta_0 & 0 & \pm i\Delta_2 & 0\\
    0 &  \pm i\Delta_0 & 0 & \Delta_2\\
    \pm i\Delta_2 & 0 & \Delta_0 & 0\\
    0 & \Delta_2 & 0 & \pm i\Delta_0
\end{array}\right),\qquad \hat{\Delta}^{(\pm1,-)_{1,1}}=\left(\begin{array}{cccc}
    0 & \Delta_3 & 0 & \pm i\Delta_1 \\
    \Delta_1 & 0 & \pm i\Delta_3 & 0 \\
    0 & \pm i\Delta_1 & 0 & \Delta_3\\
    \pm i\Delta_3 & 0 & \Delta_1 & 0 
\end{array}\right)\,.\label{q4DeltaM11}\]
\end{widetext}
These are distinct from the phases found in Ref. \cite{ZhaiOktel10}. 
We get additional phases taking \(\beta_{02}<0\), which results in \(\hat{\Delta}^{(M,\pm)_{1,N_2}}\) solutions with even \(M=0,2\) and \(N_2=1\) or \(0\) for \(\hat{\Delta}^{(L,\pm)}\) irreps respectively.
We thus find that at least one \(\mathbb{Z}_4\) symmetry remains unbroken for \(q=4\), though this may change if the even and odd \(L\) irreps are allowed to mix or higher order terms are included in the free energy. For examples of a numerical analysis of the former possibility (with some additional symmetries), see Ref. \cite{Balents05}.

\subsubsection{\(q=3\)}

The simplest odd \(q\) case is \(q=3\) (with \(p=1\) or \(2\)). There is only one MTG irrep in this case, and the free energy is
\begin{widetext}
\begin{align}
\mathcal{F}&=\alpha(|\eta_0|^2+|\eta_1|^2+|\eta_2|^2)+\beta_{00}(|\eta_0|^4+|\eta_1|^4+|\eta_2|^4)+4\beta_{11}\left(|\eta_0|^2|\eta_1|^2+|\eta_1|^2|\eta_2|^2+|\eta_2|^2|\eta_0|^2\right)+\nonumber\\
&+4|\beta_{01}||\eta_0||\eta_1||\eta_2|\left(|\eta_0|\cos\varphi_{01}^{(0)}+|\eta_1|\cos\varphi_{01}^{(1)}+|\eta_2|\cos\varphi_{01}^{(2)}\right)
\end{align}
\end{widetext}
where \(\beta_{01}=|\beta_{01}|e^{i\theta_{01}}\) is not necessarily real (unlike the \(q=4\) case) and
\[\varphi_{01}^{(L)}=2\phi_L-\phi_{L+1}-\phi_{L-1}+\theta_{01}\,.\]
Unfortunately already in this case we did not find a complete analytical solution. To make some progress, it is convenient to minimize the free energy with respect to \(\varphi_{01}^{(L)}\) instead of \(\phi_L\), noting that there is a constraint \(\sum_L\varphi_{01}^{(L)}=3\theta_{01}\). This can be enforced using a Lagrange multiplier
\[\mathcal{F}_\lambda=\lambda\sum_L\left(\varphi_{01}^{(L)}-\theta_{01}\right)\]
and minimizing \(\mathcal{F}+\mathcal{F}_\lambda\) with respect to \(|\eta_L|\), \(\varphi_{01}^{(L)}\) and \(\lambda\). Minimizing with respect to \(\varphi_{01}^{(L)}\) we find that at any local or global extremum,
\[4|\beta_{01}||\eta_0||\eta_1||\eta_2||\eta_L|\sin\varphi_{01}^{(L)}=\lambda\,.\]
In particular, \(|\eta_L|\sin\varphi_{01}^{(L)}\) are equal for all \(L\) in the ground state. This is enough to prove that \emph{if} all of \(|\eta_L|\) are in addition equal in the ground state, then the ground state has a \(\mathbb{Z}_3\) symmetry.

This is as much as we can determine analytically. Numerically, we find that for the explored parameter range at least one of the MTG symmetries is always unbroken in the ground state and either all of \(|\eta_L|\) are equal or only one is non-zero; this is another pattern that we will see generalizes to all \(q\). The sole exception are phase transitions, for example in the special case
\[\mathcal{F}=\alpha|\boldsymbol{\eta}|^2+\beta|\boldsymbol{\eta}|^4\,.\]
It is clear that the direction of the vector \(\boldsymbol{\eta}\) is arbitrary in the ground state due to the additional \(SU(3)\) symmetry, so symmetry breaking linear combinations are allowed in this case. We cannot definitively state that the MTG cannot be fully broken away from such phase transition points.

The possible distinct \(\mathbb{Z}_3\) symmetries in this case are \(\hat{T}_2\), \(\hat{T}_1\), \(\hat{T}_2\hat{T}_1\), and \(\hat{T}_2^2\hat{T}_1\) (other cases are redundant). Note that we can thus always take \(N_1=1\) in \(\boldsymbol{\eta}^{(M)_{N_1,N_2}}\). The gap functions symmetric under \(\hat{T}_2\) are again the irrep components already given in Eq. (\ref{q3irrep}):
\begin{widetext}
\[\begin{array}{ccc}
    \hat{\Delta}^{(0)}=\left(\begin{array}{ccc}
    \Delta_0 & 0 & 0 \\
    0 & 0 & \Delta_2\\
    0 & \Delta_1 & 0
\end{array}\right), &\hat{\Delta}^{(1)}=\left(\begin{array}{cccc}
    0 & 0  & \Delta_1\\
    0 & \Delta_0 & 0\\
     \Delta_2 & 0 & 0
\end{array}\right), &\hat{\Delta}^{(2)}=\left(\begin{array}{cccc}
    0 & \Delta_2  & 0\\
    \Delta_1 & 0 & 0\\
     0 & 0 & \Delta_0
\end{array}\right)
\end{array}\,,\]
\end{widetext}
where note that \(\Delta_\ell\) with \(\ell=0,1,2\) as defined in Eq. (\ref{DeltaEllOdd}) are determined by the gap equation and may be complex. Gap functions symmetric under \(\hat{T}_1\) are
\[\hat{\Delta}^{(M)_{1,0}}=\left(\begin{array}{ccc}
    \Delta_0 & \Delta_2e^{-\frac{2\pi i pM}{3}}  & \Delta_1e^{\frac{2\pi i pM}{3}}  \\
    \Delta_1e^{-\frac{2\pi i pM}{3}}  & \Delta_0e^{\frac{2\pi i pM}{3}} & \Delta_2\\
    \Delta_2e^{\frac{2\pi i pM}{3}}  & \Delta_1 & \Delta_0e^{-\frac{2\pi i pM}{3}}
\end{array}\right)\,.\label{q3DeltaM10}\]
Gap functions symmetric under \(\hat{T}_2\hat{T}_1\) are
\[\hat{\Delta}^{(M)_{1,1}}=\left(\begin{array}{ccc}
    \Delta_0 & \Delta_2e^{-\frac{2\pi i p M}{3}}  & \Delta_1e^{\frac{2\pi i p (M+2)}{3}}  \\
    \Delta_1e^{-\frac{2\pi i pM}{3}}  & \Delta_0e^{\frac{2\pi ip(M+2)}{3}} & \Delta_2\\
    \Delta_2e^{\frac{2\pi i p(M+2)}{3}}  & \Delta_1 & \Delta_0e^{-\frac{2\pi i pM}{3}}
\end{array}\right)\,.\label{q3DeltaM11}\]
Finally, gap functions symmetric under \(\hat{T}_2^2\hat{T}_1\) are
\[\hat{\Delta}^{(M)_{1,2}}=\left(\begin{array}{ccc}
    \Delta_0 & \Delta_2e^{-\frac{2\pi i p M}{3}}  & \Delta_1e^{\frac{2\pi i p (M+4)}{3}}  \\
    \Delta_1e^{-\frac{2\pi i pM}{3}}  & \Delta_0e^{\frac{2\pi ip(M+4)}{3}} & \Delta_2\\
    \Delta_2e^{\frac{2\pi i p(M+4)}{3}}  & \Delta_1 & \Delta_0e^{-\frac{2\pi i pM}{3}}
\end{array}\right)\,.\label{q3DeltaM12}\]
The phases found for the Harper-Hubbard model in Ref. \cite{ZhaiOktel10} correspond to the \(\hat{T}_2\hat{T}_1\) symmetric gaps, which will also be the phase we consider in the context of chiral pairing functions in Sec. \ref{Chiral}.

\subsection{General \(\mathbb{Z}_q\) Symmetries}
\label{otherZqSymmetries}

We now determine the general form of solutions \(\boldsymbol{\eta}^{(M)_{N_1,N_2}}\) symmetric under \(\hat{T}(\theta_0)=\hat{T}_2^{N_2}(\theta_2)\hat{T}_1^{N_1}(\theta_1)\) for some choice of \(\theta_0=N_1\theta_1+N_2\theta_2\) with \(N_1, N_2=0,\dots,q-1\). For simplicity, we will only consider the case when the order of this symmetry is \(q\). This is not the case if and only if 
\(N_1\), \(N_2\) and \(q\) all share a common divisor \(d\neq1\). The order of \(\hat{T}(\theta_0)\) in that case is \(q/d\) and the degeneracy of the solutions is \(dq\) for odd \(q\) or \(dq/2\) for even \(q\). Below we consider only \(d=1\), and treat the odd and even \(q\) separately.

\subsubsection*{Odd \(q\)}

Let us first consider the \(\hat{T}_1(\theta_1)\)-symmetric order parameters for odd \(q\), with \(e^{i\theta_1}=\omega_q^{pM}\). These can be obtained by taking \(\eta_0=\eta\) and sequentially applying \(\hat{T}_1(\theta_1)\) to \(\boldsymbol{\eta}\). This way we find the components of \(\boldsymbol{\eta}^{(M)_{1,0}}\) to be
\[\eta^{(M)_{1,0}}_{L}=\omega_q^{-p M [L/2]_q}\eta\label{etaT1}\,,\]
where \([L/2]_q=L/2\) if \(L\) is even but \([L/2]_q=(L+q)/2\) if \(\ell\) is odd.
By construction, applying \(\hat{T}_2(0)\) to \(\boldsymbol{\eta}^{(M)_{1,0}}\) maps it to \(\boldsymbol{\eta}^{(M+2)_{1,0}}\), and again there are \(q\) degenerate solutions. The corresponding gap functions
\[\hat{\Delta}^{(M)_{1,0}}=\sum_{L}\eta_L^{(M)_{1,0}} \hat{\Delta}^{(L)}\]
can be considered as an alternative set of irrep components, now symmetric with respect to \(\hat{T}_1(\theta_1)\). Eq. (\ref{q3DeltaM10}) provides an example for \(q=3\).

The procedure is essentially the same for all other \(\hat{T}(\theta_0)\) symmetries, and we obtain the solutions \(\boldsymbol{\eta}^{(M)_{N_1,N_2}}\) with components
\[\eta^{(M)_{N_1,N_2}}_{2N_1J}=\omega_{q}^{-2pN_1N_2J(J-1)-pMJ}\eta\label{etaMN1N2}\]
with \(J=0,\dots,q-1\). If the greatest common divisor \(\text{gcd}(N_1,q)\neq 1\), some components may vanish. Again for simplicity we will not consider that case; in all other cases we can take \(N_1=1\) without loss of generality since \(\hat{T}(\theta_0)\) forms a \(\mathbb{Z}_q\) subgroup. To gain a better understanding of these solutions, we consider the implications for the form of the gap functions
\[\hat{\Delta}^{(M)_{N_1,N_2}}=\sum_L \eta^{(M)_{N_1,N_2}}_{L} \hat{\Delta}^{(L)}\label{DeltaMN1N2}
\,.\]
The symmetry \(\hat{T}(\theta_0)\) constrains the elements of these matrices to be of the form
\[\hat{\Delta}^{(M)_{N_1,N_2}}_{\ell,\ell'}=\Delta_{\ell-\ell'}\exp\left[i\phi_{\ell+\ell'}^{(M)_{N_1,N_2}}\right]\label{GapOdd}
\,,\]
where \(\Delta_{\ell-\ell'}\) as given in Eq. (\ref{DeltaEllOdd}) are the same for all \(M, N_1\) and \(N_2\) and depend only on \(\ell-\ell'\) that labels which diagonal they are on. The phases \(\phi_{\ell+\ell'}^{(M)_{N_1,N_2}}\) on the other hand are different for each \(M\) and depend only on \(L=\ell+\ell'\), i.e. which anti-diagonal the element is on. From Eq. (\ref{DeltaMN1N2}) we see that for \(N_1=1\) and setting \(\phi_0^{(M)_{1,N_2}}=0\) without loss of generality,
\begin{equation}
\phi_{L}^{(M)_{1,N_2}}=\left[\frac{L}{2}\right]_{q}\left(\left[\frac{L}{2}\right]_{q}+1\right)\frac{2\pi p N_2}{q}+\frac{2\pi p M}{q}\left[\frac{L}{2}\right]_{q}\,.\label{phiL}
\end{equation}
See Eqs. (\ref{q3DeltaM10}-\ref{q3DeltaM12}) for concrete illustrations of these equations for \(q=3\). The case when \(N_1\neq 1\) is similar with the phases permuted accordingly.

\subsubsection*{Even \(q\)}

The expressions for even \(q\) are essentially the same as for odd \(q\) but with some minor modifications to keep track of the fact that there are four different irreps in this case. Note that within a given irrep, only \(\eta_L\) with the same parity of \(L\) are non-zero, so the symmetric order parameters have an additional index. For example the \(\hat{T}_1(\theta_1)\)-symmetric order parameters are \(\boldsymbol{\eta}^{(M,\pm)_{1,0}}\) with components given by the same expression as for odd \(q\), Eq. (\ref{etaT1}):
\[\eta^{(M,\pm)_{1,0}}_{L}=\omega_q^{-p M L/2}\eta\label{etaT1even}\]
but with the understanding that only even or odd \(L\) components are non-zero. In general, the order parameters with symmetries other than \(\hat{T}_2(\theta_2)\) are given by \(\boldsymbol{\eta}^{(M,\pm)_{N_1,N_2}}\).
The meaning of the labels depends on the parity of \(N_1\). If \(N_1\) is odd, \(\pm\) labels the parity of \(L\) of non-zero components while \(M\) is  determined by which of the four irreps the gap function belongs to, with \(\hat{T}_1^{q/2}(0)\) acting as \((-1)^M\) on the order parameter. If \(N_1\) is even, on the other hand, \(M\) must have the same parity as \(L\), while \(\pm\) corresponds to \(\hat{T}_1^{q/2}(0)\) acting as \(\pm1\) in the irrep.
For odd \(N_1\), Eq. (\ref{etaMN1N2}) gives \(\boldsymbol{\eta}^{(M,+)_{N_1,N_2}}\) with even \(L\) components. The odd \(L\) component combinations \(\boldsymbol{\eta}^{(M,-)_{N_1,N_2}}\) are given by essentially the same formula:
\[\eta^{(M,-)_{N_1,N_2}}_{2N_1J+1}=\omega_{q}^{-2pN_1N_2J(J-1)-pMJ}\eta\label{etaMN1N2even}\,.\]
For even \(N_1\), the RHSs of Eqs. (\ref{etaMN1N2}) and (\ref{etaMN1N2even}) give the components of \(\boldsymbol{\eta}^{(M,\pm)_{N_1,N_2}}\) for even and odd \(M\) respectively. We will consider odd \(N_1\) below for simplicity.

The parity of \(L\) of the non-zero components can be considered as the eigenvalue of \(\boldsymbol{\eta}^{(M,\pm)_{N_1,N_2}}\) under \(\hat{T}^{q/2}_2(0)\). We note that this symmetry places a particular constraint on the corresponding pairing matrices \(\hat{\Delta}\), on which it acts as
\[\hat{\Delta}\xrightarrow[]{\hat{T}_2^{q/2}(0)}\hat{\sigma}^{q/2}\hat{\Delta}\hat{\sigma}^{q/2}\]
(note that \(\hat{\sigma}^{q/2}_{\ell,\ell}=(-1)^\ell\)). This implies that the gap functions
\[\hat{\Delta}^{(M,\pm)_{N_1,N_2}}=\sum_{L}\eta_L^{(M,\pm)_{N_1,N_2}} \hat{\Delta}^{(L,(-1)^{M})}\label{DeltaMN1N2even}\]
that are even or odd under \(\hat{T}^{q/2}_2(0)\) are even or odd checkerboard matrices \cite{Jones18}: \(\hat{\Delta}^{(M,\pm)_{N_1,N_2}}_{\ell,\ell'}=0\) whenever \(\ell+\ell'\) is odd or even respectively for \(+\) and \(-\) respectively. We saw this explicitly for the \(q=2\) and \(4\) cases considered in Sec. \ref{GLsmallq}.
Note that the term \(\hat{\Delta}^{(M,+)_{N_1,N_2}\dagger}\hat{\Delta}^{(M,-)_{N_1,N_2}}\) is odd under \(\hat{T}_2^{q/2}\) so such a term is not allowed in the free energy and its trace vanishes because a product of an even and an odd checkerboard matrix is an odd checkerboard matrix with zero diagonal. This confirms once again that the four irreps for even \(q\) are not mixed at the leading second order in the free energy.

Again to understand the solutions better, we consider the form of the gap functions corresponding to the order parameters \(\boldsymbol{\eta}^{(M,\pm)_{N_1,N_2}}\).
As for odd \(q\), we can express the elements of \(\hat{\Delta}^{(M,\pm)_{N_1,N_2}}\) as
\[\hat{\Delta}^{(M,\pm)_{N_1,N_2}}_{\ell,\ell'}=\Delta_{\ell-\ell'}\exp\left[i\phi_{\ell+\ell'}^{(M,\pm)_{N_1,N_2}}\right]\label{GapEven}\]
but with a caveat that \(\ell-\ell'\) in \(\Delta_{\ell-\ell'}\) is defined modulo \(2q\) and restricted to be even or odd for \(\pm\) respectively, as in Eq. (\ref{DeltaEllEven}), and additionally satisfying \(\Delta_{\ell+q}=(-1)^M\Delta_\ell\). In this case only the relative phases between even or odd \(L\) anti-diagonals are defined, so for \(N_1=1\) we have:
\begin{align}\label{phiLeven}
\phi_{L}^{(M,+)_{1,N_2}}&=\phi_0+\frac{L(L+2)}{4}\frac{2\pi p N_2}{q}+\frac{\pi p M L}{q}\,,\\
\phi_{L}^{(M,-)_{1,N_2}}&=\phi_1+\frac{(L-1)(L+3)}{4}\frac{2\pi p N_2}{q}+\frac{\pi p M(L-1)}{q}\,.\nonumber
\end{align}
As for odd \(q\), the phase relations for \(N_1\neq 1\) are similar. See Eqs. (\ref{q2Delta}, \ref{q4DeltaM10},\ref{q4DeltaM11}) for concrete illustrations for \(q=2\) and \(4\).

\subsubsection*{\(\mathbb{Z}_q\)-symmetric and Non-Symmetric Solutions of the GL Equations}

\begin{figure*}
\centering
\includegraphics[width=0.95\textwidth]{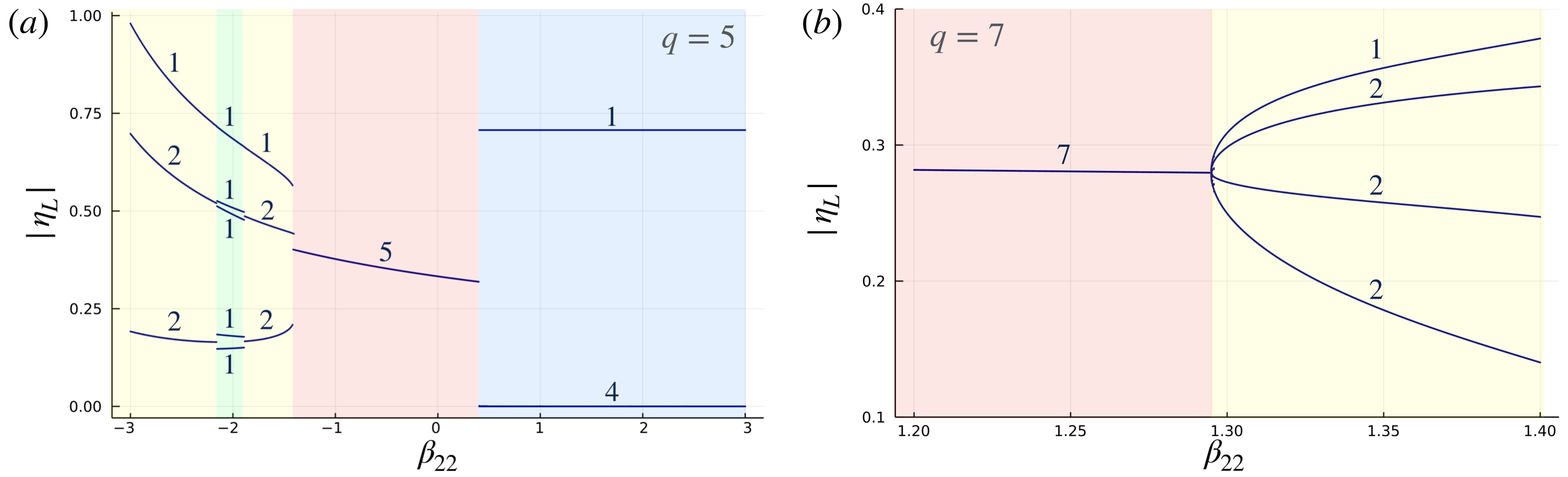}
\caption{(Color online.) 
Numerical solutions of uniform GL equations \(|\eta_L|\) for (a) \(q=5\) and (b) \(q=7\) as a function of \(\beta_{22}\) treated as a variable parameter with \(\beta_{23}=2\), \(\beta_{12}=1.7\) for \(q=5\) and \(\beta_{12}=1.6\) for \(q=7\), and all other \(\beta_{MN}=1\). Numbers indicate the number of equal \(|\eta_L|\) on each segment. For \(q=5\), we find four first-order phase transitions between four kinds of phases: a \(\hat{T}_2\) symmetric phase with a single non-zero \(\eta_L\) (blue); a \(\mathbb{Z}_5\) symmetric phase with all \(|\eta_L|\) equal (red); a phase with \(\mathcal{I}\) symmetry with pairs of equal \(|\eta_L|\) except for one (yellow); and a phase with no symmetries where none of the \(|\eta_L|\) are equal (green). For \(q=7\) we find a second-order phase transition between a \(\mathbb{Z}_7\) symmetric phase with all \(|\eta_L|\) equal (red) and a  \(\mathcal{I}\) symmetric phase with pairs of equal \(|\eta_L|\) except for one (yellow).}
\label{fig:PhaseTransitions}
\end{figure*}

So far in this subsection we have simply constructed configurations \(\boldsymbol{\eta}^{(M)_{N_1,N_2}}\) in Eqs. (\ref{etaMN1N2},\ref{etaMN1N2even}) 
that respect \(\hat{T}(\theta_0)\) symmetry. The corresponding gap functions in Eq. (\ref{DeltaMN1N2}) satisfy the defining properties
\begin{subequations}
\begin{align}
    \hat{\Delta}^{(M)_{N_1,N_2}}&\xrightarrow[]{\hat{T}(0)}\omega_q^{pM}\hat{\Delta}^{(M)_{N_1,N_2}}\,,\label{DeltaMT}\\
    \hat{\Delta}^{(M)_{N_1,N_2}}&\xrightarrow[]{\hat{T}_2(0)}\hat{\Delta}^{(M+2N_1)_{N_1,N_2}}\label{DeltaMT2}
\end{align}
\end{subequations}
for odd \(q\); the relations are the same for \(\hat{\Delta}^{(M,+)_{N_1,N_2}}\) in Eq. (\ref{DeltaMN1N2even}) for even \(q\), while for \(\hat{\Delta}^{(M,-)_{N_1,N_2}}\) the last relation becomes
\[\hat{\Delta}^{(M,-)_{N_1,N_2}}\xrightarrow[]{\hat{T}_2\left(\frac{2\pi p}{q}\right)}\hat{\Delta}^{(M+2N_1,-)_{N_1,N_2}}\,,\]
to be consistent with the fact that in this irrep \(\hat{T}^{q/2}_2(0)\) acts as \(-1\) on \(\hat{\Delta}^{(M,-)_{N_1,N_2}}\).

We still need to show that the configurations \(\boldsymbol{\eta}^{(M)_{N_1,N_2}}\)  are actually solutions of the GL equations and are therefore always at least local extrema of the free energy. To see this, note that all the symmetric configurations have the important property that all the components \(\eta_L\) have the same magnitude and differ only by a phase, \(\eta_L=\eta e^{i\phi_L}\) with real \(\eta\) (more precisely, all non-zero \(\eta_L\) have the same magnitude). Plugging this ansatz into Eq. (\ref{GLeq}) we find the equation for \(\eta\):
\[\eta=\sqrt{\frac{-\alpha }{2\sum_{MN}\left|\beta_{MN}\right|\cos\varphi_{MN}^{(L)}}}\label{GLeqSym}\]
where
\[\varphi_{MN}^{(L)}=\phi_{L+M}+\phi_{L-M}-\phi_{L+N}-\phi_{L-N}+\theta_{MN}\label{varphi}\]
with \(\beta_{MN}=\left|\beta_{MN}\right|e^{i\theta_{MN}}\). The solution exists only if the RHS of Eq. (\ref{GLeqSym}) is independent of \(L\). For symmetric gaps it is easy to check from Eq. (\ref{phiL}) and (\ref{phiLeven})
that in fact \(\varphi_{MN}^{(L)}=\varphi_{MN}^{(0)}\) for all \(L\). Therefore, as claimed above, symmetric solutions are always extrema of the free energy. The free energy at these extrema is
\[\mathcal{F}_{N_1 \neq 0,N_2}=-\frac{\alpha^2 (2q-1)}{4\sum_{MN}\left|\beta_{MN}\right|\cos\varphi_{MN}^{(L)}}\label{FN1N2}\,.\]
(Note that if both \(N_1\) and \(N_2\) divide \(q\), some \(\eta_L\) are zero, but all non-zero components have equal magnitudes.)

Although we thus conclude that the \(\hat{T}(\theta_0)\)-symmetric order parameters are possible ground states of the system, it is not true that the ground state is necessarily symmetric. Though we did find this to be the case for \(q\leq4\) in Sec. \ref{GLsmallq} (analytically for \(q=2\) and \(4\) and numerically for \(q=3\)), already for \(q=5\) we find numerically that the MTG may be fully broken in the ground state, as shown in Fig. \ref{fig:PhaseTransitions} (a). With the particular choice of parameters, we observe first-order phase transitions between phases with only one \(\eta_L\) being non-zero and all \(|\eta_L|\) equal, i.e. phases with different \(\mathbb{Z}_q\) symmetries; a phase transition into a phase with pairs of \(|\eta_L|\) being equal except for one that is symmetric under \(\eta_L\rightarrow\eta_{M-L}\) for some \(M\) (as noted above, this is an accidental symmetry in our case); and a phase transition into a phase where none of the \(|\eta_L|\) are equal and there are no symmetries. There may also be second-order phase transitions, as we find for \(q=7\) (see Fig. \ref{fig:PhaseTransitions} (b)).

\subsection{Summary}    

To summarize this section, we found that the GL equations Eq. (\ref{GLeq}) always have solutions with a \(\mathbb{Z}_q\) symmetry \(\hat{T}(\theta_0)=\hat{T}_2^{N_2}(\theta_2)\hat{T}_1^{N_1}(\theta_1)\) with \(\theta_0=N_1\theta_1+N_2\theta_2=\frac{2\pi p}{q} M\), and \(N_1, N_2=0,\dots,q-1\). \(N_1\) and \(N_2\) are determined in the ground state by which of \(\mathcal{F}_{N_1,N_2}\), given in Eqs. (\ref{F01}) and (\ref{FN1N2}) is smallest (assuming the ground state does not completely break the MTG), while each value of \(M=0,2,4,\dots\) corresponds to a degenerate solution. For even \(q\), only \(M\) of the same parity give degenerate solutions, while all \(M\) are degenerate for odd \(q\). The solutions are thus \(q\)-fold degenerate for odd \(q\) and \(q/2\)-fold degenerate for even \(q\), corresponding to the dimensions of the MTG irreps, and there is an additional \(\mathbb{Z}_2\) symmetry for even \(q\). Additional degeneracy may occur if other crystalline symmetries are present, e.g. the inversion-like symmetry \(\mathcal{I}:\eta_L\rightarrow\eta_{-L}\) that exists in presence of two-fold rotation symmetry (and is an accidental symmetry at fourth order of the free energy). Since \(\hat{T}_2(0)\) is broken by all phases except \(\boldsymbol{\eta}^{(0)_{0,1}}\), in all other cases the unit cell is extended in real space by an additional factor of \(q\) along the \(\mathbf{a}_2\) direction, resulting in a unit cell \(q\times q\) larger than the normal state unit cell in the absence of the magnetic field (note that \(\hat{T}_j^q(0)\) are always unbroken).

The symmetric solutions are given by \(\boldsymbol{\eta}^{(M)_{N_1,N_2}}\) for odd \(q\) (\(\boldsymbol{\eta}^{(M,\pm)_{N_1,N_2}}\) for even \(q\)) in Eqs. (\ref{etaT1},\ref{etaMN1N2}). As a special case, \(\boldsymbol{\eta}^{(L)_{0,1}}\) are simply vectors with a single non-zero element. These states satisfy the defining relations Eqs. (\ref{DeltaMT}-\ref{DeltaMT2}), which in particular imply that if \(N_1\neq 0\),  applying \(\hat{T}_2\) to the ground state shifts \(M\rightarrow M+2\); if \(N_1=0, N_2=1\), then applying \(\hat{T}_1(0)\) shifts \(L\rightarrow L+2\) instead. In either case the degenerate ground states are simply shifted version of each other.

An important property of the symmetric solutions is that all non-zero components \(\eta_L\) have the same magnitude \(\eta\) given by Eq. (\ref{GLeqSym}). Moreover, the relative phases \(\phi_L\) are also constrained by symmetry, which also fixes the phase relations between anti-diagonals of \(\hat{\Delta}=\sum_L\eta_L\hat{\Delta}^{(L)}\), while the phases between diagonals is fixed by the interaction via the linearized gap equation Eq. (\ref{LinGapEq}). The phase relations for \(q=3\) and \(q=4\) (and \(p=1\)) have been noted earlier in Ref. \cite{ZhaiOktel10} as a consequence of minimizing a Ginzburg-Landau free energy with the assumption of Hubbard interactions. Here we clarified that the phase relations in those cases are a consequence of the symmetry of the solutions themselves and found the phase relations for all \(q\) assuming this symmetry remains unbroken. The ground states found in Ref. \cite{ZhaiOktel10} have \(\hat{T}\) symmetry with either \(N_1=N_2=1\) or \(N_2=-1\) for \(q=3\) (the two are found to be degenerate) and \(N_1=1, N_2=0\) for \(q=4\); for even \(q\) we also need to specify whether the solutions are symmetric or anti-symmetric with respect to \(\hat{T}_2^{q/2}\), but Ref. \cite{ZhaiOktel10} finds the two cases to be degenerate. The extra degeneracy in both \(q=3\) and \(q=4\) cases can be understood as a result of extra symmetry of the square lattice. Similar partial symmetry breaking of the MTG has been seen in Ref. \cite{SohalFradkin20}, and in Refs. \cite{PowellDasSarma10,PowellDasSarma11} in the context of the related bosonic Hofstadter model. These authors did not report the phase relations between condensates with different momenta, but Refs. \cite{PowellDasSarma10,PowellDasSarma11} noted that the superfluid state necessarily at least partially breaks the MTG symmetries, essentially for the same reason that the Hofstadter SC state does as we found in this work.

Though we showed analytically that the GL equations \emph{always} have symmetric solutions, we emphasize that this only guarantees that such states are \emph{local} extrema of the free energy, not necessarily \emph{global} minima that are the true ground states. This implies that extra care must be taken when analyzing such systems numerically, as some methods are susceptible to getting stuck in local minima. Indeed, we find explicit cases for \(q=5\) for which the MTG is completely broken, see Fig. \ref{fig:PhaseTransitions}. The additional symmetry breaking can happen in two ways: first, the magnitudes of non-zero \(\eta_L\) may not all be zero; second, the phase relations determined by Eqs. (\ref{etaT1},\ref{etaMN1N2},\ref{etaT1even},\ref{etaMN1N2even}) may be violated even if the magnitudes are equal. The latter can happen due to frustration between \(\varphi_{MN}^{(L)}\) in Eq. (\ref{varphi}) for different values of \(M\neq N\). We also note that the accidental \(\mathcal{I}\) symmetry can be `spontaneously' broken. The degeneracy of the states with a completely broken MTG is \(q^2\), but note that the new unit cell in real space is the same as in the \(\mathbb{Z}_q\) symmetric phases.

Interestingly, in the limit of large \(q\) the unit cell may exceed the size of a finite sample, resulting in a phase with essentially no translational order, similar to a vortex glass phase and may melt into a vortex liquid-like phase due to thermal fluctuations \cite{Fisher89, Huse90, BlatterLarkin94, JacobsenRadzihovsky99, Radzihovsky99, SaundersRadzihovsky00, Radzihovsky21}. Note that in the limit \(q\rightarrow\infty\), the degeneracy becomes infinite for both \(\mathbb{Z}_q\)-symmetric and symmetry breaking phases, in agreement with the standard result in the continuum GL theory of both ordinary and re-entrant superconductivity of Landau levels, depending on whether the cyclotron frequency is smaller or larger than the pairing energy scale \cite{RasoltTesanovich92}. In the former case, the pairing between different Landau levels causes interference that suppresses superconductivity. More generally, in the limit of irrational flux the normal state energy bands form a Cantor set of fractal dimension that depends on the irrational value of the flux \cite{Hofstadter76,AvilaJitomirskaya09}, with pseudo-Landau levels emanating from bottoms of bands at rational fluxes \cite{HarperSimon14,wang_classification_2020}. Following the same reasoning as for ordinary Landau levels, a similar suppression of Hofstadter SC is expected at small deviations of the magnetic field from rational flux values due to the interference from pairing between the pseudo-Landau levels. This raises also an important question of whether or not Hofstadter superconductors exhibit the Meissner effect or its analogue. A more detailed microscopic analysis of these questions will be presented elsewhere.

\section{BdG Spectrum and Chiral Hofstadter SC}\label{Chiral}

Having identified the \(\mathbb{Z}_q\)-symmetric Hofstadter SC phases, we now want to consider their excitation spectrum and topological properties. For this purpose we consider the \(2q\times2q\) BdG Hamiltonian Eq. (\ref{BdG}) describing pairing of electrons in a single Hofstadter band:
\[\mathcal{H}_{BdG}(\mathbf{p})=\left(\begin{array}{cc}
     \varepsilon(\mathbf{p}) \openone_{q \times q} &  \hat{\Delta}(\mathbf{p})\\
     \hat{\Delta}^\dagger(\mathbf{p}) & -\varepsilon(-\mathbf{p}) \openone_{q \times q}
\end{array}\right)\]
with the gap function \(\hat{\Delta}\) symmetric under some order \(q\) MTG symmetry
\[\hat{T}=\hat{T}_2^{N_2}(\theta_2)\hat{T}_1^{N_1}(\theta_1)\label{T}\]
with \(\theta_0=N_1\theta_1+N_2\theta_2=\frac{2\pi p}{q}M\) and \(M,N_1,N_2=0,\dots,q-1\). For simplicity, we will set \(N_1=1\) and \(N_2=N\) in this section. Interestingly, this additional symmetry allows us to completely diagonalize the BdG Hamiltonian in the spin polarized limit for any \(q\), as we show below.

In general, symmetry can also affect the topology of the system, potentially giving rise to symmetry protected topological phases (SPTs) \cite{ChenWen13,senthil2015symmetry, Wen17,ChiuRyu16}. Familiar cases are SPTs protected by anti-unitary symmetries such as PHS and time-reversal symmetry, as well as their unitary product, that are classified according to the 10-fold way \cite{AltlandZirnbauer97, Ryu10, Ludwig15}. This classification includes helical and chiral topological superconductors (TSCs) \cite{QiHughes10, SatoAndo17}, with the best known example of the latter being the chiral \(p\)-wave SC considered in a spin polarized system \cite{ReadGreen00}, characterized by a non-trivial Chern number. Including additional unitary symmetries in the classification results in additional SPTs, for example crystalline SPTs protected by additional lattice symmetries, e.g. inversion or mirror symmetries, that also include crystalline TSCs \cite{Hughes11, ChiuRyu13, ChiuSchnyder14, ShiozakiSato14, SumitaYanase19, Cornfeld19}. This also includes translational symmetry, which is generally assumed in all other classifications but is itself known to give rise to non-trivial SPTs, as well as symmetry enriched topological (SETs) phases with topological order \cite{ChengZaletel16}. Similarly, MTG symmetries are well known to play a role in determining Chern numbers in the quantum Hall effect \cite{TKNN,Kohmoto85,Lu19}.

It is therefore natural to ask whether the MTG symmetries can also give rise to novel TSC phases realised by Hofstadter superconductors. Remarkably, we demonstrate in Sec. \ref{ChernParity} that Hofstadter superconductors described by the BdG Hamiltonian Eq. (\ref{BdG}) can indeed realize SPT phases protected by MTG symmetries where the parity of the Chern numbers is the same as the parity of $q$. This result establishes that Hofstadter superconductors with a fully gapped spectrum fall into two topological classes according to the parity of $q$; i.e., they support non-Abelian (Abelian) excitations for odd (even) $q$. We illustrate the possible topological phase transitions for a specific form of the gap functions explicitly in Sec. \ref{ChiralPhaseTransitions} for \(q=3\) and \(5\).

Symmetries, including crystalline symmetries like inversion, can also lead to topologically protected gapless excitations \cite{MatsuuraRyu13, ZhaoWang13, ChiuSchnyder14, ShiozakiSato14, KobayashiSato14, ChiuRyu16, ZhaoSchnyder16}, resulting in nodal SCs or Bogoliubov Fermi surfaces (BFSs) \cite{KobayashiSato14,ZhaoSchnyder16,Agterberg17,BrydonAgterberg18,santos_pdw_2019,YuanFu18,SumitaYanase19,MenkeBrydon19,Link20,Lapp20,Shaffer20,ZhuFu20}. We find that in Hofstadter SCs this can happen in the presence of parity symmetry \(\mathcal{P}\) that together with PHS can protect BFSs.  In particular, we show in Sec. \ref{BFS} that the \(\mathbb{Z}_2\) topological invariant \(\nu_{\mathbb{Z}_2}\) defined in \cite{Agterberg17} is trivial for even \(q\) but non-trivial for odd \(q\), implying the existence of a BFS in that case. Though it is trivial for even \(q\), we propose a new topological invariant \(\nu_{\mathbb{Z}_2,0}\) that can be defined only in the presence of the MTG and only if \(M\) and \(N_2=N\) in Eq. (\ref{T}) have the same parity, and which remains non-trivial in that case.

\subsection{Symmetry and Spectrum of the BdG Hamiltonian }\label{BdGSpec}

We first review the action of symmetries on the BdG Hamiltonian. If a general symmetry acts as a matrix \(S(\mathbf{p})\) on the fermionic annihilation operators \(d_{\mathbf{p},\ell}\), it results in the action on the Nambu spinor \(\Psi_{\mathbf{p},\ell} = (d_{\mathbf{p},\ell}, d^\dagger_{-\mathbf{p},\ell})\)
\[\tilde{S}(\mathbf{p})=\left(\begin{array}{cc}
   S(\mathbf{p})  &  0\\
    0 & S^*(-\mathbf{p})
\end{array}\right)\,.\]
For example, a \(U(1)\) symmetry \(U(\theta)\) given by $d_{\bs{p},\ell} \rightarrow e^{i\theta}\,d_{\bs{p},\ell} \equiv U(\theta) d_{\bs{p},\ell}$ is represented in the BdG formalism by
\[\tilde{U}(\theta)=\left(\begin{array}{cc}
   e^{i\theta}  &  0\\
    0 & e^{-i\theta}
\end{array}\right)\label{U1BdG}\,.\]
The symmetries act on the BdG Hamiltonian as \(\tilde{S}(\mathbf{p})\mathcal{H}_{BdG}(\mathbf{p})\tilde{S}^{-1}(\mathbf{p})\), which is compatible with the transformation of the gap function in Eq. (\ref{DeltaTrans}). As discussed in Sec. \ref{BdGSym}, since the \(U(1)\) symmetry is broken by the SC phase, we are led to consider the family of symmetries \(\tilde{S}(\mathbf{p},\theta_0) = \tilde{U}(\theta_0/2)\tilde{S}(\mathbf{p})\). Note that if \(\tilde{S}^n=1\), then \(e^{i\theta_0}\) is an \(n^{th}\) root of unity.

Furthermore, we observe that the phase \(e^{i\theta_0}\) is  encoded in the commutation relations of \(\tilde{S}\) and the PHS \(\mathcal{C}=\tau^x\mathcal{K}\) where \(\tau^j\) are Pauli matrices acting on the particle/hole sectors of the Nambu spinor and \(\mathcal{K}\) is complex conjugation:
\[\tilde{S}(\mathbf{p},\theta_0)\mathcal{C}=e^{-i\theta_0}\mathcal{C}\tilde{S}(-\mathbf{p},\theta_0)\,.\]
We remark that these commutation relations have been used to classify gapless and fully gapped crystalline topological superconducting phases for \(\mathbb{Z}_2\) symmetries with \(\tilde{S}^2=1\), in which case the two possibilities \(e^{i\theta_0}=\pm 1\) result in different topological invariants  \cite{ShiozakiSato14,ChiuSchnyder14,KobayashiSato14,ZhaoSchnyder16,Ono19}. We leave the general question of whether the same approach can lead to new topological classifications in the presence of \(\mathbb{Z}_q\) symmetry for a future study. 

Here we will only invoke the classification in Sec. \ref{BFS} for the parity symmetry \(\mathcal{P}\) that reverses \(\mathbf{p}\rightarrow-\mathbf{p}\) and acts on the BdG Hamiltonian as
\[\tilde{\mathcal{P}}(0)\mathcal{H}_{BdG}(\bm{p})\tilde{\mathcal{P}}^{\dagger}(0)=\mathcal{H}_{BdG}(-\bm{p})\,.\]
The normal state is symmetric under \(\mathcal{P}\)
consistent with the condition \(\varepsilon(\mathbf{p})=\varepsilon(-\mathbf{p})\) necessary to guarantee the pairing instability in the first place. Note that we refer to \(\mathcal{P}\) as parity symmetry since while it is similar to inversion symmetry, the full inversion symmetry also acts on the patch indices as \(\ell\rightarrow-\ell\).
The gap function of the paired state may have either even or odd parity, \(\hat{\Delta}(\mathbf{p})=\pm \hat{\Delta}(-\mathbf{p})\), in which case the BdG Hamiltonian is symmetric under \(\tilde{\mathcal{P}}(0)\) or \(\tilde{\mathcal{P}}(\pi)\) respectively. Alternatively, \(\tilde{\mathcal{P}}(\theta_0)\) may be broken for any choice of \(\theta_0\).
Note that due to the PHS relation \(\hat{\Delta}(\mathbf{p})=-\hat{\Delta}^T(-\mathbf{p})\),
even and odd parity gap functions are skew-symmetric or symmetric matrices, \(\hat{\Delta}^T(\mathbf{p})=\mp \hat{\Delta}(\mathbf{p})\) respectively. 

Finally, the MTG symmetry \(\hat{T}(\theta_0)=\hat{T}_2^{N}(\theta_2)\hat{T}_1(\theta_1)\) with \(\theta_0=\frac{2\pi p}{q}M\) and \(M, N=0,\dots,q-1\) acts in the Nambu basis via
\[\tilde{T}(\theta_0)=\left(\begin{array}{cc}
   e^{i\theta_0/2}\hat{\sigma}^{-N}\hat{\tau}  &  0\\
    0 & e^{-i\theta_0/2}\hat{\sigma}^{N}\hat{\tau}
\end{array}\right)\]
where \(\hat{\tau}\) and \(\hat{\sigma}\) are the shift and clock matrices defined in Eq. (\ref{tausigma}). With this symmetry the gap function \(\hat{\Delta}\) has the form given in Eqs. (\ref{GapOdd}) and (\ref{GapEven}) for odd and even \(q\) respectively:
\[\hat{\Delta}^{(M)}_{\ell,\ell'}=\Delta_{\ell-\ell'}\exp\left[i\phi_{\ell+\ell'}^{(M)}\right]\,;\label{DeltaM}\]
some of the indices are assumed to be fixed and thus we omit them for clarity. Due to the breaking of the MTG symmetries, there are \(q\) or \(q/2\) degenerate ground states for different values of \(M\). The phases \(\phi_L\) determined by the \(\tilde{T}\) symmetry are given in Eqs. (\ref{phiL}) and (\ref{phiLeven}). The functions \(\Delta_\ell(\mathbf{p})\), defined in Eqs. (\ref{DeltaEllOdd}) and (\ref{DeltaEllEven}) for odd and even \(q\) respectively, are ultimately determined by the microscopic interactions Eq. (\ref{Hint}) via the gap equation Eq. (\ref{LinGapEq}). In our phenomenological approach we treat them as arbitrary functions. Note that in the spin polarized case considered here, PHS additionally requires \(\Delta_\ell(\mathbf{p})=-\Delta_{-\ell}(-\mathbf{p})\); in particular, \(\Delta_0\), as well as \(\Delta_{q}\) for even \(q\) (for which \(\ell\) is defined modulo \(2q\)), have to be an odd function of \(\mathbf{p}\), which in general has to be chiral for the spectrum to be fully gapped.

As with any symmetry, we can simultaneously diagonalize the BdG Hamiltonian and \(\tilde{T}\). In particular, since \(\tilde{T}\) is two-fold degenerate, we can write the BdG Hamiltonian Eq. (\ref{BdG}) in a \(2\times2\) block-diagonal form using a basis that simultaneously diagonalizes \(\tilde{T}(0)\). This is useful because the eigenstates of \(\tilde{T}(0)\) are known: the eigenvalues of the matrix \(\hat{\sigma}^{N}\hat{\tau}\) are \(\lambda^{(n)}=\omega_q^{\frac{pN(q+1)}{2}}\omega_q^{p n}\), with corresponding eigenstates
\[f_{\mathbf{p},n}=\frac{1}{\sqrt{q}}\omega_q^{pN\frac{\ell(\ell+1)-\ell(q+1)}{2}-pn\ell}d_{\mathbf{p},\ell}\label{f}\]
with a sum over \(\ell=0,\dots,q-1\) on the right-hand side implied and with \(n=0,\dots,q-1\) being the index in the new basis. Using this new basis we can define a new Nambu spinor \(\Psi'_{\mathbf{p},n}=(f_{\mathbf{p},n},f_{-\mathbf{p},n}^\dagger)\) and the corresponding transformed BdG Hamiltonian. For odd \(q\), the gap function for the \(M^{\text{th}}\) ground state as given in Eq. (\ref{DeltaMN1N2}) and (\ref{phiL}) becomes a \(L^{\text{th}}\) anti-diagonal matrix \(\hat{\Delta}'\) with \(L=N-M\) modulo \(q\):
\[\hat{\Delta}'=\left(\begin{array}{cccccccc}
    & & & &\hat{\Delta}'_{0L} &  \\
     & & & \hat{\Delta}'_{1,L-1} & & \\
    &  \iddots & &  & & \\
    \hat{\Delta}'_{L0}  \\
    & & & & & &&\hat{\Delta}'_{L+1,q}\\
     & & & &&&  \iddots &\\
      & & &&& \hat{\Delta}'_{q,L+1}\\
\end{array}\right)\label{DeltaPrime}\]
with the only non-zero elements given by
\begin{widetext}
\[\hat{\Delta}'_{n,L-n}=\frac{1}{2}\sum_{\ell=0}^{q-1}\omega^{-N\frac{\ell^2}{4}+\ell(n+M/2)}\left[1+(-1)^\ell+(1-(-1)^\ell)(-1)^{pM}i^{pqN}\right]\Delta_\ell\equiv\Delta'_{2n-L}\label{DeltaEvals}\]
\end{widetext}
while the rest vanish; this formula holds for the even \(q\) case as well with \(\ell\) defined modulo \(2q\) and restricted to only even or only odd values depending on the irrep. In this basis, the gap functions for different values of \(M\) correspond to different anti-diagonals, just as the \(\hat{T}_2(\theta_2)\) symmetric irrep components \(\hat{\Delta}^{(L)}\) defined in Eq. (\ref{DeltaL}).

For a fixed \(M\), after a reshuffling the transformed BdG Hamiltonian splits into \(2\times2\) blocks. For example, for \(q=3\) and \(L=2\) we have
\[\mathcal{H}'_{BdG}=\left(\begin{array}{cccccc}
     \varepsilon & 0 & 0 & 0 & 0 & \Delta'_1  \\
     0 & \varepsilon & 0 & 0 & \Delta'_0 & 0  \\ 
     0 & 0 & \varepsilon & \Delta'_2 & 0 & 0  \\ 
     0 & 0 & \Delta'^{*}_2 & -\varepsilon & 0 & 0  \\  
     0 & \Delta'^{*}_0 & 0 & 0 & -\varepsilon & 0  \\
     \Delta'^{*}_1 & 0 & 0 & 0 & 0 &  -\varepsilon \\
\end{array}\right)\,.\]
It is easy to read off the \(2\times 2\) blocks, and for any \(q\) they are
\[\mathcal{H}_{BdG,2n-L}=\left(\begin{array}{cc}
\varepsilon & \Delta'_{2n-L}\\
\Delta'^{*}_{2n-L} & -\varepsilon
\end{array}\right)\label{HBdGn}\,.\]
The BdG Hamiltonian thus splits into \(q\) blocks, each being a single-band triplet SC with effective gap function \(\Delta'_{2n-L}\).
The eigenvalues of the \(2\times2\) blocks are simply
\[\label{eq: eigenenrgies} E_{n\pm}^{(M)}(\mathbf{p})=\pm\sqrt{\varepsilon^2(\mathbf{p})+\left|\Delta'_{2n-L}(\mathbf{p})\right|^2}\,,\]
and the Nambu eigenspinors of the BdG Hamiltonian are
\[\left|\Upsilon_{n\pm}^{(M)}\right\rangle=\frac{(\varepsilon+E_{n\pm}^{(M)})\hat{\mathbf{e}}_{n,u}+\Delta'_{2n-L}\hat{\mathbf{e}}_{-n-M,v}}{\sqrt{(\varepsilon+E_{n\pm}^{(M)})^2+|\Delta'_{2n-L}|^2}}\label{espinors} \,,\]
where \(\hat{\mathbf{e}}_{n,u}\) and \(\hat{\mathbf{e}}_{n,v}\) are unit basis vectors with \(u\) and \(v\) denoting particle and hole components of the spinors.

\subsection{Chern Number Parity and Phase Transitions}\label{ChernParity}

As in the regular spin polarized SC \cite{ReadGreen00}, in the spin polarized Hofstadter SC the order parameter is either gapless or chiral. 
The fact that in the presence of the \(\mathbb{Z}_q\) symmetry the BdG Hamiltonian has a \(2\times2\) block structure in Eq. (\ref{HBdGn}) implies that the total Chern number can be computed as a sum of Chern numbers of each block given by integrating  the Berry curvature \(\mathbf{F}_{n\pm}^{(M)}\) defined using the Nambu eigenspinor Eq. (\ref{espinors}):
\begin{align}
\mathbf{F}_{n\pm}^{(M)}&=-i\boldsymbol{\nabla}\times\left\langle\Upsilon_{n\pm}^{(M)}\right|\boldsymbol{\nabla}\left|\Upsilon_{n\pm}^{(M)}\right\rangle
\,.
\end{align}
Here we observe that with the chemical potential in the normal state fixed, the parity of the Chern numbers cannot change in a topological phase transition, assuming the normal state Fermi surface does not cross high-symmetry points. This is because the Chern number changes at the phase transition due to the gap closing at Dirac nodes, as happens for example in phase transitions in quantum Hall systems and Chern insulators \cite{TKNN, Simon83, Kohmoto85, wang_classification_2020}. 
In contrast to Chern insulators and quantum Hall states, in chiral SCs the Dirac nodes appear in pairs due to PHS that maps a Dirac node at \(\mathbf{p}_D\) to a second node at \(-\mathbf{p}_D\), which, under general conditions, satisfies
\(\mathbf{p}_D\neq-\mathbf{p}_D\).
The condition \(\varepsilon(\mathbf{p})=\varepsilon(-\mathbf{p})\) together with Eq. (\ref{eq: eigenenrgies}) imply that the nodes moreover appear at zero energy and at the Fermi momentum. The change in the Chern number associated with each node in the pair is the same, and therefore the total change in the Chern number is  an even integer \footnote{Note that if the Fermi surface does contain high-symmetry points with \(\mathbf{p}_D=-\mathbf{p}_D\), then \(\hat{\Delta}(\mathbf{p}_D)\) is an anti-symmetric matrix by PHS, which as we discuss in Sec. \ref{BFS} implies that \(\Delta'_n(\mathbf{p}_D)\) that appear in the eigenvalues of the BdG Hamiltonian in Eq. (\ref{eq: eigenenrgies}) satisfy \(\Delta'_{-n}(\mathbf{p}_D)=-\Delta'_n(\mathbf{p}_D)\). As a result, there are band touchings at \(\mathbf{p}_D\) and the Berry connection, along with the Chern number, is not well-defined unless we assume that the Fermi surface does not contain such high-symmetry points.}.

This implies that if the spin-polarized Hofstadter SC is not gapless, its Chern number has the same parity as $q$. To see this, consider the special case when \(\hat{\Delta}\) is diagonal, i.e. \(\Delta_\ell(\mathbf{p})\) are all zero in Eq. (\ref{DeltaM}) except for \(\Delta_0(\mathbf{p})\). For odd \(q\), the BdG Hamiltonian splits into \(q\) identical \(2\times2\) blocks; for even \(q\), the blocks only differ by an overall momentum-independent phase of the gap function. Since \(\Delta_0(\mathbf{p})\) has to be an odd function of momentum, the resulting spectrum is fully gapped only if \(\Delta_0(\mathbf{p})\propto (p_x\pm ip_y)^m\) with odd values of \(m\). The Chern number for each block is therefore the same odd number (for any parity of \(q\)). The total Chern number is the sum of the Chern numbers of each block, and is therefore even or odd for even and odd \(q\) respectively. Since the Chern number can only change by an even integer as \(\Delta_\ell(\mathbf{p})\) are varied, we conclude that the Chern number parity is always the same as the parity of \(q\) for any choice of \(\Delta_\ell(\mathbf{p})\). The assumption of \(\tilde{T}\) symmetry can be lifted, but note that the presence of \(q\) bands is a direct consequence of the MTG symmetry of the normal state.

\subsubsection*{Phase Diagrams for \(q=3\) and \(5\)}\label{ChiralPhaseTransitions}

In this subsection we illustrate the conservation of the parity of the Chern number in the special cases of \(q=3\) and \(5\), and obtain generic phase diagrams. Here we will assume that \(\tilde{T}=\tilde{T}_2(0)\tilde{T}_1(0)\), corresponding to \(N_1=N_2=1\) and \(M=0\). For \(q=3\) the corresponding gap function was given in Eq. (\ref{q3DeltaM11}):
\begin{eqnarray}
\hat{\Delta} = 
\begin{pmatrix}
\Delta_0 & \Delta_1 & \Delta_2 \mathrm{e}^{-i2\pi/3} 
\\
\Delta_2 & \Delta_0 \mathrm{e}^{-i2\pi/3}& \Delta_1  
\\
\Delta_1 \mathrm{e}^{-i2\pi/3} & \Delta_2 & \Delta_0 
\end{pmatrix}
\,.
\end{eqnarray}
We consider the functions \(\Delta_\ell(\mathbf{p})\) to be of the following general form expanded around the Fermi surface:
\begin{align}
\Delta_0(\bm{p}) & = a_0(\bm{p})(p_x+ip_y)/p_F, \\
\Delta_1(\bm{p}) &= -\Delta_2(-\bm{p}) = a_1(\bm{p})(p_x+ip_y)/p_F+b_1(\bm{p})\nonumber
\end{align}
where $p_F = \sqrt{2 m \mu}$ and \(a_0, a_1\) and \(b_1\) are complex even functions of \(\mathbf{p}\).
The pairing matrix is such that in the limit \(\Delta_1=0\) the spectrum remains fully gapped unless \(a_0=0\); in that case note that the total Chern number is just \(q\) times the winding of \(\Delta_0(\mathbf{p})\). The terms proportional to $a_0(\bs{p})$ and $a_1(\bs{p})$ constitute the odd parity components of the gap function, 
while the $b_1(\bs{p})$ term is an even parity component. We consider the possibility that parity is broken and thus consider both components coexisting.

\begin{figure}
\centering
\includegraphics[width=0.48\textwidth]{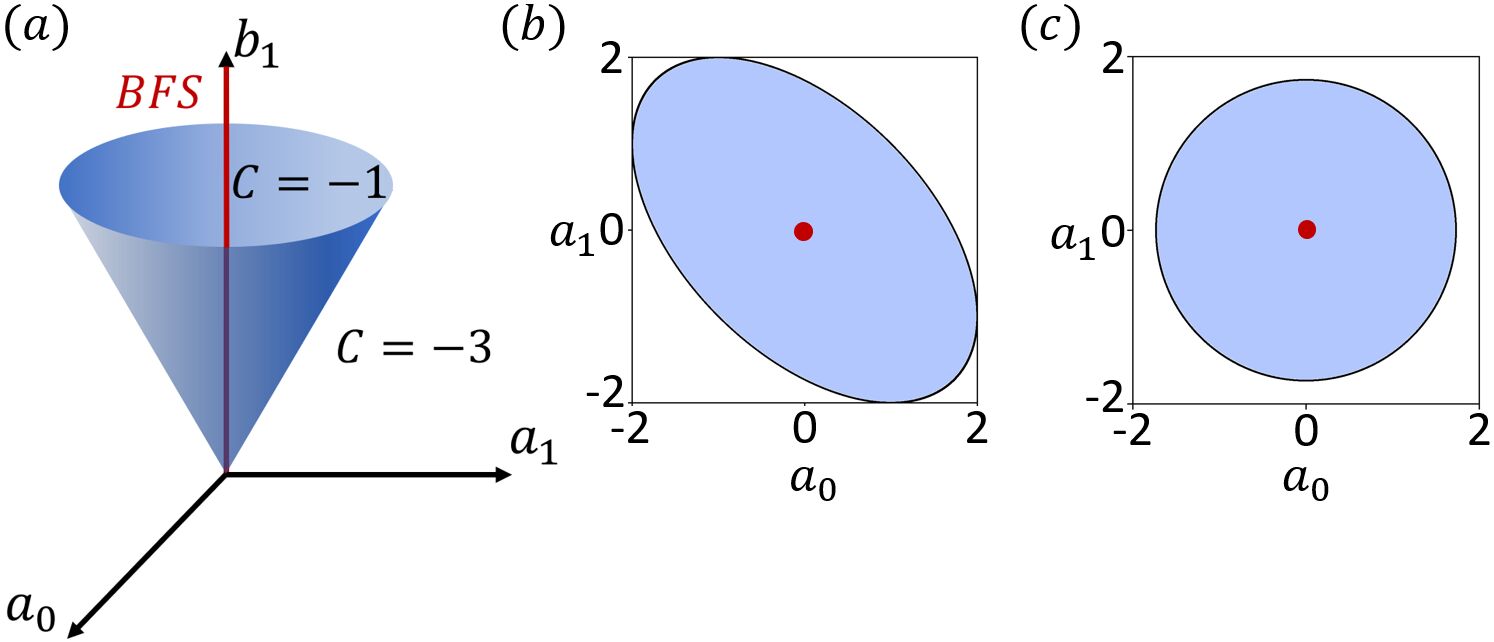}
\caption{Phase diagrams for the spin-polarized Hofstadter SC at $q=3$ in the $\hat{T}_2\hat{T}_1$ symmetric phase. (a) Full phase diagram in the space of \(a_0\), \(a_1\) and \(b_1\), with \(\vartheta=0\). The $C=-1$ and $C=-3$ phases are separated by a conical phase boundary. On the $b_1$ axis, the system is gapless and has a symmetry protected Bogoliubov Fermi surface (BFS). (b) Cut along \(b_1=1\) with $\vartheta = 0$. (c) Same as (b) but with \(\vartheta=-\pi/6\). The phase boundary is a circle in this case.}
\label{fig:q=3phaseDiagram}
\end{figure}

We first consider the simplest case when $a_0, a_1$ and $b_1$ are  constants and compute the Chern number numerically using the algorithm in \cite{Fukui2005}. We identify two phases with Chern numbers \(C=-3\) and \(C=-1\), as well as a gapless phase at \(a_0=a_1=0\). The \(C=-3\) is the expected phase for \(\Delta_1=\Delta_2=0\), but once \(b_1\) is sufficiently large there is a phase transition into the \(C=-1\) phase. The phase boundary between the two phases is an elliptic cone of eccentricity that depends on the relative phase between the $p$-wave components of the gap functions, $\vartheta=\mathrm{Arg}[a_0]-\mathrm{Arg}[a_1]$. In the phase diagram shown in Fig. \ref{fig:q=3phaseDiagram} we therefore took 
\begin{align}
\Delta_0(\bm{p}) & = a_0(p_x+ip_y)/p_F, \\
\Delta_1(\bm{p}) &= -\Delta_2(-\bm{p}) = a_1e^{i\vartheta}(p_x+ip_y)/p_F+b_1\nonumber
\end{align}
with \(a_0, a_1\) and \(b_1\) all real, with \(\vartheta=0\) in Fig. \ref{fig:q=3phaseDiagram} (b) and \(\vartheta=-\pi/6\) in Fig. \ref{fig:q=3phaseDiagram} (c).

As mentioned above, the spectrum is gapless along the \(b_1\) axis, with a Fermi surface coinciding with the normal state Fermi surface. This is a simple consequence of the matrix \(\hat{\Delta}\) being anti-symmetric along this axis and therefore having a zero eigenvalue, leading to an ungapped energy band. Below we will show that this Bogoliubov Fermi surface is not accidental and occurs for all odd \(q\). It is moreover topologically protected by the combination of PHS \(\mathcal{C}\) and parity symmetry \(\mathcal{P}\).

\begin{figure}
\centering
\includegraphics[width=0.48\textwidth]{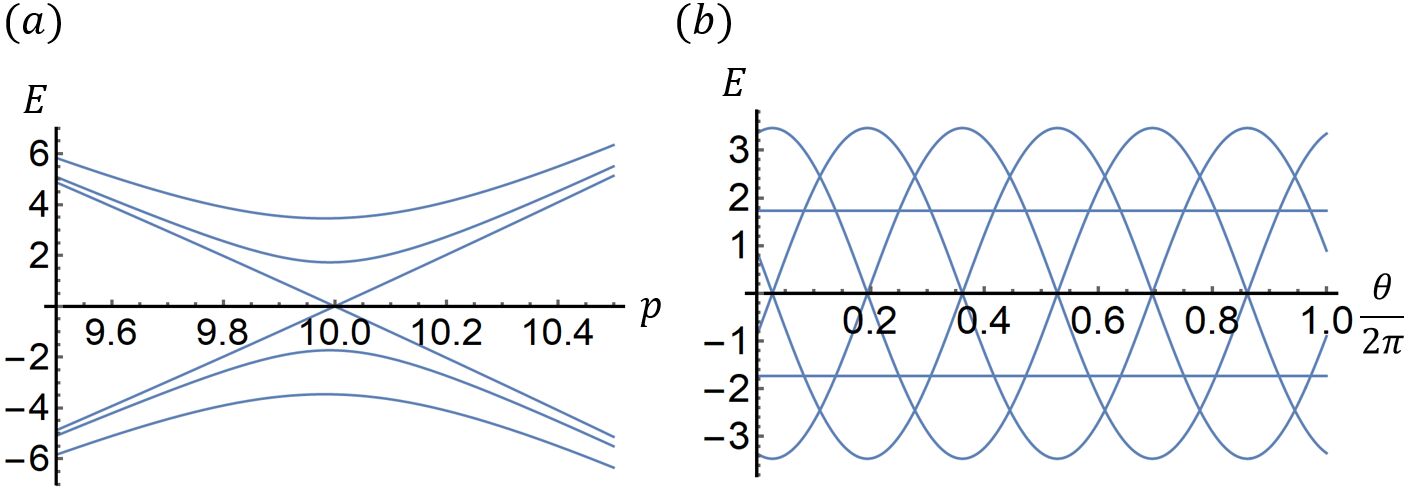}
\caption{BdG spectrum for \(q=3\) with $\Delta_0(\bm{p}) = \sqrt{3}(p_x+ip_y)^3/p_F^3$, $p_F = 10$ and $\Delta_1 = -\Delta_2 = 1$. We work in arbitrary units for the Fermi momentum and consider $p_{F} \ll \pi/a$. $6$ Dirac nodes indicate a topological phase transition with Chern number changing from $C=-9$ to $C= -3$. (a) A cut of the BdG spectrum along the $\hat{p}$ direction at the Dirac node. (b) A cut of the BdG spectrum along the Fermi momentum $p = p_F = 10$ as a function of the angle \(\theta\) of \(\mathbf{p} = p_{F} (\cos{\theta}, \sin{\theta})\), which shows the presence of $6$ Dirac touchings. Note also the unavoided crossings at non-zero energies indicative of the presence of the MTG symmetry \(\tilde{T}\).}
\label{fig:mDiraccones}
\end{figure}

While Fig. \ref{fig:q=3phaseDiagram} captures the general features of the phase transitions, we note that phase transitions between higher odd Chern numbers are in principle possible with a larger even number of Dirac nodes at the phase transitions. For example, with $\Delta_0(\bm{p}) = \sqrt{3}(p_x+ip_y)^m/p_F^m$, $m\in \mathbb{Z}$ and $\Delta_1 = -\Delta_2 = 1$, the BdG spectrum has \(2m\) Dirac nodes, corresponding to a phase transition between \(C=-3m\) and \(C=-m\) phases. Fig. \ref{fig:mDiraccones} illustrates the case of \(m=3\) with six Dirac cones. We also note that the presence of the MTG symmetry \(\tilde{T}\) generally results in unavoided crossings away from zero energy, as seen in Fig. \ref{fig:mDiraccones} (b).

\begin{figure}
\centering
\includegraphics[width=0.48\textwidth]{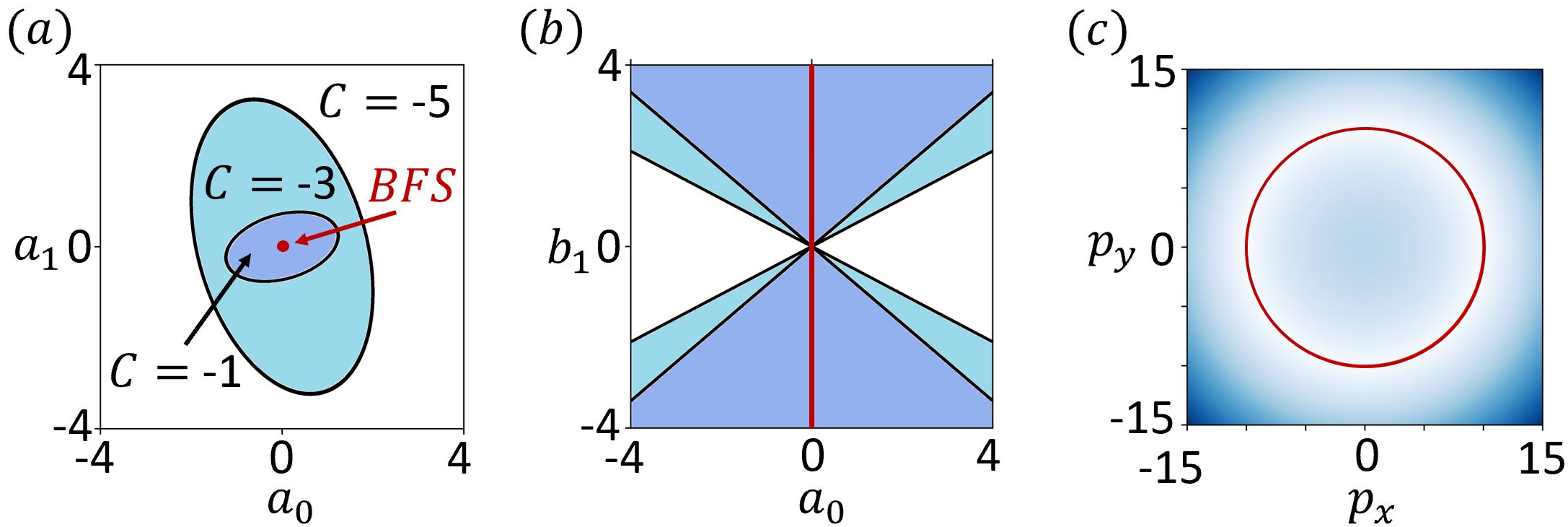}
\caption{The phase diagrams for \(q=5\)  with $\Delta_0 = a_0(p_x+ip_y)/p_F, \Delta_1 = a_1(p_1+ip_2)/p_F+b_1, \Delta_2=0$, \(\Delta_3(\mathbf{p})=-\Delta_2(-\mathbf{p})\) and \(\Delta_4(\mathbf{p})=-\Delta_1(-\mathbf{p})\). Cuts of the nested conical phase boundaries and Bogoliubov Fermi surface are shown along (a) $b_1=1$ and (b) $a_1=0$. (c) Bogoliubov Fermi surface (E=0) located at $|\bm{p}|=10$ is shown in red for $\Delta_0=0, 
\Delta_1 = 1, \Delta_2=0$. }
\label{fig:q=5phaseDiagram}
\end{figure}

More phases and phase transitions are possible for larger \(q\), but always with Chern numbers of the same parity as \(q\). We illustrate this for \(q=5\) with $\Delta_0 = a_0(p_x+ip_y)/p_F, \Delta_1 = a_1(p_x+ip_y)/p_F+b_1, \Delta_2=0$, \(\Delta_3(\mathbf{p})=-\Delta_2(-\mathbf{p})\) and \(\Delta_4(\mathbf{p})=-\Delta_1(-\mathbf{p})\). As shown in Fig. \ref{fig:q=5phaseDiagram}, there are now two phase transitions as \(b_1\) increases, with two nested conical phase boundaries. Again, there is a topologically protected BFS when \(a_0=a_1=0\), as we show in the next section.

\subsection{Symmetry-Protected Bogoliubov Fermi Surfaces}\label{BFS}

Here we show that the BFS discussed in the previous section are topologically protected in the presence of the parity symmetry \(\mathcal{P}\) and are a general feature of the phase diagram of \(\mathbb{Z}_q\) Hofstadter SCs: for odd \(q\) a BFS is always present and protected by a topological invariant that has been established in \cite{Agterberg17}; for even \(q\), a doubly degenerate BFS exists depending on the irrep the gap function belongs to and protected by a new topological invariant that can only be defined in the presence of MTG symmetries. The key observation is that the BFS appeared when the matrix \(\hat{\Delta}\) was anti-symmetric, \(\hat{\Delta}(\mathbf{p})=-\hat{\Delta}^T(\mathbf{p})\).  Since \(\hat{\Delta}\) and \(\hat{\Delta}^T\) have the same spectrum, this implies that their eigenvalues must appear in pairs with opposite signs. Assuming a \(\mathbb{Z}_q\) symmetry, these eigenvalues are \(i\Delta'_{n}\) with \(\Delta'_{n}\) given in Eq. (\ref{DeltaEvals}), which therefore satisfy \(\Delta'_{n}=-\Delta'_{-n}\). For odd \(q\) this implies that \(\Delta'_{0}=0\), and we conclude that two of the energy bands in Eq. (\ref{eq: eigenenrgies}) are \(E_{\pm}=\pm|\varepsilon(\mathbf{p})|\). In particular, they cross zero energy at the Fermi momentum, forming the BFS at the original normal state Fermi surface.

For even \(q\), note that Eq. (\ref{DeltaEvals}) implies that when \(\hat{\Delta}'\) is an \(L^\text{th}\) anti-diagonal matrix, \(n\) in \(\Delta'_n\) is either an even or odd integer modulo \(2q\) for \(L\) even or odd respectively. When \(L\) is odd, therefore, we conclude that in general none of the \(\Delta'_n\) vanish, while for even \(L\) two of them vanish, namely \(\Delta'_0\) and \(\Delta'_q\). In the latter case the BFS is doubly degenerate and formed by four bands instead of two. As we discuss below, this implies that the BFS is topologically protected for odd \(q\) but not in general for even \(q\). We conjecture that the doubly degenerate BFS is topologically protected for even \(q\) and even \(L\) as long as the \(\mathbb{Z}_q\) MTG symmetry is unbroken.

To establish whether the BFS is topologically protected, we need to compute the corresponding topological invariant.
As discussed in Sec. \ref{BdGSpec}, due to the PHS relation \(\hat{\Delta}(\mathbf{p})=-\hat{\Delta}^T(-\mathbf{p})\), \(\hat{\Delta}(\mathbf{p})\) being anti-symmetric is equivalent to it being even under \(\mathcal{P}\): \(\hat{\Delta}(\mathbf{p})=\hat{\Delta}(-\mathbf{p})\). It has been shown in \cite{KobayashiSato14} that for even parity gap functions there does indeed exist a \(\mathbb{Z}_2\) topological invariant \(\nu_{\mathbb{Z}_2}\) that can protect BFSs. This invariant was determined in \cite{Agterberg17} using the Pfaffian of the BdG Hamiltonian for \(4N\times4N\) BdG Hamiltonians, with the invariant being non-trivial if the Pfaffian changes sign as the BFS is crossed. As we will see, we can generalize this invariant for \(2q\times2q\) BdG Hamiltonian even when \(q\) is odd.

Note that while the Pfaffian is generally only defined for anti-symmetric matrices of even dimensions, the Pfaffian \(\Pf[A]\) of a matrix \(A\) can be more generally defined as a polynomial in the elements of \(A\) with integer coefficients such that \(\Pf^2[A]=\text{Det}[A]\). We therefore first want to find the determinant of \(\mathcal{H}_{BdG}\) which we can do by direct computation using the eigenvalues of the BdG Hamiltonian in Eq. (\ref{eq: eigenenrgies}):
\[\text{Det}[\mathcal{H}_{BdG}]=(-1)^q\prod_{n}\left(\varepsilon^2+|\Delta'_n|^2\right)\]
where the product is over all values of \(n\). Notice that this is positive for even \(q\) but negative for odd \(q\). We therefore compute the Pfaffian of \(\tau^z\mathcal{H}_{BdG}\) instead of \(\mathcal{H}_{BdG}\) itself, the two being equivalent for even \(q\). \(\tau^z\) is a Pauli matrix acting on the particle and hole sectors (it can be replaced with any \(2q\times2q\) matrix with integer elements and determinant equal to \((-1)^q\)).

This still does not guarantee that \(\text{Det}[\tau^z\mathcal{H}_{BdG}]\) is a square of a polynomial, and we need to invoke the anti-symmetry of \(\hat{\Delta}\). As pointed out above, when \(\hat{\Delta}\) is an anti-symmetric matrix,
\(\Delta'_{-n}=-\Delta'_{n}\). Using this fact, we conclude that when \(\hat{\Delta}'\) is an \(L^\text{th}\) anti-diagonal matrix,
\begin{widetext}
\[\Pf[\tau^z\mathcal{H}_{BdG}{(\bm{p}})]=
\left\{\begin{array}{ll}
    \prod_{n=0}^{q/2-1}\left(\varepsilon^2(\bm{p})+\left|\Delta'_{2n+1}(\bm{p})\right|^2\right), & q \text{ even, }L \text{ odd}\\
   \varepsilon^2(\bm{p})\prod_{n=1}^{q/2-1}\left(\varepsilon^2(\bm{p})+\left|\Delta'_{2n}(\bm{p})\right|^2\right), & q \text{ even, }L \text{ even}\\
   \varepsilon(\bm{p})\prod_{n=1}^{q/2-1}\left(\varepsilon^2(\bm{p})+\left|\Delta'_{n}(\bm{p})\right|^2\right),              & q \text{ odd}
\end{array}\right.\]
\end{widetext}
where the product is restricted to only include one of each pair of \(\Delta'_{\pm n}(\bm{p})\). As \(\varepsilon^2(\bm{p})+\left|\Delta'_{n}(\bm{p})\right|^2\) are positive-definite quantities, we conclude that the Pfaffian does not change sign when \(q\) is even, but necessarily changes sign at the BFS for odd \(q\) because \(\varepsilon(\bm{p})\) does.
The \(\mathbb{Z}_2\) topological invariant can be defined as
\[\nu_{\mathbb{Z}_2}=\sgn\left\{\Pf\left[\tau^z\mathcal{H}_{BdG}(\bm{p}_{+})\right]\Pf\left[\tau^z\mathcal{H}_{BdG}(\bm{p}_{-})\right]\right\}\]
where $\bm{p}_{+}$ ($\bm{p}_{-}$) is any momentum inside (outside) of the BFS; this coincides with the definition in \cite{Agterberg17} for even \(q\). The invariant is trivial when it is equal to \(1\) and non-trivial when it is equal to \(-1\), i.e. when the Pfaffian changes sign across the BFS. The topological invariant is therefore trivial for even \(q\) but non-trivial for odd \(q\), and so the BFS is topologically protected in the latter case.

Note that while we made use of the MTG \(\mathbb{Z}_q\) symmetry in the proof of the existence of the BFS for odd $q$, 
since the stability of the BFS relies only on PHS and parity symmetry \cite{KobayashiSato14,ZhaoSchnyder16}, 
the BFS established for odd $q$ remains perturbatively stable if MTG is broken 
as long \(\hat{\Delta}(\mathbf{p})\) is an even function of \(\mathbf{p}\). However, in the presence of the MTG symmetry we can simplify the invariant by noting that only the \(n=0\) block of the BdG Hamiltonian contributes to the sign change of the Pfaffian. Instead of using the Pfaffian of the whole BdG Hamiltonian we can therefore use the Pfaffian of \(\mathcal{H}_{BdG,0}\) in Eq. (\ref{HBdGn}). Note incidentally that the Pfaffian of a direct sum of two matrices is a product of the Pfaffians of the two matrices: \(\Pf[A\oplus B]=\Pf[A]\Pf[B]\), assuming the RHS exists. Since
\[\text{Det}\left[\tau^z\mathcal{H}_{BdG,0}\right]=-\text{Det}\left[\left(\begin{array}{cc}
   \varepsilon  & 0 \\
    0 & -\varepsilon
\end{array}\right)\right]=\varepsilon^2\,,\]
the Pfaffian is \(\Pf\left[\tau^z\mathcal{H}_{BdG,0}\right]=\varepsilon\). We can therefore instead define a \(\mathbb{Z}_2\) invariant
\[\nu_{\mathbb{Z}_2,0}=\sgn\left\{\Pf\left[(\tau^z\mathcal{H}_{BdG,0}(\bm{p}_{+})\right] \Pf\left[\tau^z\mathcal{H}_{BdG,0}(\bm{p}_{-})\right]\right\}\,.\label{nuZ20}\]
While for odd \(q\) we simply have \(\nu_{\mathbb{Z}_2,0}=\nu_{\mathbb{Z}_2}\), we note that interestingly we can also define \(\nu_{\mathbb{Z}_2,0}\) for even \(q\) assuming that \(L\) is even. In that case we can also analogously define \(\nu_{\mathbb{Z}_2,q}\) indicating a sign change of \(\Pf\left[\tau^z\mathcal{H}_{BdG,q}\right]=\varepsilon\). We then have \(\nu_{\mathbb{Z}_2}=\nu_{\mathbb{Z}_2,0}\nu_{\mathbb{Z}_2,q}\) is trivial, while the new invariant \(\nu_{\mathbb{Z}_2,0}\) is not, indicating the presence of a doubly degenerate BFS.

Importantly, \(\nu_{\mathbb{Z}_2,0}\) exists for even \(q\) only when \(L\) is even, and can only be defined in the presence of the \(\mathbb{Z}_q\) MTG symmetry in addition to PHS and parity \(\mathcal{P}\), unlike \(\nu_{\mathbb{Z}_2}\).
Recall that when the \(\mathbb{Z}_q\) symmetry is \(\hat{T}(\theta_0)=\hat{T}_2^N\hat{T}_1\) with \(\theta_0=\frac{2\pi p}{q}M\), \(L=N-M\) modulo \(q\) and note that \(\hat{T}(\theta_0)\) commutes or anti-commutes with \(\hat{T}_1^{q/2}(0)\) when \(L\) is even or odd. Recall also that gap functions are even or odd under \(\hat{T}_1^{q/2}(0)\) for even and odd \(M\) respectively and belong to different irreps of the MTG, as we showed in Sec. \ref{section:irreps}. The parity of \(N\), on the other hand, determines whether the \(\mathbb{Z}_q\) symmetry \(\hat{T}(\theta_0)\) commutes or anti-commute with \(\hat{T}_1^{q/2}(0)\). We therefore conclude that for even \(q\), \(\nu_{\mathbb{Z}_2,0}\) is a new topological invariant protected by the combination of PHS, parity \(\mathcal{P}\), and \(\mathbb{Z}_q\) symmetries when \(M\) and \(N\) have the same parity. We note the similarity of the definition of \(\nu_{\mathbb{Z}_2,0}\) to that of topological invariants of crystalline SPT phases defined by similarly simultaneously block-diagonalizing the Hamiltonian and the crystalline symmetry \cite{Hughes11, ChiuRyu13}. Our result therefore shows that Hamiltonians with MTG symmetries require a new classification of their topological invariants.

\section{Summary and Discussion}

In this paper, we have provided a detailed analysis of the properties of electrons undergoing pairing instabilities in time-reversal broken Hofstadter bands, which characterize the spectrum of single particle states in 2D lattices with magnetic flux $\Phi = (p/q)\Phi_0$ per unit cell.
Our approach focused on exploring the consequences of the magnetic translation symmetries on the paired state. A central result of this work is the classification of the irreducible representations of the magnetic translation group furnished by the pairing matrix $\hat{\Delta}$, which we established to have different properties from the familiar irreps furnished by single particle Bloch states. 
Furthermore, the group theory analysis shows that at least one magnetic translation symmetry is \emph{necessarily} broken in the paired state, and we find numerically that all of them can be broken at least for \(q\geq5\).

Building on the properties of the irreps of the magnetic translation group, we have formulated an effective Ginzburg-Landau theory to study the thermodynamic properties of Hofstadter superconductors at general fluxes $\Phi = (p/q)\Phi_0$ with rational \(p/q\). The theory is constructed in terms of a complex valued vector order parameter $\bs{\eta}$ of dimension $q$ ($q/2$) for odd (even) $q$.
Moreover, we found that the multi-component nature of the order parameter yields a rich phase diagram characterized by different symmetry breaking patterns of the magnetic translation group, which can be interpreted as distinct classes of ``vortex lattices."

An important class of thermodynamic phases we have identified corresponds to $\mathbb{Z}_q$-symmetric Hofstadter superconductors, in which the magnetic translation group breaks down to a $\mathbb{Z}_q$ subgroup resulting in \(q/2\)- or \(q\)-fold degenerate ground states for even and odd \(q\), respectively, with the degeneracy equal to the dimension of the irreps. 
Furthermore, we have shown that chiral $\mathbb{Z}_q$-symmetric Hofstadter superconductors provide a natural setting for the realization of topological superconductivity with tunable Chern numbers. In particular, we have established that when pairing only involves electrons in a single Hofstadter band, Hofstadter superconductors can realize SPT phases protected by magnetic translation symmetries where the parity of the Chern numbers is the same as the parity of $q$. This property establishes that Hofstadter superconductors with a fully gapped spectrum fall into two topological classes according to the parity of $q$; i.e., they support non-Abelian (Abelian) excitations for odd (even) $q$.
Moreover, we have shown that when the Hofstadter superconductor additionally possesses parity symmetry, its spectrum  \textit{necessarily} supports Bogoliubov Fermi surfaces (BFS) for odd \(q\), even when the \(\mathbb{Z}_q\) symmetry is broken by the order parameter as the associated topological invariant defined in \cite{Agterberg17} does not require it. For even \(q\), in contrast, parity alone cannot protect the BFSs, which are necessarily doubly degenerate if they exist. However, we also showed that this doubly degenerate BFS can be protected by the \(\mathbb{Z}_q\) MTG symmetry in cases when a new topological invariant, given in Eq. (\ref{nuZ20}), can be defined. This illustrates the fact that a new topological classification is required for \(\mathbb{Z}_q\)-symmetric Hofstadter superconductors.

This work raises a number of interesting questions that deserve future attention. 
First, given that the ground states we find have a large degeneracy of order \(q\), in a real system it is natural to expect domain formation. The study of such domains requires considering gradient terms in the Ginzburg Landau free energy in Eq. (\ref{F}) that we have ignored in this work. Additionally, other spatial defects (of the order parameter and/or of the underlying lattice) such as dislocations and disclinations are possible. It could therefore be fruitful to study the nature of low energy excitations of Hofstadter superconductors in the presence of such defects to seek possible realizations of defect-bound Majorana fermions \cite{Teo_Existence_2013, Benalcazar_Classification_2014}. The study of these lattice defects may also shed light on the meaning of ``gauging" the magnetic translation group, similar to the approach of gauging internal \cite{Levin_Braiding_2012} and spatial symmetries
\cite{Thorngren_Gauging_2018}.
Another potentially rich scenario could be explored by studying interfaces of Hofstadter superconductors, where different Majorana backscattering may lead to 1D SPT interfaces \cite{Santos_ParafermionicWires_2017, Santos_Symmetry-protected_2018, Santos_Parafermions_2020, Sohal_Entanglement_2020} supporting non-Abelian domain walls \cite{
alicea-fendley-2016,
BarkeshliQi-2012,
Lindner-2012,
Clarke-2013,
Cheng-2012,
Vaezi-2013,
BarkeshliJianQi-2013-a,
BarkeshliJianQi-2013-b,
Mong-2014,
khanteohughes-2014}.

Furthermore, the presence of a multi-component order parameter characterizing Hofstadter superconductors with phases relations fixed by the magnetic translation group suggests the possibility that this system may support interesting classes of Leggett modes \cite{Leggett_Number-PhaseFluctuations_1966}. In the context of pair density waves (PDW), multi-component pairing order parameters are also known to lead to fractional vortices \cite{Balents05,ChungKim07,Agterberg08, Radzihovsky09, Agterberg20, ReganSauls21}, as well as induced or vestigial orders like charge density waves or charge 4e condensates \cite{RopkeNozieres98, Babaev04, Wu05, BergFradkin09, Agterberg11, Radzihovsky11, Moon12, Lee14, FradkinKivelson15, JiangKivelson17, FernandesSchmalian19, Agterberg20, FernandesFu21}. This raises the question of what such phenomena may look like in Hofstadter SCs, for example whether vortices trapping a \(1/q\) fraction of the flux quantum may be possible \cite{Babaev02, Babaev04, Balents05, SmithsethBabaev05}. Note that charge 4e, 6e and higher charge Q orders would be classified by irreps of the MTG beyond those considered here, with irrep dimensions given by the greatest common divisor \(\text{gcd}(q,Q)\) \cite{Florek97}. We leave these open theoretical questions for future work.

Finally, it would be desirable to find direct connections between the phenomenological Ginzburg-Landau theory established on symmetry grounds and microscopic models describing moir\'e superlattices subject to a perpendicular magnetic field \cite{KangVafek18, KoshinoFu18, YuanFu18_2, AndrewsSoluyanov20}, with the purpose of shedding light on realistic parameter regimes conducive to the realization of electronic pairing in moir\'e Hofstadter bands \cite{Dean13,Ponomarenko13,Hunt13,Forsythe18,Wang15,Spanton18,Saito21}. Though in this work we focused on the spin polarized case, 
moir\'e systems are interesting in that large orbital effects (due to the large unit cell) can occur below the Chandrasekhar-Clogston-Pauli limit \cite{chandrasekhar1962note,clogston1962upper} due to small Zeeman splitting, allowing for the possibility of spin singlet pairing. Since the MTG does not act on spins, our analysis directly extends to such spin singlet condensates.

In this work we were motivated mainly by the fact that superlattice and moir\'e systems allow for both the experimental realization of Hofstadter systems \cite{Dean13,Ponomarenko13,Hunt13,Forsythe18,Wang15,Spanton18,Saito21}, as well as unconventional superconductivity \cite{cao2018unconventional}. These systems then invite an exploration on the possibility of re-entrant superconductivity in Hofstadter bands, where strong lattice effects can enable re-entrant superconductivity outside the Landau level regime \cite{Tesanovic89, AkeraMacDonald91, Tesanovic91, Rajagopal91, MacDonald92, Norman92, RasoltTesanovich92,MacDonald93, RyanRajagopal93,DukanTesanovic97,Maska02, Ueta13, ScherpelzRajagopal13,Gruenberg68,ChaudharyMacDonald21}. In light of this, it will be particularly interesting to see whether the theory of Hofstadter superconductors presented here may soon become testable in experiment.

\section*{Acknowledgments}
We acknowledge Ben Denis Shaffer for useful discussions and Lakshmi Pullasseri for collaboration in the early stages of this work.  
This research is supported by startup funds at Emory University (L.H.S.).

\appendix

\section{Details of the Microscopic Theory}\label{A}

In this appendix we fill in some of the details of the microscopic theory, including a derivation of the linearized gap equation Eq. (\ref{LinGapEq}) and the Ginzburg-Landau free energy. The starting point is the Hamiltonian Eq. (\ref{HSC})
obtained using the Hubbard-Stratonovich transformation:
\begin{align}
H&=\sum_{\ell,\mathbf{p}}\varepsilon(\mathbf{p})d^\dagger_{\mathbf{p},\ell}d_{\mathbf{p},\ell}+\frac{1}{2}\sum_{\ell,\ell',\mathbf{p}}\left[\hat{\Delta}_{\ell,\ell'}(\mathbf{p})d^\dagger_{\mathbf{p},\ell}d^\dagger_{-\mathbf{p},\ell'}+h.c.\right]+\nonumber\\
&+H_{\Delta^2}
\end{align}
where
\[H_{\Delta^2}=\sum_{\ell,n,m,\mathbf{p,p}'} \hat{\Delta}^\dagger_{\ell+m,-m}(\mathbf{p})\left[g^{-1}(\mathbf{p;p}')\right]^{(\ell)}_{mn}\hat{\Delta}_{\ell+n,-n}(\mathbf{p}')\label{HDelta2}\]
is a term quadratic in the gap function and involving the inverse of the coupling tensor:
\[\sum_{j\mathbf{q}}g^{(\ell)}_{mj}(\mathbf{p;q})\left[g^{-1}(\mathbf{q;p}')\right]^{(\ell')}_{jn}=\delta_{\ell,\ell'}\delta_{mn}\delta_{\mathbf{p p}'}\,.\]
We arranged the rest of the terms into the Bogoliubov-de Gennes (BdG) Hamiltonian:
\[H=\frac{1}{2}\sum_{\ell,\ell',\mathbf{p}}\Psi_{\ell,\mathbf{p}}^\dagger\left[\mathcal{H}_{BdG}(\mathbf{p})\right]_{\ell,\ell'}\Psi_{\ell',\mathbf{p}}+H_{\Delta^2}\]
where
\[\mathcal{H}_{BdG}(\mathbf{p})=\left(\begin{array}{cc}
     \varepsilon(\mathbf{p}) &  \hat{\Delta}(\mathbf{p})\\
     \hat{\Delta}^\dagger(\mathbf{p}) & -\varepsilon(-\mathbf{p})
\end{array}\right)\]
and \(\Psi_{\ell,\mathbf{p}}=(d_{\ell,\mathbf{p}},d_{\ell,-\mathbf{p}}^\dagger)\) are the Nambu spinors. Note that the formalism can be easily extended to include pairing between multiple bands, with the index \(\ell\) replaces with a multi-index \(I=(\ell,\alpha,\dots)\) that can include Hofstadter band indices, spin, etc.

In the mean field treatment, the gap function has to be solved for self-consistently by minimizing the free energy. The free energy is obtained from Eq. (\ref{HSC}) by integrating out the \(\Psi_{\ell,\mathbf{p}}\) fields from the partition function. The standard procedure using the Matsubara formalism yields
\[\mathcal{F}=-T\sum_{\omega,\mathbf{p}}\text{Tr}\left[\log\beta\mathcal{G}^{-1}(i\omega,\mathbf{p})\right]+H_{\Delta^2}\label{Fmicro}\]
where \(\omega=(2\pi n+1) T\) with integer \(n\) are the Matsubara frequencies and we defined the Gor'kov Green's function
\begin{align}
\mathcal{G}(i\omega,\mathbf{p})&=\left(i\omega-\mathcal{H}_{BdG}(\mathbf{p})\right)^{-1}=\nonumber\\
&=\left(\begin{array}{cc}
    \hat{G}(i\omega,\mathbf{p}) & \hat{F}(i\omega,\mathbf{p}) \\
    \hat{F}^\dagger(i\omega,\mathbf{p}) & -\hat{G}^T(-i\omega,-\mathbf{p})
\end{array}\right)\,.
\end{align}
Minimizing \(\mathcal{F}\) with respect to \(\hat{\Delta}^\dagger\) we obtain the gap equation
\[\hat{\Delta}_{\ell+n,-n}(\mathbf{p})=T\sum_{\omega\mathbf{p}'m}g^{(\ell)}_{n,m}(\mathbf{p;p}')\hat{F}_{\ell+m,-m}(i\omega,\mathbf{p}')\label{GapEq}\,.\]
Close below \(T_c\) we can expand the free energy and the Green's functions in powers of the gap function. In particular, to leading order
\[\hat{F}(i\omega,\mathbf{p})=-\hat{G}^{(0)}(i\omega,\mathbf{p})\hat{\Delta}(\mathbf{p})\hat{G}^{(0)T}(-i\omega,-\mathbf{p})\]
where \(\hat{G}^{(0)}\) is the normal Green's function in the absence of pairing, i.e. when \(\hat{\Delta}=0\). While the equations above hold for the multi-band case as well, the linearized gap equation simplifies significantly in the single-band case as the normal state Green's function becomes
\[\hat{G}^{(0)}=\frac{\mathbbm{1}}{i\omega-\varepsilon(\mathbf{p})}\]
which note is proportional to the identity matrix. Carrying out the Matsubara sum in the usual weak coupling approximation yields the linearized gap equation:
\[\hat{\Delta}_{\ell+n,-n}(\mathbf{p})=-\nu\log\frac{1.13\Lambda}{T}\sum_{\mathbf{p}'m}g^{(\ell)}_{n,m}(\mathbf{p;p}')\hat{\Delta}_{\ell+m,-m}(\mathbf{p}')\]
where \(\nu\) is the density of states at the Fermi level and \(\Lambda\) is the high energy cutoff.

In principle the parameters \(\alpha, \beta_{MN}\) and \(\kappa_{jj'}\) in Eq. (\ref{F}) can be obtained from the microscopic free energy Eq. (\ref{Fmicro}) by expanding the trace logarithm, plugging in the solutions of the gap equation Eq. (\ref{GapEq}) and summing over the momentum and patch indices (in order to obtain \(\kappa_{jj'}\) we need to additionally allow \(\hat{\Delta}\) to have spatial variations). Explicitly, assuming \(\varepsilon(\mathbf{p})\) is approximately isotropic, in the single-band case we have
\begin{widetext}
\begin{align}
    \alpha&=\sum H_{\Delta^2}-\nu\log\frac{1.13\Lambda}{T}\text{Tr}\left[\hat{\Delta}^{(L)\dagger}\hat{\Delta}^{(L)}\right]\,,\\
    \beta_{MN}&=\frac{7\zeta(3)\nu}{32\pi^2T^2}\sum\text{Tr}\left[\hat{\Delta}^{(L+M)\dagger}\hat{\Delta}^{(L-M)}\hat{\Delta}^{(L+N)\dagger}\hat{\Delta}^{(L-N)}\right]\,.\nonumber
\end{align}
\end{widetext}
Note that while only \(\alpha\) explicitly depends on the interactions \(g^{(\ell)}_{n,m}\) via the \(H_{\Delta^2}\) term, all terms depend on the interactions implicitly via \(\hat{\Delta}^{(L)}\). All the parameters also depend on the band structure and temperature. In the multi-band case the traces include summations over the additional indices and the Green's functions can no longer be pulled out of the trace, resulting in more involved expressions.

\section{Action of Lattice Symmetries}\label{B}

In the main text we have neglected point group symmetries in order to keep the discussion as general as possible and applicable to any kind of lattice. For completeness, here we consider the action of some point group symmetries and discuss their consequences. Importantly, since the vector potential \(\mathbf{A}\) breaks some of the point group symmetries, they have to be combined with gauge transformations in order to produce a symmetry of the system, just as the regular translations had to be combined with a gauge transformation to produce the MTG symmetries \(\hat{T}_1\) and \(\hat{T}_2\). We thus call such symmetries magnetic rotations, etc.  Another significant complication is that some point group symmetries act non-trivially on the MTG symmetries \(\hat{T}_1\) and \(\hat{T}_2\).

As an example also considered in \cite{Balents05}, on a square lattice a four fold rotation \(C_4\), we have a magnetic rotation \(\hat{C}_4\) that satisfies
\begin{gather}
    \hat{C}_4\hat{T}_1\hat{C}_4^{-1}=\hat{T}_2\,,\nonumber\\
    \hat{C}_4\hat{T}_2\hat{C}_4^{-1}=\hat{T}_1^{-1}\,,\nonumber\\
    \hat{C}_4^4=1\,.\label{C4}
\end{gather}
In particular, \(\hat{C}_4\) takes eigenvectors of \(\hat{T}_1\) with eigenvalues \(\omega_q^{pM}\) to eigenvectors of \(\hat{T}_2\) with the same eigenvalues \(\omega_q^{pM}\), and similarly eigenvectors of \(\hat{T}_2\) with eigenvalues \(\omega_q^{pL}\) to eigenvectors of \(\hat{T}_1^{-1}=\hat{T}_1^{q-1}\) with eigenvalues \(\omega_q^{-pL}\). This fixes the action of \(\hat{C}_4\), including its action on the gap function \(\hat{\Delta}(\mathbf{p})=\sum_L\eta_L\hat{\Delta}^{(L)}(\mathbf{p})\) which can be separated into an action on \(\mathbf{p}\) (which is the same as the action of \(C_4\)) and an action on the order parameter \(\boldsymbol{\eta}\). The latter can be deduced from Eqs. (\ref{etaT1}) or (\ref{etaT1even}) by requiring
\begin{align}
    \boldsymbol{\eta}^{(M)_{1,0}}\xrightarrow[]{\hat{C}_4}\boldsymbol{\eta}^{(M)_{0,1}}\,,\nonumber\\
    \boldsymbol{\eta}^{(L)_{0,1}}\xrightarrow[]{\hat{C}_4}\boldsymbol{\eta}^{(-L)_{1,0}}\,.
\end{align}
The condition \(\hat{C}_4^4=1\) in principle allows us to take instead \(\boldsymbol{\eta}^{(M)_{1,0}}\xrightarrow[]{\hat{C}_4}i^N\boldsymbol{\eta}^{(M)_{0,1}}\) with \(N=0,1,2,3\), but this can be obtained by combining \(\hat{C}_4\) with a \(U(1)\) transformation so we can take Eq. (\ref{C4}) as the canonical definition of \(\hat{C}_4\) (we can also define \(\hat{C}_4(i^N)\) in the same way we defined \(\hat{T}_j(\theta)\)). Note that \(\hat{C}_4^2=\mathcal{I}\), the accidental symmetry that we defined in Sec. \ref{GL}.

We can similarly define a three fold magnetic rotation
\begin{gather}\label{C3}
    \hat{C}_3\hat{T}_1\hat{C}_3^{-1}=\hat{T}_2\,,\nonumber\\
    \hat{C}_3\hat{T}_2\hat{C}_3^{-1}=\hat{T}_2^{-1}\hat{T}_1^{-1}\,,\nonumber\\
    \hat{C}_3\hat{T}_2^{-1}\hat{T}_1^{-1}\hat{C}_3^{-1}=\hat{T}_1\,,\nonumber\\
    \hat{C}_3^3=1\,,
\end{gather}
as well as magnetic reflections, e.g. a reflection in the \(xz\) plane \(\mathcal{M}_y\) that yields the magnetic reflection \(\hat{\mathcal{M}}_y\)
\begin{gather}
    \hat{\mathcal{M}}_y\hat{T}_1\hat{\mathcal{M}}_y^{-1}=\hat{T}_1\,,\nonumber\\
    \hat{\mathcal{M}}_y\hat{T}_2\hat{\mathcal{M}}_y^{-1}=\hat{T}_2^{-1}\,,\nonumber\\
    \hat{\mathcal{M}}_y^2=1\,.\label{My}
\end{gather}
Note that the magnetic field itself breaks the out-of-plane mirror symmetries, while the in-plane mirror symmetry commutes with both \(\hat{T}_1\) and \(\hat{T}_2\) and so acts trivially on the patch indices. The rest of the magnetic point group symmetries do not commute with the MTG symmetries. This means in particular that the \(\mathbb{Z}_q\) phases discussed in the main text generally break the point groups symmetries. On the other hand, we could consider phases that respect the magnetic point group symmetries, but which therefore in general break all of the MTG symmetries.

Note also that the relations in Eqs. (\ref{C4}-\ref{My}) imply that the MTG is a normal subgroup of the total symmetry group that includes the point group (i.e. any element of the MTG conjugated by a point group symmetry is again an element of the MTG). This implies by Mackey's irreducibility criterion that the representations of the total symmetry group induced by the irreps of the MTG presented in Sec. \ref{section:irreps} are themselves irreducible. In other words, in most cases the irreps of the MTG can essentially be considered as irreps of the total symmetry group, though additional irreps (in which the MTG is completely broken) may be possible.

\bibliographystyle{apsrev4-1}
\bibliography{literature_note_bibliography}

\begin{thebibliography}{183}%
\makeatletter
\providecommand \@ifxundefined [1]{%
 \@ifx{#1\undefined}
}%
\providecommand \@ifnum [1]{%
 \ifnum #1\expandafter \@firstoftwo
 \else \expandafter \@secondoftwo
 \fi
}%
\providecommand \@ifx [1]{%
 \ifx #1\expandafter \@firstoftwo
 \else \expandafter \@secondoftwo
 \fi
}%
\providecommand \natexlab [1]{#1}%
\providecommand \enquote  [1]{``#1''}%
\providecommand \bibnamefont  [1]{#1}%
\providecommand \bibfnamefont [1]{#1}%
\providecommand \citenamefont [1]{#1}%
\providecommand \href@noop [0]{\@secondoftwo}%
\providecommand \href [0]{\begingroup \@sanitize@url \@href}%
\providecommand \@href[1]{\@@startlink{#1}\@@href}%
\providecommand \@@href[1]{\endgroup#1\@@endlink}%
\providecommand \@sanitize@url [0]{\catcode `\\12\catcode `\$12\catcode
  `\&12\catcode `\#12\catcode `\^12\catcode `\_12\catcode `\%12\relax}%
\providecommand \@@startlink[1]{}%
\providecommand \@@endlink[0]{}%
\providecommand \url  [0]{\begingroup\@sanitize@url \@url }%
\providecommand \@url [1]{\endgroup\@href {#1}{\urlprefix }}%
\providecommand \urlprefix  [0]{URL }%
\providecommand \Eprint [0]{\href }%
\providecommand \doibase [0]{http://dx.doi.org/}%
\providecommand \selectlanguage [0]{\@gobble}%
\providecommand \bibinfo  [0]{\@secondoftwo}%
\providecommand \bibfield  [0]{\@secondoftwo}%
\providecommand \translation [1]{[#1]}%
\providecommand \BibitemOpen [0]{}%
\providecommand \bibitemStop [0]{}%
\providecommand \bibitemNoStop [0]{.\EOS\space}%
\providecommand \EOS [0]{\spacefactor3000\relax}%
\providecommand \BibitemShut  [1]{\csname bibitem#1\endcsname}%
\let\auto@bib@innerbib\@empty
\bibitem [{\citenamefont {Harper}(1955)}]{Harper55}%
  \BibitemOpen
  \bibfield  {author} {\bibinfo {author} {\bibfnamefont {P.~G.}\ \bibnamefont
  {Harper}},\ }\href {\doibase 10.1088/0370-1298/68/10/305} {\bibfield
  {journal} {\bibinfo  {journal} {Proceedings of the Physical Society. Section
  A}\ }\textbf {\bibinfo {volume} {68}},\ \bibinfo {pages} {879} (\bibinfo
  {year} {1955})}\BibitemShut {NoStop}%
\bibitem [{\citenamefont {Azbel}(1964)}]{Azbel64}%
  \BibitemOpen
  \bibfield  {author} {\bibinfo {author} {\bibfnamefont {M.~Y.}\ \bibnamefont
  {Azbel}},\ }\href@noop {} {\bibfield  {journal} {\bibinfo  {journal} {Sov.
  Phys. JETP}\ }\textbf {\bibinfo {volume} {19}},\ \bibinfo {pages} {634}
  (\bibinfo {year} {1964})}\BibitemShut {NoStop}%
\bibitem [{\citenamefont {Zak}(1964{\natexlab{a}})}]{Zak64_1}%
  \BibitemOpen
  \bibfield  {author} {\bibinfo {author} {\bibfnamefont {J.}~\bibnamefont
  {Zak}},\ }\href {\doibase 10.1103/PhysRev.134.A1602} {\bibfield  {journal}
  {\bibinfo  {journal} {Physical Review}\ }\textbf {\bibinfo {volume} {134}},\
  \bibinfo {pages} {A1602} (\bibinfo {year} {1964}{\natexlab{a}})},\ \bibinfo
  {note} {publisher: American Physical Society}\BibitemShut {NoStop}%
\bibitem [{\citenamefont {Brown}(1964)}]{Brown64}%
  \BibitemOpen
  \bibfield  {author} {\bibinfo {author} {\bibfnamefont {E.}~\bibnamefont
  {Brown}},\ }\href {\doibase 10.1103/PhysRev.133.A1038} {\bibfield  {journal}
  {\bibinfo  {journal} {Physical Review}\ }\textbf {\bibinfo {volume} {133}},\
  \bibinfo {pages} {A1038} (\bibinfo {year} {1964})},\ \bibinfo {note}
  {publisher: American Physical Society}\BibitemShut {NoStop}%
\bibitem [{\citenamefont {Hofstadter}(1976)}]{Hofstadter76}%
  \BibitemOpen
  \bibfield  {author} {\bibinfo {author} {\bibfnamefont {D.~R.}\ \bibnamefont
  {Hofstadter}},\ }\href {\doibase 10.1103/PhysRevB.14.2239} {\bibfield
  {journal} {\bibinfo  {journal} {Physical Review B}\ }\textbf {\bibinfo
  {volume} {14}},\ \bibinfo {pages} {2239} (\bibinfo {year} {1976})},\ \bibinfo
  {note} {publisher: American Physical Society}\BibitemShut {NoStop}%
\bibitem [{\citenamefont {Thouless}\ \emph {et~al.}(1982)\citenamefont
  {Thouless}, \citenamefont {Kohmoto}, \citenamefont {Nightingale},\ and\
  \citenamefont {den Nijs}}]{TKNN}%
  \BibitemOpen
  \bibfield  {author} {\bibinfo {author} {\bibfnamefont {D.~J.}\ \bibnamefont
  {Thouless}}, \bibinfo {author} {\bibfnamefont {M.}~\bibnamefont {Kohmoto}},
  \bibinfo {author} {\bibfnamefont {M.~P.}\ \bibnamefont {Nightingale}}, \ and\
  \bibinfo {author} {\bibfnamefont {M.}~\bibnamefont {den Nijs}},\ }\href
  {\doibase 10.1103/PhysRevLett.49.405} {\bibfield  {journal} {\bibinfo
  {journal} {Physical Review Letters}\ }\textbf {\bibinfo {volume} {49}},\
  \bibinfo {pages} {405} (\bibinfo {year} {1982})},\ \bibinfo {note}
  {publisher: American Physical Society}\BibitemShut {NoStop}%
\bibitem [{\citenamefont {Kohmoto}(1985)}]{Kohmoto85}%
  \BibitemOpen
  \bibfield  {author} {\bibinfo {author} {\bibfnamefont {M.}~\bibnamefont
  {Kohmoto}},\ }\href {\doibase 10.1016/0003-4916(85)90148-4} {\bibfield
  {journal} {\bibinfo  {journal} {Annals of Physics}\ }\textbf {\bibinfo
  {volume} {160}},\ \bibinfo {pages} {343} (\bibinfo {year}
  {1985})}\BibitemShut {NoStop}%
\bibitem [{\citenamefont {Haldane}(1988)}]{Haldane1988}%
  \BibitemOpen
  \bibfield  {author} {\bibinfo {author} {\bibfnamefont {F.~D.~M.}\
  \bibnamefont {Haldane}},\ }\href {\doibase 10.1103/PhysRevLett.61.2015}
  {\bibfield  {journal} {\bibinfo  {journal} {Phys. Rev. Lett.}\ }\textbf
  {\bibinfo {volume} {61}},\ \bibinfo {pages} {2015} (\bibinfo {year}
  {1988})}\BibitemShut {NoStop}%
\bibitem [{\citenamefont {Neupert}\ \emph {et~al.}(2011)\citenamefont
  {Neupert}, \citenamefont {Santos}, \citenamefont {Chamon},\ and\
  \citenamefont {Mudry}}]{Neupert-2011}%
  \BibitemOpen
  \bibfield  {author} {\bibinfo {author} {\bibfnamefont {T.}~\bibnamefont
  {Neupert}}, \bibinfo {author} {\bibfnamefont {L.}~\bibnamefont {Santos}},
  \bibinfo {author} {\bibfnamefont {C.}~\bibnamefont {Chamon}}, \ and\ \bibinfo
  {author} {\bibfnamefont {C.}~\bibnamefont {Mudry}},\ }\href {\doibase
  10.1103/PhysRevLett.106.236804} {\bibfield  {journal} {\bibinfo  {journal}
  {Phys. Rev. Lett.}\ }\textbf {\bibinfo {volume} {106}},\ \bibinfo {pages}
  {236804} (\bibinfo {year} {2011})}\BibitemShut {NoStop}%
\bibitem [{\citenamefont {Sheng}\ \emph {et~al.}(2011)\citenamefont {Sheng},
  \citenamefont {Gu}, \citenamefont {Sun},\ and\ \citenamefont
  {Sheng}}]{Sheng-2011}%
  \BibitemOpen
  \bibfield  {author} {\bibinfo {author} {\bibfnamefont {D.~N.}\ \bibnamefont
  {Sheng}}, \bibinfo {author} {\bibfnamefont {Z.-C.}\ \bibnamefont {Gu}},
  \bibinfo {author} {\bibfnamefont {K.}~\bibnamefont {Sun}}, \ and\ \bibinfo
  {author} {\bibfnamefont {L.}~\bibnamefont {Sheng}},\ }\href {\doibase
  10.1038/ncomms1380} {\bibfield  {journal} {\bibinfo  {journal} {Nature
  Communications}\ }\textbf {\bibinfo {volume} {2}},\ \bibinfo {pages} {389 EP
  } (\bibinfo {year} {2011})}\BibitemShut {NoStop}%
\bibitem [{\citenamefont {Tang}\ \emph {et~al.}(2011)\citenamefont {Tang},
  \citenamefont {Mei},\ and\ \citenamefont {Wen}}]{Tang-2011}%
  \BibitemOpen
  \bibfield  {author} {\bibinfo {author} {\bibfnamefont {E.}~\bibnamefont
  {Tang}}, \bibinfo {author} {\bibfnamefont {J.-W.}\ \bibnamefont {Mei}}, \
  and\ \bibinfo {author} {\bibfnamefont {X.-G.}\ \bibnamefont {Wen}},\ }\href
  {\doibase 10.1103/PhysRevLett.106.236802} {\bibfield  {journal} {\bibinfo
  {journal} {Phys. Rev. Lett.}\ }\textbf {\bibinfo {volume} {106}},\ \bibinfo
  {pages} {236802} (\bibinfo {year} {2011})}\BibitemShut {NoStop}%
\bibitem [{\citenamefont {Sun}\ \emph {et~al.}(2011)\citenamefont {Sun},
  \citenamefont {Gu}, \citenamefont {Katsura},\ and\ \citenamefont
  {Das~Sarma}}]{Sun-2011}%
  \BibitemOpen
  \bibfield  {author} {\bibinfo {author} {\bibfnamefont {K.}~\bibnamefont
  {Sun}}, \bibinfo {author} {\bibfnamefont {Z.}~\bibnamefont {Gu}}, \bibinfo
  {author} {\bibfnamefont {H.}~\bibnamefont {Katsura}}, \ and\ \bibinfo
  {author} {\bibfnamefont {S.}~\bibnamefont {Das~Sarma}},\ }\href {\doibase
  10.1103/PhysRevLett.106.236803} {\bibfield  {journal} {\bibinfo  {journal}
  {Phys. Rev. Lett.}\ }\textbf {\bibinfo {volume} {106}},\ \bibinfo {pages}
  {236803} (\bibinfo {year} {2011})}\BibitemShut {NoStop}%
\bibitem [{\citenamefont {Regnault}\ and\ \citenamefont
  {Bernevig}(2011)}]{Regnault2011}%
  \BibitemOpen
  \bibfield  {author} {\bibinfo {author} {\bibfnamefont {N.}~\bibnamefont
  {Regnault}}\ and\ \bibinfo {author} {\bibfnamefont {B.~A.}\ \bibnamefont
  {Bernevig}},\ }\href {\doibase 10.1103/PhysRevX.1.021014} {\bibfield
  {journal} {\bibinfo  {journal} {Phys. Rev. X}\ }\textbf {\bibinfo {volume}
  {1}},\ \bibinfo {pages} {021014} (\bibinfo {year} {2011})}\BibitemShut
  {NoStop}%
\bibitem [{\citenamefont {Parameswaran}\ \emph {et~al.}(2013)\citenamefont
  {Parameswaran}, \citenamefont {Roy},\ and\ \citenamefont
  {Sondhi}}]{parameswaran2013fractional}%
  \BibitemOpen
  \bibfield  {author} {\bibinfo {author} {\bibfnamefont {S.~A.}\ \bibnamefont
  {Parameswaran}}, \bibinfo {author} {\bibfnamefont {R.}~\bibnamefont {Roy}}, \
  and\ \bibinfo {author} {\bibfnamefont {S.~L.}\ \bibnamefont {Sondhi}},\
  }\href@noop {} {\bibfield  {journal} {\bibinfo  {journal} {Comptes Rendus
  Physique}\ }\textbf {\bibinfo {volume} {14}},\ \bibinfo {pages} {816}
  (\bibinfo {year} {2013})}\BibitemShut {NoStop}%
\bibitem [{\citenamefont {Kol}\ and\ \citenamefont {Read}(1993)}]{Kol1993}%
  \BibitemOpen
  \bibfield  {author} {\bibinfo {author} {\bibfnamefont {A.}~\bibnamefont
  {Kol}}\ and\ \bibinfo {author} {\bibfnamefont {N.}~\bibnamefont {Read}},\
  }\href {\doibase 10.1103/PhysRevB.48.8890} {\bibfield  {journal} {\bibinfo
  {journal} {Phys. Rev. B}\ }\textbf {\bibinfo {volume} {48}},\ \bibinfo
  {pages} {8890} (\bibinfo {year} {1993})}\BibitemShut {NoStop}%
\bibitem [{\citenamefont {M\"oller}\ and\ \citenamefont
  {Cooper}(2015)}]{Moller2015}%
  \BibitemOpen
  \bibfield  {author} {\bibinfo {author} {\bibfnamefont {G.}~\bibnamefont
  {M\"oller}}\ and\ \bibinfo {author} {\bibfnamefont {N.~R.}\ \bibnamefont
  {Cooper}},\ }\href {\doibase 10.1103/PhysRevLett.115.126401} {\bibfield
  {journal} {\bibinfo  {journal} {Phys. Rev. Lett.}\ }\textbf {\bibinfo
  {volume} {115}},\ \bibinfo {pages} {126401} (\bibinfo {year}
  {2015})}\BibitemShut {NoStop}%
\bibitem [{\citenamefont {Murthy}\ and\ \citenamefont
  {Shankar}(2012)}]{murthyshankar2012}%
  \BibitemOpen
  \bibfield  {author} {\bibinfo {author} {\bibfnamefont {G.}~\bibnamefont
  {Murthy}}\ and\ \bibinfo {author} {\bibfnamefont {R.}~\bibnamefont
  {Shankar}},\ }\href {\doibase 10.1103/PhysRevB.86.195146} {\bibfield
  {journal} {\bibinfo  {journal} {Phys. Rev. B}\ }\textbf {\bibinfo {volume}
  {86}},\ \bibinfo {pages} {195146} (\bibinfo {year} {2012})}\BibitemShut
  {NoStop}%
\bibitem [{\citenamefont {Sohal}\ \emph {et~al.}(2018)\citenamefont {Sohal},
  \citenamefont {Santos},\ and\ \citenamefont {Fradkin}}]{Sohal-2018}%
  \BibitemOpen
  \bibfield  {author} {\bibinfo {author} {\bibfnamefont {R.}~\bibnamefont
  {Sohal}}, \bibinfo {author} {\bibfnamefont {L.~H.}\ \bibnamefont {Santos}}, \
  and\ \bibinfo {author} {\bibfnamefont {E.}~\bibnamefont {Fradkin}},\ }\href
  {\doibase 10.1103/PhysRevB.97.125131} {\bibfield  {journal} {\bibinfo
  {journal} {Phys. Rev. B}\ }\textbf {\bibinfo {volume} {97}},\ \bibinfo
  {pages} {125131} (\bibinfo {year} {2018})}\BibitemShut {NoStop}%
\bibitem [{\citenamefont {Andrews}\ \emph {et~al.}(2021)\citenamefont
  {Andrews}, \citenamefont {Neupert},\ and\ \citenamefont
  {M\"oller}}]{AndrewsNeupert21}%
  \BibitemOpen
  \bibfield  {author} {\bibinfo {author} {\bibfnamefont {B.}~\bibnamefont
  {Andrews}}, \bibinfo {author} {\bibfnamefont {T.}~\bibnamefont {Neupert}}, \
  and\ \bibinfo {author} {\bibfnamefont {G.}~\bibnamefont {M\"oller}},\ }\href
  {\doibase 10.1103/PhysRevB.104.125107} {\bibfield  {journal} {\bibinfo
  {journal} {Phys. Rev. B}\ }\textbf {\bibinfo {volume} {104}},\ \bibinfo
  {pages} {125107} (\bibinfo {year} {2021})}\BibitemShut {NoStop}%
\bibitem [{\citenamefont {Te\v{s}anovi\'{c}}\ \emph {et~al.}(1989)\citenamefont
  {Te\v{s}anovi\'{c}}, \citenamefont {Rasolt},\ and\ \citenamefont
  {Xing}}]{Tesanovic89}%
  \BibitemOpen
  \bibfield  {author} {\bibinfo {author} {\bibfnamefont {Z.}~\bibnamefont
  {Te\v{s}anovi\'{c}}}, \bibinfo {author} {\bibfnamefont {M.}~\bibnamefont
  {Rasolt}}, \ and\ \bibinfo {author} {\bibfnamefont {L.}~\bibnamefont
  {Xing}},\ }\href {\doibase 10.1103/PhysRevLett.63.2425} {\bibfield  {journal}
  {\bibinfo  {journal} {Physical Review Letters}\ }\textbf {\bibinfo {volume}
  {63}},\ \bibinfo {pages} {2425} (\bibinfo {year} {1989})},\ \bibinfo {note}
  {publisher: American Physical Society}\BibitemShut {NoStop}%
\bibitem [{\citenamefont {Akera}\ \emph {et~al.}(1991)\citenamefont {Akera},
  \citenamefont {MacDonald}, \citenamefont {Girvin},\ and\ \citenamefont
  {Norman}}]{AkeraMacDonald91}%
  \BibitemOpen
  \bibfield  {author} {\bibinfo {author} {\bibfnamefont {H.}~\bibnamefont
  {Akera}}, \bibinfo {author} {\bibfnamefont {A.~H.}\ \bibnamefont
  {MacDonald}}, \bibinfo {author} {\bibfnamefont {S.~M.}\ \bibnamefont
  {Girvin}}, \ and\ \bibinfo {author} {\bibfnamefont {M.~R.}\ \bibnamefont
  {Norman}},\ }\href {\doibase 10.1103/PhysRevLett.67.2375} {\bibfield
  {journal} {\bibinfo  {journal} {Physical Review Letters}\ }\textbf {\bibinfo
  {volume} {67}},\ \bibinfo {pages} {2375} (\bibinfo {year} {1991})},\ \bibinfo
  {note} {publisher: American Physical Society}\BibitemShut {NoStop}%
\bibitem [{\citenamefont {Te\v{s}anovi\'{c}}\ \emph {et~al.}(1991)\citenamefont
  {Te\v{s}anovi\'{c}}, \citenamefont {Rasolt},\ and\ \citenamefont
  {Xing}}]{Tesanovic91}%
  \BibitemOpen
  \bibfield  {author} {\bibinfo {author} {\bibfnamefont {Z.}~\bibnamefont
  {Te\v{s}anovi\'{c}}}, \bibinfo {author} {\bibfnamefont {M.}~\bibnamefont
  {Rasolt}}, \ and\ \bibinfo {author} {\bibfnamefont {L.}~\bibnamefont
  {Xing}},\ }\href {\doibase 10.1103/PhysRevB.43.288} {\bibfield  {journal}
  {\bibinfo  {journal} {Physical Review B}\ }\textbf {\bibinfo {volume} {43}},\
  \bibinfo {pages} {288} (\bibinfo {year} {1991})},\ \bibinfo {note}
  {publisher: American Physical Society}\BibitemShut {NoStop}%
\bibitem [{\citenamefont {Rajagopal}\ and\ \citenamefont
  {Vasudevan}(1991)}]{Rajagopal91}%
  \BibitemOpen
  \bibfield  {author} {\bibinfo {author} {\bibfnamefont {A.~K.}\ \bibnamefont
  {Rajagopal}}\ and\ \bibinfo {author} {\bibfnamefont {R.}~\bibnamefont
  {Vasudevan}},\ }\href {\doibase 10.1103/PhysRevB.44.2807} {\bibfield
  {journal} {\bibinfo  {journal} {Physical Review B}\ }\textbf {\bibinfo
  {volume} {44}},\ \bibinfo {pages} {2807} (\bibinfo {year} {1991})},\ \bibinfo
  {note} {publisher: American Physical Society}\BibitemShut {NoStop}%
\bibitem [{\citenamefont {MacDonald}\ \emph {et~al.}(1992)\citenamefont
  {MacDonald}, \citenamefont {Akera},\ and\ \citenamefont
  {Norman}}]{MacDonald92}%
  \BibitemOpen
  \bibfield  {author} {\bibinfo {author} {\bibfnamefont {A.~H.}\ \bibnamefont
  {MacDonald}}, \bibinfo {author} {\bibfnamefont {H.}~\bibnamefont {Akera}}, \
  and\ \bibinfo {author} {\bibfnamefont {M.~R.}\ \bibnamefont {Norman}},\
  }\href {\doibase 10.1103/PhysRevB.45.10147} {\bibfield  {journal} {\bibinfo
  {journal} {Physical Review B}\ }\textbf {\bibinfo {volume} {45}},\ \bibinfo
  {pages} {10147} (\bibinfo {year} {1992})},\ \bibinfo {note} {publisher:
  American Physical Society}\BibitemShut {NoStop}%
\bibitem [{\citenamefont {Norman}\ \emph {et~al.}(1992)\citenamefont {Norman},
  \citenamefont {Akera},\ and\ \citenamefont {MacDonald}}]{Norman92}%
  \BibitemOpen
  \bibfield  {author} {\bibinfo {author} {\bibfnamefont {M.~R.}\ \bibnamefont
  {Norman}}, \bibinfo {author} {\bibfnamefont {H.}~\bibnamefont {Akera}}, \
  and\ \bibinfo {author} {\bibfnamefont {A.~H.}\ \bibnamefont {MacDonald}},\
  }\href {\doibase 10.1016/0921-4534(92)90135-Y} {\bibfield  {journal}
  {\bibinfo  {journal} {Physica C: Superconductivity}\ }\textbf {\bibinfo
  {volume} {196}},\ \bibinfo {pages} {43} (\bibinfo {year} {1992})}\BibitemShut
  {NoStop}%
\bibitem [{\citenamefont {Rasolt}\ and\ \citenamefont
  {Te\v{s}anovi\'{c}}(1992)}]{RasoltTesanovich92}%
  \BibitemOpen
  \bibfield  {author} {\bibinfo {author} {\bibfnamefont {M.}~\bibnamefont
  {Rasolt}}\ and\ \bibinfo {author} {\bibfnamefont {Z.}~\bibnamefont
  {Te\v{s}anovi\'{c}}},\ }\href {\doibase 10.1103/RevModPhys.64.709} {\bibfield
   {journal} {\bibinfo  {journal} {Reviews of Modern Physics}\ }\textbf
  {\bibinfo {volume} {64}},\ \bibinfo {pages} {709} (\bibinfo {year} {1992})},\
  \bibinfo {note} {publisher: American Physical Society}\BibitemShut {NoStop}%
\bibitem [{\citenamefont {MacDonald}\ \emph {et~al.}(1993)\citenamefont
  {MacDonald}, \citenamefont {Akera},\ and\ \citenamefont
  {Norman}}]{MacDonald93}%
  \BibitemOpen
  \bibfield  {author} {\bibinfo {author} {\bibfnamefont {A.~H.}\ \bibnamefont
  {MacDonald}}, \bibinfo {author} {\bibfnamefont {H.}~\bibnamefont {Akera}}, \
  and\ \bibinfo {author} {\bibfnamefont {M.~R.}\ \bibnamefont {Norman}},\
  }\href {\doibase 10.1071/ph930333} {\bibfield  {journal} {\bibinfo  {journal}
  {Australian Journal of Physics}\ }\textbf {\bibinfo {volume} {46}},\ \bibinfo
  {pages} {333} (\bibinfo {year} {1993})},\ \bibinfo {note} {publisher: CSIRO
  PUBLISHING}\BibitemShut {NoStop}%
\bibitem [{\citenamefont {Ryan}\ and\ \citenamefont
  {Rajagopal}(1993)}]{RyanRajagopal93}%
  \BibitemOpen
  \bibfield  {author} {\bibinfo {author} {\bibfnamefont {J.~C.}\ \bibnamefont
  {Ryan}}\ and\ \bibinfo {author} {\bibfnamefont {A.~K.}\ \bibnamefont
  {Rajagopal}},\ }\href {\doibase 10.1103/PhysRevB.47.8843} {\bibfield
  {journal} {\bibinfo  {journal} {Physical Review B}\ }\textbf {\bibinfo
  {volume} {47}},\ \bibinfo {pages} {8843} (\bibinfo {year} {1993})},\ \bibinfo
  {note} {publisher: American Physical Society}\BibitemShut {NoStop}%
\bibitem [{\citenamefont {Dukan}\ and\ \citenamefont {Te\ifmmode \check{s}\else
  \v{s}\fi{}anovi\ifmmode~\acute{c}\else \'{c}\fi{}}(1997)}]{DukanTesanovic97}%
  \BibitemOpen
  \bibfield  {author} {\bibinfo {author} {\bibfnamefont {S.~c.~v.}\
  \bibnamefont {Dukan}}\ and\ \bibinfo {author} {\bibfnamefont
  {Z.}~\bibnamefont {Te\ifmmode \check{s}\else
  \v{s}\fi{}anovi\ifmmode~\acute{c}\else \'{c}\fi{}}},\ }\href {\doibase
  10.1103/PhysRevB.56.838} {\bibfield  {journal} {\bibinfo  {journal} {Phys.
  Rev. B}\ }\textbf {\bibinfo {volume} {56}},\ \bibinfo {pages} {838} (\bibinfo
  {year} {1997})}\BibitemShut {NoStop}%
\bibitem [{\citenamefont {Ma\ifmmode~\acute{s}\else
  \'{s}\fi{}ka}(2002)}]{Maska02}%
  \BibitemOpen
  \bibfield  {author} {\bibinfo {author} {\bibfnamefont {M.~M.}\ \bibnamefont
  {Ma\ifmmode~\acute{s}\else \'{s}\fi{}ka}},\ }\href {\doibase
  10.1103/PhysRevB.66.054533} {\bibfield  {journal} {\bibinfo  {journal}
  {Physical Review B}\ }\textbf {\bibinfo {volume} {66}},\ \bibinfo {pages}
  {054533} (\bibinfo {year} {2002})},\ \bibinfo {note} {publisher: American
  Physical Society}\BibitemShut {NoStop}%
\bibitem [{\citenamefont {Ueta}\ and\ \citenamefont {Hioki}(2013)}]{Ueta13}%
  \BibitemOpen
  \bibfield  {author} {\bibinfo {author} {\bibfnamefont {T.}~\bibnamefont
  {Ueta}}\ and\ \bibinfo {author} {\bibfnamefont {T.}~\bibnamefont {Hioki}},\
  }\href {\doibase 10.1007/s10948-012-1872-y} {\bibfield  {journal} {\bibinfo
  {journal} {Journal of Superconductivity and Novel Magnetism}\ }\textbf
  {\bibinfo {volume} {26}},\ \bibinfo {pages} {1921} (\bibinfo {year}
  {2013})}\BibitemShut {NoStop}%
\bibitem [{\citenamefont {Scherpelz}\ \emph {et~al.}(2013)\citenamefont
  {Scherpelz}, \citenamefont {Wulin}, \citenamefont {\ifmmode~\check{S}\else
  \v{S}\fi{}op\'{\i}k}, \citenamefont {Levin},\ and\ \citenamefont
  {Rajagopal}}]{ScherpelzRajagopal13}%
  \BibitemOpen
  \bibfield  {author} {\bibinfo {author} {\bibfnamefont {P.}~\bibnamefont
  {Scherpelz}}, \bibinfo {author} {\bibfnamefont {D.}~\bibnamefont {Wulin}},
  \bibinfo {author} {\bibfnamefont {B.~c.~v.}\ \bibnamefont
  {\ifmmode~\check{S}\else \v{S}\fi{}op\'{\i}k}}, \bibinfo {author}
  {\bibfnamefont {K.}~\bibnamefont {Levin}}, \ and\ \bibinfo {author}
  {\bibfnamefont {A.~K.}\ \bibnamefont {Rajagopal}},\ }\href {\doibase
  10.1103/PhysRevB.87.024516} {\bibfield  {journal} {\bibinfo  {journal} {Phys.
  Rev. B}\ }\textbf {\bibinfo {volume} {87}},\ \bibinfo {pages} {024516}
  (\bibinfo {year} {2013})}\BibitemShut {NoStop}%
\bibitem [{\citenamefont {Gruenberg}\ and\ \citenamefont
  {Gunther}(1968)}]{Gruenberg68}%
  \BibitemOpen
  \bibfield  {author} {\bibinfo {author} {\bibfnamefont {L.~W.}\ \bibnamefont
  {Gruenberg}}\ and\ \bibinfo {author} {\bibfnamefont {L.}~\bibnamefont
  {Gunther}},\ }\href {\doibase 10.1103/PhysRev.176.606} {\bibfield  {journal}
  {\bibinfo  {journal} {Physical Review}\ }\textbf {\bibinfo {volume} {176}},\
  \bibinfo {pages} {606} (\bibinfo {year} {1968})},\ \bibinfo {note}
  {publisher: American Physical Society}\BibitemShut {NoStop}%
\bibitem [{\citenamefont {Alexander}(1983)}]{Alexander83}%
  \BibitemOpen
  \bibfield  {author} {\bibinfo {author} {\bibfnamefont {S.}~\bibnamefont
  {Alexander}},\ }\href {\doibase 10.1103/PhysRevB.27.1541} {\bibfield
  {journal} {\bibinfo  {journal} {Physical Review B}\ }\textbf {\bibinfo
  {volume} {27}},\ \bibinfo {pages} {1541} (\bibinfo {year} {1983})},\ \bibinfo
  {note} {publisher: American Physical Society}\BibitemShut {NoStop}%
\bibitem [{\citenamefont {Pannetier}\ \emph {et~al.}(1984)\citenamefont
  {Pannetier}, \citenamefont {Chaussy}, \citenamefont {Rammal},\ and\
  \citenamefont {Villegier}}]{Pannetier84}%
  \BibitemOpen
  \bibfield  {author} {\bibinfo {author} {\bibfnamefont {B.}~\bibnamefont
  {Pannetier}}, \bibinfo {author} {\bibfnamefont {J.}~\bibnamefont {Chaussy}},
  \bibinfo {author} {\bibfnamefont {R.}~\bibnamefont {Rammal}}, \ and\ \bibinfo
  {author} {\bibfnamefont {J.~C.}\ \bibnamefont {Villegier}},\ }\href {\doibase
  10.1103/PhysRevLett.53.1845} {\bibfield  {journal} {\bibinfo  {journal}
  {Physical Review Letters}\ }\textbf {\bibinfo {volume} {53}},\ \bibinfo
  {pages} {1845} (\bibinfo {year} {1984})},\ \bibinfo {note} {publisher:
  American Physical Society}\BibitemShut {NoStop}%
\bibitem [{\citenamefont {Niu}\ and\ \citenamefont {Nori}(1989)}]{Niu89}%
  \BibitemOpen
  \bibfield  {author} {\bibinfo {author} {\bibfnamefont {Q.}~\bibnamefont
  {Niu}}\ and\ \bibinfo {author} {\bibfnamefont {F.}~\bibnamefont {Nori}},\
  }\href {\doibase 10.1103/PhysRevB.39.2134} {\bibfield  {journal} {\bibinfo
  {journal} {Physical Review B}\ }\textbf {\bibinfo {volume} {39}},\ \bibinfo
  {pages} {2134} (\bibinfo {year} {1989})},\ \bibinfo {note} {publisher:
  American Physical Society}\BibitemShut {NoStop}%
\bibitem [{\citenamefont {Kato}\ and\ \citenamefont {Sato}(2013)}]{Kato13}%
  \BibitemOpen
  \bibfield  {author} {\bibinfo {author} {\bibfnamefont {M.}~\bibnamefont
  {Kato}}\ and\ \bibinfo {author} {\bibfnamefont {O.}~\bibnamefont {Sato}},\
  }\href {\doibase 10.1088/0953-2048/26/3/033001} {\bibfield  {journal}
  {\bibinfo  {journal} {Superconductor Science and Technology}\ }\textbf
  {\bibinfo {volume} {26}},\ \bibinfo {pages} {033001} (\bibinfo {year}
  {2013})},\ \bibinfo {note} {publisher: IOP Publishing}\BibitemShut {NoStop}%
\bibitem [{\citenamefont {Dean}\ \emph {et~al.}(2013)\citenamefont {Dean},
  \citenamefont {Wang}, \citenamefont {Maher}, \citenamefont {Forsythe},
  \citenamefont {Ghahari}, \citenamefont {Gao}, \citenamefont {Katoch},
  \citenamefont {Ishigami}, \citenamefont {Moon}, \citenamefont {Koshino},
  \citenamefont {Taniguchi}, \citenamefont {Watanabe}, \citenamefont {Shepard},
  \citenamefont {Hone},\ and\ \citenamefont {Kim}}]{Dean13}%
  \BibitemOpen
  \bibfield  {author} {\bibinfo {author} {\bibfnamefont {C.~R.}\ \bibnamefont
  {Dean}}, \bibinfo {author} {\bibfnamefont {L.}~\bibnamefont {Wang}}, \bibinfo
  {author} {\bibfnamefont {P.}~\bibnamefont {Maher}}, \bibinfo {author}
  {\bibfnamefont {C.}~\bibnamefont {Forsythe}}, \bibinfo {author}
  {\bibfnamefont {F.}~\bibnamefont {Ghahari}}, \bibinfo {author} {\bibfnamefont
  {Y.}~\bibnamefont {Gao}}, \bibinfo {author} {\bibfnamefont {J.}~\bibnamefont
  {Katoch}}, \bibinfo {author} {\bibfnamefont {M.}~\bibnamefont {Ishigami}},
  \bibinfo {author} {\bibfnamefont {P.}~\bibnamefont {Moon}}, \bibinfo {author}
  {\bibfnamefont {M.}~\bibnamefont {Koshino}}, \bibinfo {author} {\bibfnamefont
  {T.}~\bibnamefont {Taniguchi}}, \bibinfo {author} {\bibfnamefont
  {K.}~\bibnamefont {Watanabe}}, \bibinfo {author} {\bibfnamefont {K.~L.}\
  \bibnamefont {Shepard}}, \bibinfo {author} {\bibfnamefont {J.}~\bibnamefont
  {Hone}}, \ and\ \bibinfo {author} {\bibfnamefont {P.}~\bibnamefont {Kim}},\
  }\href {\doibase 10.1038/nature12186} {\bibfield  {journal} {\bibinfo
  {journal} {Nature}\ }\textbf {\bibinfo {volume} {497}},\ \bibinfo {pages}
  {598} (\bibinfo {year} {2013})},\ \bibinfo {note} {number: 7451 Publisher:
  Nature Publishing Group}\BibitemShut {NoStop}%
\bibitem [{\citenamefont {Ponomarenko}\ \emph {et~al.}(2013)\citenamefont
  {Ponomarenko}, \citenamefont {Gorbachev}, \citenamefont {Yu}, \citenamefont
  {Elias}, \citenamefont {Jalil}, \citenamefont {Patel}, \citenamefont
  {Mishchenko}, \citenamefont {Mayorov}, \citenamefont {Woods}, \citenamefont
  {Wallbank}, \citenamefont {Mucha-Kruczynski}, \citenamefont {Piot},
  \citenamefont {Potemski}, \citenamefont {Grigorieva}, \citenamefont
  {Novoselov}, \citenamefont {Guinea}, \citenamefont {Fal'ko},\ and\
  \citenamefont {Geim}}]{Ponomarenko13}%
  \BibitemOpen
  \bibfield  {author} {\bibinfo {author} {\bibfnamefont {L.~A.}\ \bibnamefont
  {Ponomarenko}}, \bibinfo {author} {\bibfnamefont {R.~V.}\ \bibnamefont
  {Gorbachev}}, \bibinfo {author} {\bibfnamefont {G.~L.}\ \bibnamefont {Yu}},
  \bibinfo {author} {\bibfnamefont {D.~C.}\ \bibnamefont {Elias}}, \bibinfo
  {author} {\bibfnamefont {R.}~\bibnamefont {Jalil}}, \bibinfo {author}
  {\bibfnamefont {A.~A.}\ \bibnamefont {Patel}}, \bibinfo {author}
  {\bibfnamefont {A.}~\bibnamefont {Mishchenko}}, \bibinfo {author}
  {\bibfnamefont {A.~S.}\ \bibnamefont {Mayorov}}, \bibinfo {author}
  {\bibfnamefont {C.~R.}\ \bibnamefont {Woods}}, \bibinfo {author}
  {\bibfnamefont {J.~R.}\ \bibnamefont {Wallbank}}, \bibinfo {author}
  {\bibfnamefont {M.}~\bibnamefont {Mucha-Kruczynski}}, \bibinfo {author}
  {\bibfnamefont {B.~A.}\ \bibnamefont {Piot}}, \bibinfo {author}
  {\bibfnamefont {M.}~\bibnamefont {Potemski}}, \bibinfo {author}
  {\bibfnamefont {I.~V.}\ \bibnamefont {Grigorieva}}, \bibinfo {author}
  {\bibfnamefont {K.~S.}\ \bibnamefont {Novoselov}}, \bibinfo {author}
  {\bibfnamefont {F.}~\bibnamefont {Guinea}}, \bibinfo {author} {\bibfnamefont
  {V.~I.}\ \bibnamefont {Fal'ko}}, \ and\ \bibinfo {author} {\bibfnamefont
  {A.~K.}\ \bibnamefont {Geim}},\ }\href {\doibase 10.1038/nature12187}
  {\bibfield  {journal} {\bibinfo  {journal} {Nature}\ }\textbf {\bibinfo
  {volume} {497}},\ \bibinfo {pages} {594} (\bibinfo {year}
  {2013})}\BibitemShut {NoStop}%
\bibitem [{\citenamefont {Hunt}\ \emph {et~al.}(2013)\citenamefont {Hunt},
  \citenamefont {Sanchez-Yamagishi}, \citenamefont {Young}, \citenamefont
  {Yankowitz}, \citenamefont {LeRoy}, \citenamefont {Watanabe}, \citenamefont
  {Taniguchi}, \citenamefont {Moon}, \citenamefont {Koshino}, \citenamefont
  {Jarillo-Herrero},\ and\ \citenamefont {Ashoori}}]{Hunt13}%
  \BibitemOpen
  \bibfield  {author} {\bibinfo {author} {\bibfnamefont {B.}~\bibnamefont
  {Hunt}}, \bibinfo {author} {\bibfnamefont {J.~D.}\ \bibnamefont
  {Sanchez-Yamagishi}}, \bibinfo {author} {\bibfnamefont {A.~F.}\ \bibnamefont
  {Young}}, \bibinfo {author} {\bibfnamefont {M.}~\bibnamefont {Yankowitz}},
  \bibinfo {author} {\bibfnamefont {B.~J.}\ \bibnamefont {LeRoy}}, \bibinfo
  {author} {\bibfnamefont {K.}~\bibnamefont {Watanabe}}, \bibinfo {author}
  {\bibfnamefont {T.}~\bibnamefont {Taniguchi}}, \bibinfo {author}
  {\bibfnamefont {P.}~\bibnamefont {Moon}}, \bibinfo {author} {\bibfnamefont
  {M.}~\bibnamefont {Koshino}}, \bibinfo {author} {\bibfnamefont
  {P.}~\bibnamefont {Jarillo-Herrero}}, \ and\ \bibinfo {author} {\bibfnamefont
  {R.~C.}\ \bibnamefont {Ashoori}},\ }\href {\doibase 10.1126/science.1237240}
  {\bibfield  {journal} {\bibinfo  {journal} {Science}\ }\textbf {\bibinfo
  {volume} {340}},\ \bibinfo {pages} {1427} (\bibinfo {year} {2013})},\
  \bibinfo {note} {publisher: American Association for the Advancement of
  Science Section: Report}\BibitemShut {NoStop}%
\bibitem [{\citenamefont {Forsythe}\ \emph {et~al.}(2018)\citenamefont
  {Forsythe}, \citenamefont {Zhou}, \citenamefont {Watanabe}, \citenamefont
  {Taniguchi}, \citenamefont {Pasupathy}, \citenamefont {Moon}, \citenamefont
  {Koshino}, \citenamefont {Kim},\ and\ \citenamefont {Dean}}]{Forsythe18}%
  \BibitemOpen
  \bibfield  {author} {\bibinfo {author} {\bibfnamefont {C.}~\bibnamefont
  {Forsythe}}, \bibinfo {author} {\bibfnamefont {X.}~\bibnamefont {Zhou}},
  \bibinfo {author} {\bibfnamefont {K.}~\bibnamefont {Watanabe}}, \bibinfo
  {author} {\bibfnamefont {T.}~\bibnamefont {Taniguchi}}, \bibinfo {author}
  {\bibfnamefont {A.}~\bibnamefont {Pasupathy}}, \bibinfo {author}
  {\bibfnamefont {P.}~\bibnamefont {Moon}}, \bibinfo {author} {\bibfnamefont
  {M.}~\bibnamefont {Koshino}}, \bibinfo {author} {\bibfnamefont
  {P.}~\bibnamefont {Kim}}, \ and\ \bibinfo {author} {\bibfnamefont {C.~R.}\
  \bibnamefont {Dean}},\ }\href {\doibase 10.1038/s41565-018-0138-7} {\bibfield
   {journal} {\bibinfo  {journal} {Nature Nanotechnology}\ }\textbf {\bibinfo
  {volume} {13}},\ \bibinfo {pages} {566} (\bibinfo {year} {2018})},\ \bibinfo
  {note} {number: 7 Publisher: Nature Publishing Group}\BibitemShut {NoStop}%
\bibitem [{\citenamefont {Wang}\ \emph {et~al.}(2015)\citenamefont {Wang},
  \citenamefont {Gao}, \citenamefont {Wen}, \citenamefont {Han}, \citenamefont
  {Taniguchi}, \citenamefont {Watanabe}, \citenamefont {Koshino}, \citenamefont
  {Hone},\ and\ \citenamefont {Dean}}]{Wang15}%
  \BibitemOpen
  \bibfield  {author} {\bibinfo {author} {\bibfnamefont {L.}~\bibnamefont
  {Wang}}, \bibinfo {author} {\bibfnamefont {Y.}~\bibnamefont {Gao}}, \bibinfo
  {author} {\bibfnamefont {B.}~\bibnamefont {Wen}}, \bibinfo {author}
  {\bibfnamefont {Z.}~\bibnamefont {Han}}, \bibinfo {author} {\bibfnamefont
  {T.}~\bibnamefont {Taniguchi}}, \bibinfo {author} {\bibfnamefont
  {K.}~\bibnamefont {Watanabe}}, \bibinfo {author} {\bibfnamefont
  {M.}~\bibnamefont {Koshino}}, \bibinfo {author} {\bibfnamefont
  {J.}~\bibnamefont {Hone}}, \ and\ \bibinfo {author} {\bibfnamefont {C.~R.}\
  \bibnamefont {Dean}},\ }\href {\doibase 10.1126/science.aad2102} {\bibfield
  {journal} {\bibinfo  {journal} {Science}\ }\textbf {\bibinfo {volume}
  {350}},\ \bibinfo {pages} {1231} (\bibinfo {year} {2015})},\ \bibinfo {note}
  {publisher: American Association for the Advancement of Science Section:
  Report}\BibitemShut {NoStop}%
\bibitem [{\citenamefont {Spanton}\ \emph {et~al.}(2018)\citenamefont
  {Spanton}, \citenamefont {Zibrov}, \citenamefont {Zhou}, \citenamefont
  {Taniguchi}, \citenamefont {Watanabe}, \citenamefont {Zaletel},\ and\
  \citenamefont {Young}}]{Spanton18}%
  \BibitemOpen
  \bibfield  {author} {\bibinfo {author} {\bibfnamefont {E.~M.}\ \bibnamefont
  {Spanton}}, \bibinfo {author} {\bibfnamefont {A.~A.}\ \bibnamefont {Zibrov}},
  \bibinfo {author} {\bibfnamefont {H.}~\bibnamefont {Zhou}}, \bibinfo {author}
  {\bibfnamefont {T.}~\bibnamefont {Taniguchi}}, \bibinfo {author}
  {\bibfnamefont {K.}~\bibnamefont {Watanabe}}, \bibinfo {author}
  {\bibfnamefont {M.~P.}\ \bibnamefont {Zaletel}}, \ and\ \bibinfo {author}
  {\bibfnamefont {A.~F.}\ \bibnamefont {Young}},\ }\href {\doibase
  10.1126/science.aan8458} {\bibfield  {journal} {\bibinfo  {journal}
  {Science}\ }\textbf {\bibinfo {volume} {360}},\ \bibinfo {pages} {62}
  (\bibinfo {year} {2018})},\ \bibinfo {note} {publisher: American Association
  for the Advancement of Science Section: Report}\BibitemShut {NoStop}%
\bibitem [{\citenamefont {Saito}\ \emph {et~al.}(2021)\citenamefont {Saito},
  \citenamefont {Ge}, \citenamefont {Rademaker}, \citenamefont {Watanabe},
  \citenamefont {Taniguchi}, \citenamefont {Abanin},\ and\ \citenamefont
  {Young}}]{Saito21}%
  \BibitemOpen
  \bibfield  {author} {\bibinfo {author} {\bibfnamefont {Y.}~\bibnamefont
  {Saito}}, \bibinfo {author} {\bibfnamefont {J.}~\bibnamefont {Ge}}, \bibinfo
  {author} {\bibfnamefont {L.}~\bibnamefont {Rademaker}}, \bibinfo {author}
  {\bibfnamefont {K.}~\bibnamefont {Watanabe}}, \bibinfo {author}
  {\bibfnamefont {T.}~\bibnamefont {Taniguchi}}, \bibinfo {author}
  {\bibfnamefont {D.~A.}\ \bibnamefont {Abanin}}, \ and\ \bibinfo {author}
  {\bibfnamefont {A.~F.}\ \bibnamefont {Young}},\ }\href {\doibase
  10.1038/s41567-020-01129-4} {\bibfield  {journal} {\bibinfo  {journal}
  {Nature Physics}\ }\textbf {\bibinfo {volume} {17}},\ \bibinfo {pages} {478}
  (\bibinfo {year} {2021})},\ \bibinfo {note} {number: 4 Publisher: Nature
  Publishing Group}\BibitemShut {NoStop}%
\bibitem [{\citenamefont {Mueller}(2004)}]{Mueller04}%
  \BibitemOpen
  \bibfield  {author} {\bibinfo {author} {\bibfnamefont {E.~J.}\ \bibnamefont
  {Mueller}},\ }\href {\doibase 10.1103/PhysRevA.70.041603} {\bibfield
  {journal} {\bibinfo  {journal} {Physical Review A}\ }\textbf {\bibinfo
  {volume} {70}},\ \bibinfo {pages} {041603} (\bibinfo {year} {2004})},\
  \bibinfo {note} {publisher: American Physical Society}\BibitemShut {NoStop}%
\bibitem [{\citenamefont {Gerbier}\ and\ \citenamefont
  {Dalibard}(2010)}]{Gerbier10}%
  \BibitemOpen
  \bibfield  {author} {\bibinfo {author} {\bibfnamefont {F.}~\bibnamefont
  {Gerbier}}\ and\ \bibinfo {author} {\bibfnamefont {J.}~\bibnamefont
  {Dalibard}},\ }\href {\doibase 10.1088/1367-2630/12/3/033007} {\bibfield
  {journal} {\bibinfo  {journal} {New Journal of Physics}\ }\textbf {\bibinfo
  {volume} {12}},\ \bibinfo {pages} {033007} (\bibinfo {year} {2010})},\
  \bibinfo {note} {publisher: IOP Publishing}\BibitemShut {NoStop}%
\bibitem [{\citenamefont {Aidelsburger}\ \emph {et~al.}(2011)\citenamefont
  {Aidelsburger}, \citenamefont {Atala}, \citenamefont {Nascimb\`ene},
  \citenamefont {Trotzky}, \citenamefont {Chen},\ and\ \citenamefont
  {Bloch}}]{Aidelsburger11}%
  \BibitemOpen
  \bibfield  {author} {\bibinfo {author} {\bibfnamefont {M.}~\bibnamefont
  {Aidelsburger}}, \bibinfo {author} {\bibfnamefont {M.}~\bibnamefont {Atala}},
  \bibinfo {author} {\bibfnamefont {S.}~\bibnamefont {Nascimb\`ene}}, \bibinfo
  {author} {\bibfnamefont {S.}~\bibnamefont {Trotzky}}, \bibinfo {author}
  {\bibfnamefont {Y.-A.}\ \bibnamefont {Chen}}, \ and\ \bibinfo {author}
  {\bibfnamefont {I.}~\bibnamefont {Bloch}},\ }\href {\doibase
  10.1103/PhysRevLett.107.255301} {\bibfield  {journal} {\bibinfo  {journal}
  {Phys. Rev. Lett.}\ }\textbf {\bibinfo {volume} {107}},\ \bibinfo {pages}
  {255301} (\bibinfo {year} {2011})}\BibitemShut {NoStop}%
\bibitem [{\citenamefont {Hauke}\ \emph {et~al.}(2012)\citenamefont {Hauke},
  \citenamefont {Tieleman}, \citenamefont {Celi}, \citenamefont
  {\"Olschl\"ager}, \citenamefont {Simonet}, \citenamefont {Struck},
  \citenamefont {Weinberg}, \citenamefont {Windpassinger}, \citenamefont
  {Sengstock}, \citenamefont {Lewenstein},\ and\ \citenamefont
  {Eckardt}}]{Hauke12}%
  \BibitemOpen
  \bibfield  {author} {\bibinfo {author} {\bibfnamefont {P.}~\bibnamefont
  {Hauke}}, \bibinfo {author} {\bibfnamefont {O.}~\bibnamefont {Tieleman}},
  \bibinfo {author} {\bibfnamefont {A.}~\bibnamefont {Celi}}, \bibinfo {author}
  {\bibfnamefont {C.}~\bibnamefont {\"Olschl\"ager}}, \bibinfo {author}
  {\bibfnamefont {J.}~\bibnamefont {Simonet}}, \bibinfo {author} {\bibfnamefont
  {J.}~\bibnamefont {Struck}}, \bibinfo {author} {\bibfnamefont
  {M.}~\bibnamefont {Weinberg}}, \bibinfo {author} {\bibfnamefont
  {P.}~\bibnamefont {Windpassinger}}, \bibinfo {author} {\bibfnamefont
  {K.}~\bibnamefont {Sengstock}}, \bibinfo {author} {\bibfnamefont
  {M.}~\bibnamefont {Lewenstein}}, \ and\ \bibinfo {author} {\bibfnamefont
  {A.}~\bibnamefont {Eckardt}},\ }\href {\doibase
  10.1103/PhysRevLett.109.145301} {\bibfield  {journal} {\bibinfo  {journal}
  {Physical Review Letters}\ }\textbf {\bibinfo {volume} {109}},\ \bibinfo
  {pages} {145301} (\bibinfo {year} {2012})},\ \bibinfo {note} {publisher:
  American Physical Society}\BibitemShut {NoStop}%
\bibitem [{\citenamefont {Celi}\ \emph {et~al.}(2014)\citenamefont {Celi},
  \citenamefont {Massignan}, \citenamefont {Ruseckas}, \citenamefont {Goldman},
  \citenamefont {Spielman}, \citenamefont
  {Juzeli\ifmmode\bar{u}\else\={u}\fi{}nas},\ and\ \citenamefont
  {Lewenstein}}]{Celi14}%
  \BibitemOpen
  \bibfield  {author} {\bibinfo {author} {\bibfnamefont {A.}~\bibnamefont
  {Celi}}, \bibinfo {author} {\bibfnamefont {P.}~\bibnamefont {Massignan}},
  \bibinfo {author} {\bibfnamefont {J.}~\bibnamefont {Ruseckas}}, \bibinfo
  {author} {\bibfnamefont {N.}~\bibnamefont {Goldman}}, \bibinfo {author}
  {\bibfnamefont {I.~B.}\ \bibnamefont {Spielman}}, \bibinfo {author}
  {\bibfnamefont {G.}~\bibnamefont {Juzeli\ifmmode\bar{u}\else\={u}\fi{}nas}},
  \ and\ \bibinfo {author} {\bibfnamefont {M.}~\bibnamefont {Lewenstein}},\
  }\href {\doibase 10.1103/PhysRevLett.112.043001} {\bibfield  {journal}
  {\bibinfo  {journal} {Physical Review Letters}\ }\textbf {\bibinfo {volume}
  {112}},\ \bibinfo {pages} {043001} (\bibinfo {year} {2014})},\ \bibinfo
  {note} {publisher: American Physical Society}\BibitemShut {NoStop}%
\bibitem [{\citenamefont {Ran}\ \emph {et~al.}(2019)\citenamefont {Ran},
  \citenamefont {Liu}, \citenamefont {Eo}, \citenamefont {Campbell},
  \citenamefont {Neves}, \citenamefont {Fuhrman}, \citenamefont {Saha},
  \citenamefont {Eckberg}, \citenamefont {Kim}, \citenamefont {Graf},
  \citenamefont {Balakirev}, \citenamefont {Singleton}, \citenamefont
  {Paglione},\ and\ \citenamefont {Butch}}]{Ran19}%
  \BibitemOpen
  \bibfield  {author} {\bibinfo {author} {\bibfnamefont {S.}~\bibnamefont
  {Ran}}, \bibinfo {author} {\bibfnamefont {I.-L.}\ \bibnamefont {Liu}},
  \bibinfo {author} {\bibfnamefont {Y.~S.}\ \bibnamefont {Eo}}, \bibinfo
  {author} {\bibfnamefont {D.~J.}\ \bibnamefont {Campbell}}, \bibinfo {author}
  {\bibfnamefont {P.~M.}\ \bibnamefont {Neves}}, \bibinfo {author}
  {\bibfnamefont {W.~T.}\ \bibnamefont {Fuhrman}}, \bibinfo {author}
  {\bibfnamefont {S.~R.}\ \bibnamefont {Saha}}, \bibinfo {author}
  {\bibfnamefont {C.}~\bibnamefont {Eckberg}}, \bibinfo {author} {\bibfnamefont
  {H.}~\bibnamefont {Kim}}, \bibinfo {author} {\bibfnamefont {D.}~\bibnamefont
  {Graf}}, \bibinfo {author} {\bibfnamefont {F.}~\bibnamefont {Balakirev}},
  \bibinfo {author} {\bibfnamefont {J.}~\bibnamefont {Singleton}}, \bibinfo
  {author} {\bibfnamefont {J.}~\bibnamefont {Paglione}}, \ and\ \bibinfo
  {author} {\bibfnamefont {N.~P.}\ \bibnamefont {Butch}},\ }\href {\doibase
  10.1038/s41567-019-0670-x} {\bibfield  {journal} {\bibinfo  {journal} {Nature
  Physics}\ }\textbf {\bibinfo {volume} {15}},\ \bibinfo {pages} {1250}
  (\bibinfo {year} {2019})},\ \bibinfo {note} {number: 12 Publisher: Nature
  Publishing Group}\BibitemShut {NoStop}%
\bibitem [{\citenamefont {Mineev}(2020)}]{Mineev20}%
  \BibitemOpen
  \bibfield  {author} {\bibinfo {author} {\bibfnamefont {V.~P.}\ \bibnamefont
  {Mineev}},\ }\href {\doibase 10.1134/S0021364020120036} {\bibfield  {journal}
  {\bibinfo  {journal} {JETP Letters}\ }\textbf {\bibinfo {volume} {111}},\
  \bibinfo {pages} {715} (\bibinfo {year} {2020})}\BibitemShut {NoStop}%
\bibitem [{\citenamefont {Lebed}(2020)}]{Lebed20}%
  \BibitemOpen
  \bibfield  {author} {\bibinfo {author} {\bibfnamefont {A.~G.}\ \bibnamefont
  {Lebed}},\ }\href {\doibase 10.1142/S0217984920300070} {\bibfield  {journal}
  {\bibinfo  {journal} {Modern Physics Letters B}\ }\textbf {\bibinfo {volume}
  {34}},\ \bibinfo {pages} {2030007} (\bibinfo {year} {2020})}\BibitemShut
  {NoStop}%
\bibitem [{\citenamefont {Cao}\ \emph {et~al.}(2021)\citenamefont {Cao},
  \citenamefont {Park}, \citenamefont {Watanabe}, \citenamefont {Taniguchi},\
  and\ \citenamefont {Jarillo-Herrero}}]{CaoTTG21}%
  \BibitemOpen
  \bibfield  {author} {\bibinfo {author} {\bibfnamefont {Y.}~\bibnamefont
  {Cao}}, \bibinfo {author} {\bibfnamefont {J.~M.}\ \bibnamefont {Park}},
  \bibinfo {author} {\bibfnamefont {K.}~\bibnamefont {Watanabe}}, \bibinfo
  {author} {\bibfnamefont {T.}~\bibnamefont {Taniguchi}}, \ and\ \bibinfo
  {author} {\bibfnamefont {P.}~\bibnamefont {Jarillo-Herrero}},\ }\href
  {http://arxiv.org/abs/2103.12083} {\bibfield  {journal} {\bibinfo  {journal}
  {arXiv:2103.12083 [cond-mat]}\ } (\bibinfo {year} {2021})},\ \bibinfo {note}
  {arXiv: 2103.12083}\BibitemShut {NoStop}%
\bibitem [{\citenamefont {Park}\ \emph {et~al.}(2020)\citenamefont {Park},
  \citenamefont {Kim},\ and\ \citenamefont {Lee}}]{Park20}%
  \BibitemOpen
  \bibfield  {author} {\bibinfo {author} {\bibfnamefont {M.~J.}\ \bibnamefont
  {Park}}, \bibinfo {author} {\bibfnamefont {Y.~B.}\ \bibnamefont {Kim}}, \
  and\ \bibinfo {author} {\bibfnamefont {S.}~\bibnamefont {Lee}},\ }\href
  {http://arxiv.org/abs/2007.16205} {\bibfield  {journal} {\bibinfo  {journal}
  {arXiv:2007.16205 [cond-mat]}\ } (\bibinfo {year} {2020})},\ \bibinfo {note}
  {arXiv: 2007.16205}\BibitemShut {NoStop}%
\bibitem [{\citenamefont {Wang}\ and\ \citenamefont
  {Santos}(2020)}]{wang_classification_2020}%
  \BibitemOpen
  \bibfield  {author} {\bibinfo {author} {\bibfnamefont {J.}~\bibnamefont
  {Wang}}\ and\ \bibinfo {author} {\bibfnamefont {L.~H.}\ \bibnamefont
  {Santos}},\ }\href {\doibase 10.1103/PhysRevLett.125.236805} {\bibfield
  {journal} {\bibinfo  {journal} {Phys. Rev. Lett.}\ }\textbf {\bibinfo
  {volume} {125}},\ \bibinfo {pages} {236805} (\bibinfo {year}
  {2020})}\BibitemShut {NoStop}%
\bibitem [{\citenamefont {Lee}\ \emph {et~al.}(2018)\citenamefont {Lee},
  \citenamefont {Wang}, \citenamefont {Zaletel}, \citenamefont {Vishwanath},\
  and\ \citenamefont {He}}]{Lee-PRX-2018}%
  \BibitemOpen
  \bibfield  {author} {\bibinfo {author} {\bibfnamefont {J.~Y.}\ \bibnamefont
  {Lee}}, \bibinfo {author} {\bibfnamefont {C.}~\bibnamefont {Wang}}, \bibinfo
  {author} {\bibfnamefont {M.~P.}\ \bibnamefont {Zaletel}}, \bibinfo {author}
  {\bibfnamefont {A.}~\bibnamefont {Vishwanath}}, \ and\ \bibinfo {author}
  {\bibfnamefont {Y.-C.}\ \bibnamefont {He}},\ }\href {\doibase
  10.1103/PhysRevX.8.031015} {\bibfield  {journal} {\bibinfo  {journal} {Phys.
  Rev. X}\ }\textbf {\bibinfo {volume} {8}},\ \bibinfo {pages} {031015}
  (\bibinfo {year} {2018})}\BibitemShut {NoStop}%
\bibitem [{\citenamefont {Zak}(1964{\natexlab{b}})}]{Zak64_2}%
  \BibitemOpen
  \bibfield  {author} {\bibinfo {author} {\bibfnamefont {J.}~\bibnamefont
  {Zak}},\ }\href {\doibase 10.1103/PhysRev.134.A1607} {\bibfield  {journal}
  {\bibinfo  {journal} {Physical Review}\ }\textbf {\bibinfo {volume} {134}},\
  \bibinfo {pages} {A1607} (\bibinfo {year} {1964}{\natexlab{b}})},\ \bibinfo
  {note} {publisher: American Physical Society}\BibitemShut {NoStop}%
\bibitem [{\citenamefont {Florek}(1997)}]{Florek97}%
  \BibitemOpen
  \bibfield  {author} {\bibinfo {author} {\bibfnamefont {W.}~\bibnamefont
  {Florek}},\ }\href {\doibase 10.1103/PhysRevB.55.1449} {\bibfield  {journal}
  {\bibinfo  {journal} {Phys. Rev. B}\ }\textbf {\bibinfo {volume} {55}},\
  \bibinfo {pages} {1449} (\bibinfo {year} {1997})}\BibitemShut {NoStop}%
\bibitem [{\citenamefont {Balents}\ \emph {et~al.}(2005)\citenamefont
  {Balents}, \citenamefont {Bartosch}, \citenamefont {Burkov}, \citenamefont
  {Sachdev},\ and\ \citenamefont {Sengupta}}]{Balents05}%
  \BibitemOpen
  \bibfield  {author} {\bibinfo {author} {\bibfnamefont {L.}~\bibnamefont
  {Balents}}, \bibinfo {author} {\bibfnamefont {L.}~\bibnamefont {Bartosch}},
  \bibinfo {author} {\bibfnamefont {A.}~\bibnamefont {Burkov}}, \bibinfo
  {author} {\bibfnamefont {S.}~\bibnamefont {Sachdev}}, \ and\ \bibinfo
  {author} {\bibfnamefont {K.}~\bibnamefont {Sengupta}},\ }\href {\doibase
  10.1103/PhysRevB.71.144508} {\bibfield  {journal} {\bibinfo  {journal}
  {Physical Review B}\ }\textbf {\bibinfo {volume} {71}},\ \bibinfo {pages}
  {144508} (\bibinfo {year} {2005})},\ \bibinfo {note} {publisher: American
  Physical Society}\BibitemShut {NoStop}%
\bibitem [{\citenamefont {Powell}\ \emph {et~al.}(2010)\citenamefont {Powell},
  \citenamefont {Barnett}, \citenamefont {Sensarma},\ and\ \citenamefont
  {Das~Sarma}}]{PowellDasSarma10}%
  \BibitemOpen
  \bibfield  {author} {\bibinfo {author} {\bibfnamefont {S.}~\bibnamefont
  {Powell}}, \bibinfo {author} {\bibfnamefont {R.}~\bibnamefont {Barnett}},
  \bibinfo {author} {\bibfnamefont {R.}~\bibnamefont {Sensarma}}, \ and\
  \bibinfo {author} {\bibfnamefont {S.}~\bibnamefont {Das~Sarma}},\ }\href
  {\doibase 10.1103/PhysRevLett.104.255303} {\bibfield  {journal} {\bibinfo
  {journal} {Physical Review Letters}\ }\textbf {\bibinfo {volume} {104}},\
  \bibinfo {pages} {255303} (\bibinfo {year} {2010})},\ \bibinfo {note}
  {publisher: American Physical Society}\BibitemShut {NoStop}%
\bibitem [{\citenamefont {Powell}\ \emph {et~al.}(2011)\citenamefont {Powell},
  \citenamefont {Barnett}, \citenamefont {Sensarma},\ and\ \citenamefont
  {Das~Sarma}}]{PowellDasSarma11}%
  \BibitemOpen
  \bibfield  {author} {\bibinfo {author} {\bibfnamefont {S.}~\bibnamefont
  {Powell}}, \bibinfo {author} {\bibfnamefont {R.}~\bibnamefont {Barnett}},
  \bibinfo {author} {\bibfnamefont {R.}~\bibnamefont {Sensarma}}, \ and\
  \bibinfo {author} {\bibfnamefont {S.}~\bibnamefont {Das~Sarma}},\ }\href
  {\doibase 10.1103/PhysRevA.83.013612} {\bibfield  {journal} {\bibinfo
  {journal} {Physical Review A}\ }\textbf {\bibinfo {volume} {83}},\ \bibinfo
  {pages} {013612} (\bibinfo {year} {2011})},\ \bibinfo {note} {publisher:
  American Physical Society}\BibitemShut {NoStop}%
\bibitem [{\citenamefont {Natu}\ \emph {et~al.}(2016)\citenamefont {Natu},
  \citenamefont {Mueller},\ and\ \citenamefont {Das~Sarma}}]{NatuDasSarma16}%
  \BibitemOpen
  \bibfield  {author} {\bibinfo {author} {\bibfnamefont {S.~S.}\ \bibnamefont
  {Natu}}, \bibinfo {author} {\bibfnamefont {E.~J.}\ \bibnamefont {Mueller}}, \
  and\ \bibinfo {author} {\bibfnamefont {S.}~\bibnamefont {Das~Sarma}},\ }\href
  {\doibase 10.1103/PhysRevA.93.063610} {\bibfield  {journal} {\bibinfo
  {journal} {Physical Review A}\ }\textbf {\bibinfo {volume} {93}},\ \bibinfo
  {pages} {063610} (\bibinfo {year} {2016})},\ \bibinfo {note} {publisher:
  American Physical Society}\BibitemShut {NoStop}%
\bibitem [{\citenamefont {Song}\ and\ \citenamefont {Yang}(2019)}]{Song19}%
  \BibitemOpen
  \bibfield  {author} {\bibinfo {author} {\bibfnamefont {Y.-F.}\ \bibnamefont
  {Song}}\ and\ \bibinfo {author} {\bibfnamefont {S.-J.}\ \bibnamefont
  {Yang}},\ }\href {\doibase 10.1088/1361-6455/ab08da} {\bibfield  {journal}
  {\bibinfo  {journal} {Journal of Physics B: Atomic, Molecular and Optical
  Physics}\ }\textbf {\bibinfo {volume} {52}},\ \bibinfo {pages} {118001}
  (\bibinfo {year} {2019})},\ \bibinfo {note} {publisher: IOP
  Publishing}\BibitemShut {NoStop}%
\bibitem [{\citenamefont {Sigrist}\ and\ \citenamefont
  {Ueda}(1991)}]{SigristUeda91}%
  \BibitemOpen
  \bibfield  {author} {\bibinfo {author} {\bibfnamefont {M.}~\bibnamefont
  {Sigrist}}\ and\ \bibinfo {author} {\bibfnamefont {K.}~\bibnamefont {Ueda}},\
  }\href {\doibase 10.1103/RevModPhys.63.239} {\bibfield  {journal} {\bibinfo
  {journal} {Reviews of Modern Physics}\ }\textbf {\bibinfo {volume} {63}},\
  \bibinfo {pages} {239} (\bibinfo {year} {1991})},\ \bibinfo {note}
  {publisher: American Physical Society}\BibitemShut {NoStop}%
\bibitem [{\citenamefont {Zhai}\ \emph {et~al.}(2010)\citenamefont {Zhai},
  \citenamefont {Umucal\ifmmode\imath\else\i\fi{}lar},\ and\ \citenamefont
  {Oktel}}]{ZhaiOktel10}%
  \BibitemOpen
  \bibfield  {author} {\bibinfo {author} {\bibfnamefont {H.}~\bibnamefont
  {Zhai}}, \bibinfo {author} {\bibfnamefont {R.~O.}\ \bibnamefont
  {Umucal\ifmmode\imath\else\i\fi{}lar}}, \ and\ \bibinfo {author}
  {\bibfnamefont {M.~O.}\ \bibnamefont {Oktel}},\ }\href {\doibase
  10.1103/PhysRevLett.104.145301} {\bibfield  {journal} {\bibinfo  {journal}
  {Physical Review Letters}\ }\textbf {\bibinfo {volume} {104}},\ \bibinfo
  {pages} {145301} (\bibinfo {year} {2010})},\ \bibinfo {note} {publisher:
  American Physical Society}\BibitemShut {NoStop}%
\bibitem [{\citenamefont {Iskin}(2015{\natexlab{a}})}]{Iskin15_1}%
  \BibitemOpen
  \bibfield  {author} {\bibinfo {author} {\bibfnamefont {M.}~\bibnamefont
  {Iskin}},\ }\href {\doibase 10.1103/PhysRevA.91.011601} {\bibfield  {journal}
  {\bibinfo  {journal} {Physical Review A}\ }\textbf {\bibinfo {volume} {91}},\
  \bibinfo {pages} {011601} (\bibinfo {year} {2015}{\natexlab{a}})},\ \bibinfo
  {note} {publisher: American Physical Society}\BibitemShut {NoStop}%
\bibitem [{\citenamefont {Iskin}(2015{\natexlab{b}})}]{Iskin15_2}%
  \BibitemOpen
  \bibfield  {author} {\bibinfo {author} {\bibfnamefont {M.}~\bibnamefont
  {Iskin}},\ }\href {\doibase 10.1103/PhysRevA.91.053606} {\bibfield  {journal}
  {\bibinfo  {journal} {Physical Review A}\ }\textbf {\bibinfo {volume} {91}},\
  \bibinfo {pages} {053606} (\bibinfo {year} {2015}{\natexlab{b}})},\ \bibinfo
  {note} {publisher: American Physical Society}\BibitemShut {NoStop}%
\bibitem [{\citenamefont {Umucal\ifmmode\imath\else\i\fi{}lar}\ and\
  \citenamefont {Iskin}(2016)}]{UmucalilarIskin16}%
  \BibitemOpen
  \bibfield  {author} {\bibinfo {author} {\bibfnamefont {R.~O.}\ \bibnamefont
  {Umucal\ifmmode\imath\else\i\fi{}lar}}\ and\ \bibinfo {author} {\bibfnamefont
  {M.}~\bibnamefont {Iskin}},\ }\href {\doibase 10.1103/PhysRevA.94.023611}
  {\bibfield  {journal} {\bibinfo  {journal} {Physical Review A}\ }\textbf
  {\bibinfo {volume} {94}},\ \bibinfo {pages} {023611} (\bibinfo {year}
  {2016})},\ \bibinfo {note} {publisher: American Physical Society}\BibitemShut
  {NoStop}%
\bibitem [{\citenamefont {Sohal}\ and\ \citenamefont
  {Fradkin}(2020)}]{SohalFradkin20}%
  \BibitemOpen
  \bibfield  {author} {\bibinfo {author} {\bibfnamefont {R.}~\bibnamefont
  {Sohal}}\ and\ \bibinfo {author} {\bibfnamefont {E.}~\bibnamefont
  {Fradkin}},\ }\href {\doibase 10.1103/PhysRevB.101.245154} {\bibfield
  {journal} {\bibinfo  {journal} {Physical Review B}\ }\textbf {\bibinfo
  {volume} {101}},\ \bibinfo {pages} {245154} (\bibinfo {year} {2020})},\
  \bibinfo {note} {publisher: American Physical Society}\BibitemShut {NoStop}%
\bibitem [{\citenamefont {Read}\ and\ \citenamefont
  {Green}(2000)}]{ReadGreen00}%
  \BibitemOpen
  \bibfield  {author} {\bibinfo {author} {\bibfnamefont {N.}~\bibnamefont
  {Read}}\ and\ \bibinfo {author} {\bibfnamefont {D.}~\bibnamefont {Green}},\
  }\href {\doibase 10.1103/PhysRevB.61.10267} {\bibfield  {journal} {\bibinfo
  {journal} {Phys. Rev. B}\ }\textbf {\bibinfo {volume} {61}},\ \bibinfo
  {pages} {10267} (\bibinfo {year} {2000})}\BibitemShut {NoStop}%
\bibitem [{\citenamefont {Agterberg}\ \emph {et~al.}(2017)\citenamefont
  {Agterberg}, \citenamefont {Brydon},\ and\ \citenamefont
  {Timm}}]{Agterberg17}%
  \BibitemOpen
  \bibfield  {author} {\bibinfo {author} {\bibfnamefont {D.~F.}\ \bibnamefont
  {Agterberg}}, \bibinfo {author} {\bibfnamefont {P.~M.~R.}\ \bibnamefont
  {Brydon}}, \ and\ \bibinfo {author} {\bibfnamefont {C.}~\bibnamefont
  {Timm}},\ }\href {\doibase 10.1103/PhysRevLett.118.127001} {\bibfield
  {journal} {\bibinfo  {journal} {Physical Review Letters}\ }\textbf {\bibinfo
  {volume} {118}},\ \bibinfo {pages} {127001} (\bibinfo {year} {2017})},\
  \bibinfo {note} {publisher: American Physical Society}\BibitemShut {NoStop}%
\bibitem [{\citenamefont {Brydon}\ \emph {et~al.}(2018)\citenamefont {Brydon},
  \citenamefont {Agterberg}, \citenamefont {Menke},\ and\ \citenamefont
  {Timm}}]{BrydonAgterberg18}%
  \BibitemOpen
  \bibfield  {author} {\bibinfo {author} {\bibfnamefont {P.~M.~R.}\
  \bibnamefont {Brydon}}, \bibinfo {author} {\bibfnamefont {D.~F.}\
  \bibnamefont {Agterberg}}, \bibinfo {author} {\bibfnamefont {H.}~\bibnamefont
  {Menke}}, \ and\ \bibinfo {author} {\bibfnamefont {C.}~\bibnamefont {Timm}},\
  }\href {\doibase 10.1103/PhysRevB.98.224509} {\bibfield  {journal} {\bibinfo
  {journal} {Physical Review B}\ }\textbf {\bibinfo {volume} {98}},\ \bibinfo
  {pages} {224509} (\bibinfo {year} {2018})},\ \bibinfo {note} {publisher:
  American Physical Society}\BibitemShut {NoStop}%
\bibitem [{\citenamefont {Santos}\ \emph {et~al.}(2019)\citenamefont {Santos},
  \citenamefont {Wang},\ and\ \citenamefont {Fradkin}}]{santos_pdw_2019}%
  \BibitemOpen
  \bibfield  {author} {\bibinfo {author} {\bibfnamefont {L.~H.}\ \bibnamefont
  {Santos}}, \bibinfo {author} {\bibfnamefont {Y.}~\bibnamefont {Wang}}, \ and\
  \bibinfo {author} {\bibfnamefont {E.}~\bibnamefont {Fradkin}},\ }\href
  {\doibase 10.1103/PhysRevX.9.021047} {\bibfield  {journal} {\bibinfo
  {journal} {Phys. Rev. X}\ }\textbf {\bibinfo {volume} {9}},\ \bibinfo {pages}
  {021047} (\bibinfo {year} {2019})}\BibitemShut {NoStop}%
\bibitem [{\citenamefont {Yuan}\ and\ \citenamefont
  {Fu}(2018{\natexlab{a}})}]{YuanFu18}%
  \BibitemOpen
  \bibfield  {author} {\bibinfo {author} {\bibfnamefont {N.~F.~Q.}\
  \bibnamefont {Yuan}}\ and\ \bibinfo {author} {\bibfnamefont {L.}~\bibnamefont
  {Fu}},\ }\href {\doibase 10.1103/PhysRevB.97.115139} {\bibfield  {journal}
  {\bibinfo  {journal} {Physical Review B}\ }\textbf {\bibinfo {volume} {97}},\
  \bibinfo {pages} {115139} (\bibinfo {year} {2018}{\natexlab{a}})},\ \bibinfo
  {note} {publisher: American Physical Society}\BibitemShut {NoStop}%
\bibitem [{\citenamefont {Sumita}\ \emph {et~al.}(2019)\citenamefont {Sumita},
  \citenamefont {Nomoto}, \citenamefont {Shiozaki},\ and\ \citenamefont
  {Yanase}}]{SumitaYanase19}%
  \BibitemOpen
  \bibfield  {author} {\bibinfo {author} {\bibfnamefont {S.}~\bibnamefont
  {Sumita}}, \bibinfo {author} {\bibfnamefont {T.}~\bibnamefont {Nomoto}},
  \bibinfo {author} {\bibfnamefont {K.}~\bibnamefont {Shiozaki}}, \ and\
  \bibinfo {author} {\bibfnamefont {Y.}~\bibnamefont {Yanase}},\ }\href
  {\doibase 10.1103/PhysRevB.99.134513} {\bibfield  {journal} {\bibinfo
  {journal} {Physical Review B}\ }\textbf {\bibinfo {volume} {99}},\ \bibinfo
  {pages} {134513} (\bibinfo {year} {2019})},\ \bibinfo {note} {publisher:
  American Physical Society}\BibitemShut {NoStop}%
\bibitem [{\citenamefont {Menke}\ \emph {et~al.}(2019)\citenamefont {Menke},
  \citenamefont {Timm},\ and\ \citenamefont {Brydon}}]{MenkeBrydon19}%
  \BibitemOpen
  \bibfield  {author} {\bibinfo {author} {\bibfnamefont {H.}~\bibnamefont
  {Menke}}, \bibinfo {author} {\bibfnamefont {C.}~\bibnamefont {Timm}}, \ and\
  \bibinfo {author} {\bibfnamefont {P.~M.~R.}\ \bibnamefont {Brydon}},\ }\href
  {\doibase 10.1103/PhysRevB.100.224505} {\bibfield  {journal} {\bibinfo
  {journal} {Physical Review B}\ }\textbf {\bibinfo {volume} {100}},\ \bibinfo
  {pages} {224505} (\bibinfo {year} {2019})},\ \bibinfo {note} {publisher:
  American Physical Society}\BibitemShut {NoStop}%
\bibitem [{\citenamefont {Link}\ and\ \citenamefont {Herbut}(2020)}]{Link20}%
  \BibitemOpen
  \bibfield  {author} {\bibinfo {author} {\bibfnamefont {J.~M.}\ \bibnamefont
  {Link}}\ and\ \bibinfo {author} {\bibfnamefont {I.~F.}\ \bibnamefont
  {Herbut}},\ }\href {\doibase 10.1103/PhysRevLett.125.237004} {\bibfield
  {journal} {\bibinfo  {journal} {Physical Review Letters}\ }\textbf {\bibinfo
  {volume} {125}},\ \bibinfo {pages} {237004} (\bibinfo {year} {2020})},\
  \bibinfo {note} {publisher: American Physical Society}\BibitemShut {NoStop}%
\bibitem [{\citenamefont {Lapp}\ \emph {et~al.}(2020)\citenamefont {Lapp},
  \citenamefont {B\"orner},\ and\ \citenamefont {Timm}}]{Lapp20}%
  \BibitemOpen
  \bibfield  {author} {\bibinfo {author} {\bibfnamefont {C.~J.}\ \bibnamefont
  {Lapp}}, \bibinfo {author} {\bibfnamefont {G.}~\bibnamefont {B\"orner}}, \
  and\ \bibinfo {author} {\bibfnamefont {C.}~\bibnamefont {Timm}},\ }\href
  {\doibase 10.1103/PhysRevB.101.024505} {\bibfield  {journal} {\bibinfo
  {journal} {Physical Review B}\ }\textbf {\bibinfo {volume} {101}},\ \bibinfo
  {pages} {024505} (\bibinfo {year} {2020})},\ \bibinfo {note} {publisher:
  American Physical Society}\BibitemShut {NoStop}%
\bibitem [{\citenamefont {Shaffer}\ \emph {et~al.}(2020)\citenamefont
  {Shaffer}, \citenamefont {Kang}, \citenamefont {Burnell},\ and\ \citenamefont
  {Fernandes}}]{Shaffer20}%
  \BibitemOpen
  \bibfield  {author} {\bibinfo {author} {\bibfnamefont {D.}~\bibnamefont
  {Shaffer}}, \bibinfo {author} {\bibfnamefont {J.}~\bibnamefont {Kang}},
  \bibinfo {author} {\bibfnamefont {F.~J.}\ \bibnamefont {Burnell}}, \ and\
  \bibinfo {author} {\bibfnamefont {R.~M.}\ \bibnamefont {Fernandes}},\ }\href
  {\doibase 10.1103/PhysRevB.101.224503} {\bibfield  {journal} {\bibinfo
  {journal} {Physical Review B}\ }\textbf {\bibinfo {volume} {101}},\ \bibinfo
  {pages} {224503} (\bibinfo {year} {2020})},\ \bibinfo {note} {publisher:
  American Physical Society}\BibitemShut {NoStop}%
\bibitem [{\citenamefont {Zhu}\ \emph {et~al.}(2020)\citenamefont {Zhu},
  \citenamefont {Papaj}, \citenamefont {Nie}, \citenamefont {Xu}, \citenamefont
  {Gu}, \citenamefont {Yang}, \citenamefont {Guan}, \citenamefont {Wang},
  \citenamefont {Li}, \citenamefont {Liu}, \citenamefont {Luo}, \citenamefont
  {Xu}, \citenamefont {Zheng}, \citenamefont {Fu},\ and\ \citenamefont
  {Jia}}]{ZhuFu20}%
  \BibitemOpen
  \bibfield  {author} {\bibinfo {author} {\bibfnamefont {Z.}~\bibnamefont
  {Zhu}}, \bibinfo {author} {\bibfnamefont {M.}~\bibnamefont {Papaj}}, \bibinfo
  {author} {\bibfnamefont {X.-A.}\ \bibnamefont {Nie}}, \bibinfo {author}
  {\bibfnamefont {H.-K.}\ \bibnamefont {Xu}}, \bibinfo {author} {\bibfnamefont
  {Y.-S.}\ \bibnamefont {Gu}}, \bibinfo {author} {\bibfnamefont
  {X.}~\bibnamefont {Yang}}, \bibinfo {author} {\bibfnamefont {D.}~\bibnamefont
  {Guan}}, \bibinfo {author} {\bibfnamefont {S.}~\bibnamefont {Wang}}, \bibinfo
  {author} {\bibfnamefont {Y.}~\bibnamefont {Li}}, \bibinfo {author}
  {\bibfnamefont {C.}~\bibnamefont {Liu}}, \bibinfo {author} {\bibfnamefont
  {J.}~\bibnamefont {Luo}}, \bibinfo {author} {\bibfnamefont {Z.-A.}\
  \bibnamefont {Xu}}, \bibinfo {author} {\bibfnamefont {H.}~\bibnamefont
  {Zheng}}, \bibinfo {author} {\bibfnamefont {L.}~\bibnamefont {Fu}}, \ and\
  \bibinfo {author} {\bibfnamefont {J.-F.}\ \bibnamefont {Jia}},\ }\href
  {http://arxiv.org/abs/2010.02216} {\bibfield  {journal} {\bibinfo  {journal}
  {arXiv:2010.02216 [cond-mat]}\ } (\bibinfo {year} {2020})},\ \bibinfo {note}
  {arXiv: 2010.02216}\BibitemShut {NoStop}%
\bibitem [{\citenamefont {Kobayashi}\ \emph {et~al.}(2014)\citenamefont
  {Kobayashi}, \citenamefont {Shiozaki}, \citenamefont {Tanaka},\ and\
  \citenamefont {Sato}}]{KobayashiSato14}%
  \BibitemOpen
  \bibfield  {author} {\bibinfo {author} {\bibfnamefont {S.}~\bibnamefont
  {Kobayashi}}, \bibinfo {author} {\bibfnamefont {K.}~\bibnamefont {Shiozaki}},
  \bibinfo {author} {\bibfnamefont {Y.}~\bibnamefont {Tanaka}}, \ and\ \bibinfo
  {author} {\bibfnamefont {M.}~\bibnamefont {Sato}},\ }\href {\doibase
  10.1103/PhysRevB.90.024516} {\bibfield  {journal} {\bibinfo  {journal}
  {Physical Review B}\ }\textbf {\bibinfo {volume} {90}},\ \bibinfo {pages}
  {024516} (\bibinfo {year} {2014})},\ \bibinfo {note} {publisher: American
  Physical Society}\BibitemShut {NoStop}%
\bibitem [{\citenamefont {Zhao}\ \emph {et~al.}(2016)\citenamefont {Zhao},
  \citenamefont {Schnyder},\ and\ \citenamefont {Wang}}]{ZhaoSchnyder16}%
  \BibitemOpen
  \bibfield  {author} {\bibinfo {author} {\bibfnamefont {Y.}~\bibnamefont
  {Zhao}}, \bibinfo {author} {\bibfnamefont {A.~P.}\ \bibnamefont {Schnyder}},
  \ and\ \bibinfo {author} {\bibfnamefont {Z.}~\bibnamefont {Wang}},\ }\href
  {\doibase 10.1103/PhysRevLett.116.156402} {\bibfield  {journal} {\bibinfo
  {journal} {Physical Review Letters}\ }\textbf {\bibinfo {volume} {116}},\
  \bibinfo {pages} {156402} (\bibinfo {year} {2016})},\ \bibinfo {note}
  {publisher: American Physical Society}\BibitemShut {NoStop}%
\bibitem [{\citenamefont {Zocher}\ and\ \citenamefont
  {Rosenow}(2016)}]{Zocher16}%
  \BibitemOpen
  \bibfield  {author} {\bibinfo {author} {\bibfnamefont {B.}~\bibnamefont
  {Zocher}}\ and\ \bibinfo {author} {\bibfnamefont {B.}~\bibnamefont
  {Rosenow}},\ }\href {\doibase 10.1103/PhysRevB.93.214504} {\bibfield
  {journal} {\bibinfo  {journal} {Physical Review B}\ }\textbf {\bibinfo
  {volume} {93}},\ \bibinfo {pages} {214504} (\bibinfo {year} {2016})},\
  \bibinfo {note} {publisher: American Physical Society}\BibitemShut {NoStop}%
\bibitem [{\citenamefont {Sahu}\ \emph {et~al.}(2018)\citenamefont {Sahu},
  \citenamefont {Liu}, \citenamefont {Paul}, \citenamefont {Das}, \citenamefont
  {Raychaudhuri}, \citenamefont {Jain},\ and\ \citenamefont
  {Das}}]{SahuJain18}%
  \BibitemOpen
  \bibfield  {author} {\bibinfo {author} {\bibfnamefont {M.~R.}\ \bibnamefont
  {Sahu}}, \bibinfo {author} {\bibfnamefont {X.}~\bibnamefont {Liu}}, \bibinfo
  {author} {\bibfnamefont {A.~K.}\ \bibnamefont {Paul}}, \bibinfo {author}
  {\bibfnamefont {S.}~\bibnamefont {Das}}, \bibinfo {author} {\bibfnamefont
  {P.}~\bibnamefont {Raychaudhuri}}, \bibinfo {author} {\bibfnamefont
  {J.}~\bibnamefont {Jain}}, \ and\ \bibinfo {author} {\bibfnamefont
  {A.}~\bibnamefont {Das}},\ }\href {\doibase 10.1103/PhysRevLett.121.086809}
  {\bibfield  {journal} {\bibinfo  {journal} {Physical Review Letters}\
  }\textbf {\bibinfo {volume} {121}},\ \bibinfo {pages} {086809} (\bibinfo
  {year} {2018})},\ \bibinfo {note} {publisher: American Physical
  Society}\BibitemShut {NoStop}%
\bibitem [{\citenamefont {Schirmer}\ \emph {et~al.}(2020)\citenamefont
  {Schirmer}, \citenamefont {Kumar}, \citenamefont {Bagwe}, \citenamefont
  {Raychaudhuri}, \citenamefont {Taniguchi}, \citenamefont {Watanabe},
  \citenamefont {Liu}, \citenamefont {Das},\ and\ \citenamefont
  {Jain}}]{SchirmerJain20}%
  \BibitemOpen
  \bibfield  {author} {\bibinfo {author} {\bibfnamefont {J.}~\bibnamefont
  {Schirmer}}, \bibinfo {author} {\bibfnamefont {R.}~\bibnamefont {Kumar}},
  \bibinfo {author} {\bibfnamefont {V.}~\bibnamefont {Bagwe}}, \bibinfo
  {author} {\bibfnamefont {P.}~\bibnamefont {Raychaudhuri}}, \bibinfo {author}
  {\bibfnamefont {T.}~\bibnamefont {Taniguchi}}, \bibinfo {author}
  {\bibfnamefont {K.}~\bibnamefont {Watanabe}}, \bibinfo {author}
  {\bibfnamefont {C.-X.}\ \bibnamefont {Liu}}, \bibinfo {author} {\bibfnamefont
  {A.}~\bibnamefont {Das}}, \ and\ \bibinfo {author} {\bibfnamefont {J.~K.}\
  \bibnamefont {Jain}},\ }\href {\doibase 10.1209/0295-5075/132/37002}
  {\bibfield  {journal} {\bibinfo  {journal} {EPL (Europhysics Letters)}\
  }\textbf {\bibinfo {volume} {132}},\ \bibinfo {pages} {37002} (\bibinfo
  {year} {2020})},\ \bibinfo {note} {publisher: IOP Publishing}\BibitemShut
  {NoStop}%
\bibitem [{\citenamefont {Chaudhary}\ and\ \citenamefont
  {MacDonald}(2020)}]{ChaudharyMacDonald20}%
  \BibitemOpen
  \bibfield  {author} {\bibinfo {author} {\bibfnamefont {G.}~\bibnamefont
  {Chaudhary}}\ and\ \bibinfo {author} {\bibfnamefont {A.~H.}\ \bibnamefont
  {MacDonald}},\ }\href {\doibase 10.1103/PhysRevB.101.024516} {\bibfield
  {journal} {\bibinfo  {journal} {Physical Review B}\ }\textbf {\bibinfo
  {volume} {101}},\ \bibinfo {pages} {024516} (\bibinfo {year} {2020})},\
  \bibinfo {note} {publisher: American Physical Society}\BibitemShut {NoStop}%
\bibitem [{\citenamefont {Chaudhary}\ \emph {et~al.}(2021)\citenamefont
  {Chaudhary}, \citenamefont {MacDonald},\ and\ \citenamefont
  {Norman}}]{ChaudharyMacDonald21}%
  \BibitemOpen
  \bibfield  {author} {\bibinfo {author} {\bibfnamefont {G.}~\bibnamefont
  {Chaudhary}}, \bibinfo {author} {\bibfnamefont {A.~H.}\ \bibnamefont
  {MacDonald}}, \ and\ \bibinfo {author} {\bibfnamefont {M.~R.}\ \bibnamefont
  {Norman}},\ }\href {\doibase 10.1103/PhysRevResearch.3.033260} {\bibfield
  {journal} {\bibinfo  {journal} {Phys. Rev. Research}\ }\textbf {\bibinfo
  {volume} {3}},\ \bibinfo {pages} {033260} (\bibinfo {year}
  {2021})}\BibitemShut {NoStop}%
\bibitem [{\citenamefont {Weeks}\ \emph {et~al.}(2007)\citenamefont {Weeks},
  \citenamefont {Rosenberg}, \citenamefont {Seradjeh},\ and\ \citenamefont
  {Franz}}]{WeeksFranz07}%
  \BibitemOpen
  \bibfield  {author} {\bibinfo {author} {\bibfnamefont {C.}~\bibnamefont
  {Weeks}}, \bibinfo {author} {\bibfnamefont {G.}~\bibnamefont {Rosenberg}},
  \bibinfo {author} {\bibfnamefont {B.}~\bibnamefont {Seradjeh}}, \ and\
  \bibinfo {author} {\bibfnamefont {M.}~\bibnamefont {Franz}},\ }\href
  {\doibase 10.1038/nphys730} {\bibfield  {journal} {\bibinfo  {journal}
  {Nature Physics}\ }\textbf {\bibinfo {volume} {3}},\ \bibinfo {pages} {796}
  (\bibinfo {year} {2007})},\ \bibinfo {note} {publisher: Springer
  Nature}\BibitemShut {NoStop}%
\bibitem [{\citenamefont {Jeon}\ \emph {et~al.}(2019)\citenamefont {Jeon},
  \citenamefont {Jain},\ and\ \citenamefont {Liu}}]{JeonJain19}%
  \BibitemOpen
  \bibfield  {author} {\bibinfo {author} {\bibfnamefont {G.~S.}\ \bibnamefont
  {Jeon}}, \bibinfo {author} {\bibfnamefont {J.~K.}\ \bibnamefont {Jain}}, \
  and\ \bibinfo {author} {\bibfnamefont {C.-X.}\ \bibnamefont {Liu}},\ }\href
  {\doibase 10.1103/PhysRevB.99.094509} {\bibfield  {journal} {\bibinfo
  {journal} {Physical Review B}\ }\textbf {\bibinfo {volume} {99}},\ \bibinfo
  {pages} {094509} (\bibinfo {year} {2019})},\ \bibinfo {note} {publisher:
  American Physical Society}\BibitemShut {NoStop}%
\bibitem [{\citenamefont {Wang}\ \emph {et~al.}(2014)\citenamefont {Wang},
  \citenamefont {Hung},\ and\ \citenamefont {Troyer}}]{Wang14}%
  \BibitemOpen
  \bibfield  {author} {\bibinfo {author} {\bibfnamefont {L.}~\bibnamefont
  {Wang}}, \bibinfo {author} {\bibfnamefont {H.-H.}\ \bibnamefont {Hung}}, \
  and\ \bibinfo {author} {\bibfnamefont {M.}~\bibnamefont {Troyer}},\ }\href
  {\doibase 10.1103/PhysRevB.90.205111} {\bibfield  {journal} {\bibinfo
  {journal} {Physical Review B}\ }\textbf {\bibinfo {volume} {90}},\ \bibinfo
  {pages} {205111} (\bibinfo {year} {2014})},\ \bibinfo {note} {publisher:
  American Physical Society}\BibitemShut {NoStop}%
\bibitem [{\citenamefont {Peotta}\ and\ \citenamefont
  {T\"orm\"a}(2015)}]{Peotta15}%
  \BibitemOpen
  \bibfield  {author} {\bibinfo {author} {\bibfnamefont {S.}~\bibnamefont
  {Peotta}}\ and\ \bibinfo {author} {\bibfnamefont {P.}~\bibnamefont
  {T\"orm\"a}},\ }\href {\doibase 10.1038/ncomms9944} {\bibfield  {journal}
  {\bibinfo  {journal} {Nature Communications}\ }\textbf {\bibinfo {volume}
  {6}},\ \bibinfo {pages} {8944} (\bibinfo {year} {2015})},\ \bibinfo {note}
  {number: 1 Publisher: Nature Publishing Group}\BibitemShut {NoStop}%
\bibitem [{\citenamefont {Umucal\ifmmode\imath\else\i\fi{}lar}\ and\
  \citenamefont {Iskin}(2017)}]{UmucalilarIskin17}%
  \BibitemOpen
  \bibfield  {author} {\bibinfo {author} {\bibfnamefont {R.~O.}\ \bibnamefont
  {Umucal\ifmmode\imath\else\i\fi{}lar}}\ and\ \bibinfo {author} {\bibfnamefont
  {M.}~\bibnamefont {Iskin}},\ }\href {\doibase 10.1103/PhysRevLett.119.085301}
  {\bibfield  {journal} {\bibinfo  {journal} {Physical Review Letters}\
  }\textbf {\bibinfo {volume} {119}},\ \bibinfo {pages} {085301} (\bibinfo
  {year} {2017})},\ \bibinfo {note} {publisher: American Physical
  Society}\BibitemShut {NoStop}%
\bibitem [{\citenamefont {Iskin}(2018)}]{Iskin18}%
  \BibitemOpen
  \bibfield  {author} {\bibinfo {author} {\bibfnamefont {M.}~\bibnamefont
  {Iskin}},\ }\href {\doibase 10.1103/PhysRevA.97.013618} {\bibfield  {journal}
  {\bibinfo  {journal} {Physical Review A}\ }\textbf {\bibinfo {volume} {97}},\
  \bibinfo {pages} {013618} (\bibinfo {year} {2018})},\ \bibinfo {note}
  {publisher: American Physical Society}\BibitemShut {NoStop}%
\bibitem [{\citenamefont {Iskin}(2019)}]{Iskin19}%
  \BibitemOpen
  \bibfield  {author} {\bibinfo {author} {\bibfnamefont {M.}~\bibnamefont
  {Iskin}},\ }\href {\doibase 10.1103/PhysRevA.99.023608} {\bibfield  {journal}
  {\bibinfo  {journal} {Physical Review A}\ }\textbf {\bibinfo {volume} {99}},\
  \bibinfo {pages} {023608} (\bibinfo {year} {2019})},\ \bibinfo {note}
  {publisher: American Physical Society}\BibitemShut {NoStop}%
\bibitem [{\citenamefont {Fulde}\ and\ \citenamefont {Ferrell}(1964)}]{FF}%
  \BibitemOpen
  \bibfield  {author} {\bibinfo {author} {\bibfnamefont {P.}~\bibnamefont
  {Fulde}}\ and\ \bibinfo {author} {\bibfnamefont {R.~A.}\ \bibnamefont
  {Ferrell}},\ }\href {\doibase 10.1103/PhysRev.135.A550} {\bibfield  {journal}
  {\bibinfo  {journal} {Phys. Rev.}\ }\textbf {\bibinfo {volume} {135}},\
  \bibinfo {pages} {A550} (\bibinfo {year} {1964})}\BibitemShut {NoStop}%
\bibitem [{\citenamefont {Larkin}\ and\ \citenamefont
  {Ovchinnikov}(1965)}]{LO}%
  \BibitemOpen
  \bibfield  {author} {\bibinfo {author} {\bibfnamefont {A.}~\bibnamefont
  {Larkin}}\ and\ \bibinfo {author} {\bibfnamefont {I.}~\bibnamefont
  {Ovchinnikov}},\ }\href@noop {} {\bibfield  {journal} {\bibinfo  {journal}
  {Soviet Physics-JETP}\ }\textbf {\bibinfo {volume} {20}},\ \bibinfo {pages}
  {762} (\bibinfo {year} {1965})}\BibitemShut {NoStop}%
\bibitem [{\citenamefont {Hu}\ and\ \citenamefont {Liu}(2006)}]{Hu06}%
  \BibitemOpen
  \bibfield  {author} {\bibinfo {author} {\bibfnamefont {H.}~\bibnamefont
  {Hu}}\ and\ \bibinfo {author} {\bibfnamefont {X.-J.}\ \bibnamefont {Liu}},\
  }\href {\doibase 10.1103/PhysRevA.73.051603} {\bibfield  {journal} {\bibinfo
  {journal} {Phys. Rev. A}\ }\textbf {\bibinfo {volume} {73}},\ \bibinfo
  {pages} {051603} (\bibinfo {year} {2006})}\BibitemShut {NoStop}%
\bibitem [{\citenamefont {Agterberg}\ and\ \citenamefont
  {Kaur}(2007)}]{AgterbergKaur07}%
  \BibitemOpen
  \bibfield  {author} {\bibinfo {author} {\bibfnamefont {D.~F.}\ \bibnamefont
  {Agterberg}}\ and\ \bibinfo {author} {\bibfnamefont {R.~P.}\ \bibnamefont
  {Kaur}},\ }\href {\doibase 10.1103/PhysRevB.75.064511} {\bibfield  {journal}
  {\bibinfo  {journal} {Physical Review B}\ }\textbf {\bibinfo {volume} {75}},\
  \bibinfo {pages} {064511} (\bibinfo {year} {2007})},\ \bibinfo {note}
  {publisher: American Physical Society}\BibitemShut {NoStop}%
\bibitem [{\citenamefont {Radzihovsky}\ and\ \citenamefont
  {Vishwanath}(2009)}]{Radzihovsky09}%
  \BibitemOpen
  \bibfield  {author} {\bibinfo {author} {\bibfnamefont {L.}~\bibnamefont
  {Radzihovsky}}\ and\ \bibinfo {author} {\bibfnamefont {A.}~\bibnamefont
  {Vishwanath}},\ }\href {\doibase 10.1103/PhysRevLett.103.010404} {\bibfield
  {journal} {\bibinfo  {journal} {Physical Review Letters}\ }\textbf {\bibinfo
  {volume} {103}},\ \bibinfo {pages} {010404} (\bibinfo {year} {2009})},\
  \bibinfo {note} {publisher: American Physical Society}\BibitemShut {NoStop}%
\bibitem [{\citenamefont {Radzihovsky}(2011)}]{Radzihovsky11}%
  \BibitemOpen
  \bibfield  {author} {\bibinfo {author} {\bibfnamefont {L.}~\bibnamefont
  {Radzihovsky}},\ }\href {\doibase 10.1103/PhysRevA.84.023611} {\bibfield
  {journal} {\bibinfo  {journal} {Physical Review A}\ }\textbf {\bibinfo
  {volume} {84}},\ \bibinfo {pages} {023611} (\bibinfo {year} {2011})},\
  \bibinfo {note} {publisher: American Physical Society}\BibitemShut {NoStop}%
\bibitem [{\citenamefont {Cho}\ \emph {et~al.}(2012)\citenamefont {Cho},
  \citenamefont {Bardarson}, \citenamefont {Lu},\ and\ \citenamefont
  {Moore}}]{ChoMoore12}%
  \BibitemOpen
  \bibfield  {author} {\bibinfo {author} {\bibfnamefont {G.~Y.}\ \bibnamefont
  {Cho}}, \bibinfo {author} {\bibfnamefont {J.~H.}\ \bibnamefont {Bardarson}},
  \bibinfo {author} {\bibfnamefont {Y.-M.}\ \bibnamefont {Lu}}, \ and\ \bibinfo
  {author} {\bibfnamefont {J.~E.}\ \bibnamefont {Moore}},\ }\href {\doibase
  10.1103/PhysRevB.86.214514} {\bibfield  {journal} {\bibinfo  {journal}
  {Physical Review B}\ }\textbf {\bibinfo {volume} {86}},\ \bibinfo {pages}
  {214514} (\bibinfo {year} {2012})},\ \bibinfo {note} {publisher: American
  Physical Society}\BibitemShut {NoStop}%
\bibitem [{\citenamefont {Zheng}\ \emph {et~al.}(2013)\citenamefont {Zheng},
  \citenamefont {Gong}, \citenamefont {Zou}, \citenamefont {Zhang},\ and\
  \citenamefont {Guo}}]{Zheng13}%
  \BibitemOpen
  \bibfield  {author} {\bibinfo {author} {\bibfnamefont {Z.}~\bibnamefont
  {Zheng}}, \bibinfo {author} {\bibfnamefont {M.}~\bibnamefont {Gong}},
  \bibinfo {author} {\bibfnamefont {X.}~\bibnamefont {Zou}}, \bibinfo {author}
  {\bibfnamefont {C.}~\bibnamefont {Zhang}}, \ and\ \bibinfo {author}
  {\bibfnamefont {G.}~\bibnamefont {Guo}},\ }\href {\doibase
  10.1103/PhysRevA.87.031602} {\bibfield  {journal} {\bibinfo  {journal}
  {Physical Review A}\ }\textbf {\bibinfo {volume} {87}},\ \bibinfo {pages}
  {031602} (\bibinfo {year} {2013})},\ \bibinfo {note} {publisher: American
  Physical Society}\BibitemShut {NoStop}%
\bibitem [{\citenamefont {Agterberg}\ \emph {et~al.}(2020)\citenamefont
  {Agterberg}, \citenamefont {Davis}, \citenamefont {Edkins}, \citenamefont
  {Fradkin}, \citenamefont {Van~Harlingen}, \citenamefont {Kivelson},
  \citenamefont {Lee}, \citenamefont {Radzihovsky}, \citenamefont {Tranquada},\
  and\ \citenamefont {Wang}}]{Agterberg20}%
  \BibitemOpen
  \bibfield  {author} {\bibinfo {author} {\bibfnamefont {D.~F.}\ \bibnamefont
  {Agterberg}}, \bibinfo {author} {\bibfnamefont {J.~S.}\ \bibnamefont
  {Davis}}, \bibinfo {author} {\bibfnamefont {S.~D.}\ \bibnamefont {Edkins}},
  \bibinfo {author} {\bibfnamefont {E.}~\bibnamefont {Fradkin}}, \bibinfo
  {author} {\bibfnamefont {D.~J.}\ \bibnamefont {Van~Harlingen}}, \bibinfo
  {author} {\bibfnamefont {S.~A.}\ \bibnamefont {Kivelson}}, \bibinfo {author}
  {\bibfnamefont {P.~A.}\ \bibnamefont {Lee}}, \bibinfo {author} {\bibfnamefont
  {L.}~\bibnamefont {Radzihovsky}}, \bibinfo {author} {\bibfnamefont {J.~M.}\
  \bibnamefont {Tranquada}}, \ and\ \bibinfo {author} {\bibfnamefont
  {Y.}~\bibnamefont {Wang}},\ }\href {\doibase
  10.1146/annurev-conmatphys-031119-050711} {\bibfield  {journal} {\bibinfo
  {journal} {Annual Review of Condensed Matter Physics}\ }\textbf {\bibinfo
  {volume} {11}},\ \bibinfo {pages} {231} (\bibinfo {year} {2020})},\ \bibinfo
  {note} {publisher: Annual Reviews}\BibitemShut {NoStop}%
\bibitem [{Note1()}]{Note1}%
  \BibitemOpen
  \bibinfo {note} {A similar result was found in \cite {PowellDasSarma11} for
  charge 1e condensates in synthetic magnetic fields}\BibitemShut {NoStop}%
\bibitem [{\citenamefont {Hasegawa}\ \emph {et~al.}(1989)\citenamefont
  {Hasegawa}, \citenamefont {Lederer}, \citenamefont {Rice},\ and\
  \citenamefont {Wiegmann}}]{HasegawaWiegmann89}%
  \BibitemOpen
  \bibfield  {author} {\bibinfo {author} {\bibfnamefont {Y.}~\bibnamefont
  {Hasegawa}}, \bibinfo {author} {\bibfnamefont {P.}~\bibnamefont {Lederer}},
  \bibinfo {author} {\bibfnamefont {T.~M.}\ \bibnamefont {Rice}}, \ and\
  \bibinfo {author} {\bibfnamefont {P.~B.}\ \bibnamefont {Wiegmann}},\ }\href
  {\doibase 10.1103/PhysRevLett.63.907} {\bibfield  {journal} {\bibinfo
  {journal} {Phys. Rev. Lett.}\ }\textbf {\bibinfo {volume} {63}},\ \bibinfo
  {pages} {907} (\bibinfo {year} {1989})}\BibitemShut {NoStop}%
\bibitem [{Note2()}]{Note2}%
  \BibitemOpen
  \bibinfo {note} {The irreducible representations of the MTG can also be
  considered as projective irreducible representations of the regular
  translation group.}\BibitemShut {Stop}%
\bibitem [{Note3()}]{Note3}%
  \BibitemOpen
  \bibinfo {note} {There are additional constraints from hermiticity:
  \(g^{(\ell )}_{n,m}(\protect \mathbf {p;k})=g^{(\ell )*}_{m,n}(\protect
  \mathbf {k;p})\). Moreover, anti-commutation relations imply that we can
  further take \(g^{(\ell )}_{n,m}(\protect \mathbf {p;k})=-g^{(\ell )}_{-\ell
  -n,m}(\protect \mathbf {-p;k})=-g^{(\ell )}_{n,-\ell -m}(\protect \mathbf
  {p;-k})=g^{(\ell )}_{-\ell -n,-\ell -m}(\protect \mathbf
  {-p;-k})\).}\BibitemShut {Stop}%
\bibitem [{\citenamefont {Sato}\ and\ \citenamefont {Ando}(2017)}]{SatoAndo17}%
  \BibitemOpen
  \bibfield  {author} {\bibinfo {author} {\bibfnamefont {M.}~\bibnamefont
  {Sato}}\ and\ \bibinfo {author} {\bibfnamefont {Y.}~\bibnamefont {Ando}},\
  }\href {\doibase 10.1088/1361-6633/aa6ac7} {\bibfield  {journal} {\bibinfo
  {journal} {Reports on Progress in Physics}\ }\textbf {\bibinfo {volume}
  {80}},\ \bibinfo {pages} {076501} (\bibinfo {year} {2017})},\ \bibinfo {note}
  {publisher: IOP Publishing}\BibitemShut {NoStop}%
\bibitem [{\citenamefont {Ono}\ \emph {et~al.}(2019)\citenamefont {Ono},
  \citenamefont {Yanase},\ and\ \citenamefont {Watanabe}}]{Ono19}%
  \BibitemOpen
  \bibfield  {author} {\bibinfo {author} {\bibfnamefont {S.}~\bibnamefont
  {Ono}}, \bibinfo {author} {\bibfnamefont {Y.}~\bibnamefont {Yanase}}, \ and\
  \bibinfo {author} {\bibfnamefont {H.}~\bibnamefont {Watanabe}},\ }\href
  {\doibase 10.1103/PhysRevResearch.1.013012} {\bibfield  {journal} {\bibinfo
  {journal} {Phys. Rev. Research}\ }\textbf {\bibinfo {volume} {1}},\ \bibinfo
  {pages} {013012} (\bibinfo {year} {2019})}\BibitemShut {NoStop}%
\bibitem [{Note4()}]{Note4}%
  \BibitemOpen
  \bibinfo {note} {For example, for $q=3$, the $L = 0, 1, 2$ values, under the
  action of $T_{1}$, are cycled as $0 \rightarrow 2 \rightarrow 1 \rightarrow 0
  \rightarrow ...$; for $q=5$, $0 \rightarrow 2 \rightarrow 4 \rightarrow 1
  \rightarrow 3 \rightarrow 0 \rightarrow ..., etc.$}\BibitemShut {NoStop}%
\bibitem [{\citenamefont {Blatter}\ \emph {et~al.}(1994)\citenamefont
  {Blatter}, \citenamefont {Feigel'man}, \citenamefont {Geshkenbein},
  \citenamefont {Larkin},\ and\ \citenamefont {Vinokur}}]{BlatterLarkin94}%
  \BibitemOpen
  \bibfield  {author} {\bibinfo {author} {\bibfnamefont {G.}~\bibnamefont
  {Blatter}}, \bibinfo {author} {\bibfnamefont {M.~V.}\ \bibnamefont
  {Feigel'man}}, \bibinfo {author} {\bibfnamefont {V.~B.}\ \bibnamefont
  {Geshkenbein}}, \bibinfo {author} {\bibfnamefont {A.~I.}\ \bibnamefont
  {Larkin}}, \ and\ \bibinfo {author} {\bibfnamefont {V.~M.}\ \bibnamefont
  {Vinokur}},\ }\href {\doibase 10.1103/RevModPhys.66.1125} {\bibfield
  {journal} {\bibinfo  {journal} {Reviews of Modern Physics}\ }\textbf
  {\bibinfo {volume} {66}},\ \bibinfo {pages} {1125} (\bibinfo {year}
  {1994})},\ \bibinfo {note} {publisher: American Physical Society}\BibitemShut
  {NoStop}%
\bibitem [{Note5()}]{Note5}%
  \BibitemOpen
  \bibinfo {note} {Eq. (\ref {etaLT1}) needs to be modified to \(\eta
  _L\protect \xrightarrow []{\protect \mathaccentV {hat}05E{T}_1(2\pi p/q)}\eta
  _{L+2}\) for \(\protect \mathaccentV {hat}05E{\Delta }^{(L,-)}\) irrep
  components for even \(q\) since \(\protect \mathaccentV
  {hat}05E{T}^{q/2}_1(0)\) acts as \(-1\) on that irrep.}\BibitemShut {Stop}%
\bibitem [{\citenamefont {Jones}\ and\ \citenamefont {Willms}(2018)}]{Jones18}%
  \BibitemOpen
  \bibfield  {author} {\bibinfo {author} {\bibfnamefont {T.~H.}\ \bibnamefont
  {Jones}}\ and\ \bibinfo {author} {\bibfnamefont {N.~B.}\ \bibnamefont
  {Willms}},\ }\href {\doibase 10.1088/1742-6596/1047/1/012016} {\bibfield
  {journal} {\bibinfo  {journal} {Journal of Physics: Conference Series}\
  }\textbf {\bibinfo {volume} {1047}},\ \bibinfo {pages} {012016} (\bibinfo
  {year} {2018})},\ \bibinfo {note} {publisher: IOP Publishing}\BibitemShut
  {NoStop}%
\bibitem [{\citenamefont {Fisher}(1989)}]{Fisher89}%
  \BibitemOpen
  \bibfield  {author} {\bibinfo {author} {\bibfnamefont {M.~P.~A.}\
  \bibnamefont {Fisher}},\ }\href {\doibase 10.1103/PhysRevLett.62.1415}
  {\bibfield  {journal} {\bibinfo  {journal} {Physical Review Letters}\
  }\textbf {\bibinfo {volume} {62}},\ \bibinfo {pages} {1415} (\bibinfo {year}
  {1989})},\ \bibinfo {note} {publisher: American Physical Society}\BibitemShut
  {NoStop}%
\bibitem [{\citenamefont {Huse}\ and\ \citenamefont {Seung}(1990)}]{Huse90}%
  \BibitemOpen
  \bibfield  {author} {\bibinfo {author} {\bibfnamefont {D.~A.}\ \bibnamefont
  {Huse}}\ and\ \bibinfo {author} {\bibfnamefont {H.~S.}\ \bibnamefont
  {Seung}},\ }\href {\doibase 10.1103/PhysRevB.42.1059} {\bibfield  {journal}
  {\bibinfo  {journal} {Physical Review B}\ }\textbf {\bibinfo {volume} {42}},\
  \bibinfo {pages} {1059} (\bibinfo {year} {1990})},\ \bibinfo {note}
  {publisher: American Physical Society}\BibitemShut {NoStop}%
\bibitem [{\citenamefont {Jacobsen}\ \emph {et~al.}(1999)\citenamefont
  {Jacobsen}, \citenamefont {Saunders}, \citenamefont {Radzihovsky},\ and\
  \citenamefont {Toner}}]{JacobsenRadzihovsky99}%
  \BibitemOpen
  \bibfield  {author} {\bibinfo {author} {\bibfnamefont {B.}~\bibnamefont
  {Jacobsen}}, \bibinfo {author} {\bibfnamefont {K.}~\bibnamefont {Saunders}},
  \bibinfo {author} {\bibfnamefont {L.}~\bibnamefont {Radzihovsky}}, \ and\
  \bibinfo {author} {\bibfnamefont {J.}~\bibnamefont {Toner}},\ }\href
  {\doibase 10.1103/PhysRevLett.83.1363} {\bibfield  {journal} {\bibinfo
  {journal} {Physical Review Letters}\ }\textbf {\bibinfo {volume} {83}},\
  \bibinfo {pages} {1363} (\bibinfo {year} {1999})},\ \bibinfo {note}
  {publisher: American Physical Society}\BibitemShut {NoStop}%
\bibitem [{\citenamefont {Radzihovsky}\ and\ \citenamefont
  {Toner}(1999)}]{Radzihovsky99}%
  \BibitemOpen
  \bibfield  {author} {\bibinfo {author} {\bibfnamefont {L.}~\bibnamefont
  {Radzihovsky}}\ and\ \bibinfo {author} {\bibfnamefont {J.}~\bibnamefont
  {Toner}},\ }\href {\doibase 10.1103/PhysRevB.60.206} {\bibfield  {journal}
  {\bibinfo  {journal} {Physical Review B}\ }\textbf {\bibinfo {volume} {60}},\
  \bibinfo {pages} {206} (\bibinfo {year} {1999})},\ \bibinfo {note}
  {publisher: American Physical Society}\BibitemShut {NoStop}%
\bibitem [{\citenamefont {Saunders}\ \emph {et~al.}(2000)\citenamefont
  {Saunders}, \citenamefont {Jacobsen}, \citenamefont {Radzihovsky},\ and\
  \citenamefont {Toner}}]{SaundersRadzihovsky00}%
  \BibitemOpen
  \bibfield  {author} {\bibinfo {author} {\bibfnamefont {K.}~\bibnamefont
  {Saunders}}, \bibinfo {author} {\bibfnamefont {B.}~\bibnamefont {Jacobsen}},
  \bibinfo {author} {\bibfnamefont {L.}~\bibnamefont {Radzihovsky}}, \ and\
  \bibinfo {author} {\bibfnamefont {J.}~\bibnamefont {Toner}},\ }\href
  {\doibase 10.1088/0953-8984/12/8A/326} {\bibfield  {journal} {\bibinfo
  {journal} {Journal of Physics: Condensed Matter}\ }\textbf {\bibinfo {volume}
  {12}},\ \bibinfo {pages} {A215} (\bibinfo {year} {2000})},\ \bibinfo {note}
  {publisher: IOP Publishing}\BibitemShut {NoStop}%
\bibitem [{\citenamefont {Radzihovsky}(2021)}]{Radzihovsky21}%
  \BibitemOpen
  \bibfield  {author} {\bibinfo {author} {\bibfnamefont {L.}~\bibnamefont
  {Radzihovsky}},\ }\href {\doibase 10.1103/PhysRevB.104.024510} {\bibfield
  {journal} {\bibinfo  {journal} {Phys. Rev. B}\ }\textbf {\bibinfo {volume}
  {104}},\ \bibinfo {pages} {024510} (\bibinfo {year} {2021})}\BibitemShut
  {NoStop}%
\bibitem [{\citenamefont {Avila}\ and\ \citenamefont
  {Jitomirskaya}(2009)}]{AvilaJitomirskaya09}%
  \BibitemOpen
  \bibfield  {author} {\bibinfo {author} {\bibfnamefont {A.}~\bibnamefont
  {Avila}}\ and\ \bibinfo {author} {\bibfnamefont {S.}~\bibnamefont
  {Jitomirskaya}},\ }\href {\doibase 10.4007/annals.2009.170.303} {\bibfield
  {journal} {\bibinfo  {journal} {Annals of Mathematics. Second Series}\
  }\textbf {\bibinfo {volume} {170}},\ \bibinfo {pages} {303} (\bibinfo {year}
  {2009})},\ \bibinfo {note} {publisher: Princeton University, Mathematics
  Department, Princeton, NJ MSC2010: 47B37 = Linear operators on special spaces
  (weighted shifts, operators on sequence spaces, etc.) MSC2010: 47A10 =
  Spectrum, resolvent}\BibitemShut {NoStop}%
\bibitem [{\citenamefont {Harper}\ \emph {et~al.}(2014)\citenamefont {Harper},
  \citenamefont {Simon},\ and\ \citenamefont {Roy}}]{HarperSimon14}%
  \BibitemOpen
  \bibfield  {author} {\bibinfo {author} {\bibfnamefont {F.}~\bibnamefont
  {Harper}}, \bibinfo {author} {\bibfnamefont {S.~H.}\ \bibnamefont {Simon}}, \
  and\ \bibinfo {author} {\bibfnamefont {R.}~\bibnamefont {Roy}},\ }\href
  {\doibase 10.1103/PhysRevB.90.075104} {\bibfield  {journal} {\bibinfo
  {journal} {Physical Review B}\ }\textbf {\bibinfo {volume} {90}},\ \bibinfo
  {pages} {075104} (\bibinfo {year} {2014})},\ \bibinfo {note} {publisher:
  American Physical Society}\BibitemShut {NoStop}%
\bibitem [{\citenamefont {Chen}\ \emph {et~al.}(2013)\citenamefont {Chen},
  \citenamefont {Gu}, \citenamefont {Liu},\ and\ \citenamefont
  {Wen}}]{ChenWen13}%
  \BibitemOpen
  \bibfield  {author} {\bibinfo {author} {\bibfnamefont {X.}~\bibnamefont
  {Chen}}, \bibinfo {author} {\bibfnamefont {Z.-C.}\ \bibnamefont {Gu}},
  \bibinfo {author} {\bibfnamefont {Z.-X.}\ \bibnamefont {Liu}}, \ and\
  \bibinfo {author} {\bibfnamefont {X.-G.}\ \bibnamefont {Wen}},\ }\href
  {\doibase 10.1103/PhysRevB.87.155114} {\bibfield  {journal} {\bibinfo
  {journal} {Physical Review B}\ }\textbf {\bibinfo {volume} {87}},\ \bibinfo
  {pages} {155114} (\bibinfo {year} {2013})},\ \bibinfo {note} {publisher:
  American Physical Society}\BibitemShut {NoStop}%
\bibitem [{\citenamefont {Senthil}(2015)}]{senthil2015symmetry}%
  \BibitemOpen
  \bibfield  {author} {\bibinfo {author} {\bibfnamefont {T.}~\bibnamefont
  {Senthil}},\ }\href {\doibase 10.1146/annurev-conmatphys-031214-014740}
  {\bibfield  {journal} {\bibinfo  {journal} {Annual Review of Condensed Matter
  Physics}\ }\textbf {\bibinfo {volume} {6}},\ \bibinfo {pages} {299} (\bibinfo
  {year} {2015})},\ \Eprint
  {http://arxiv.org/abs/https://doi.org/10.1146/annurev-conmatphys-031214-014740}
  {https://doi.org/10.1146/annurev-conmatphys-031214-014740} \BibitemShut
  {NoStop}%
\bibitem [{\citenamefont {Wen}(2017)}]{Wen17}%
  \BibitemOpen
  \bibfield  {author} {\bibinfo {author} {\bibfnamefont {X.-G.}\ \bibnamefont
  {Wen}},\ }\href {\doibase 10.1103/RevModPhys.89.041004} {\bibfield  {journal}
  {\bibinfo  {journal} {Reviews of Modern Physics}\ }\textbf {\bibinfo {volume}
  {89}},\ \bibinfo {pages} {041004} (\bibinfo {year} {2017})},\ \bibinfo {note}
  {publisher: American Physical Society}\BibitemShut {NoStop}%
\bibitem [{\citenamefont {Chiu}\ \emph {et~al.}(2016)\citenamefont {Chiu},
  \citenamefont {Teo}, \citenamefont {Schnyder},\ and\ \citenamefont
  {Ryu}}]{ChiuRyu16}%
  \BibitemOpen
  \bibfield  {author} {\bibinfo {author} {\bibfnamefont {C.-K.}\ \bibnamefont
  {Chiu}}, \bibinfo {author} {\bibfnamefont {J.~C.}\ \bibnamefont {Teo}},
  \bibinfo {author} {\bibfnamefont {A.~P.}\ \bibnamefont {Schnyder}}, \ and\
  \bibinfo {author} {\bibfnamefont {S.}~\bibnamefont {Ryu}},\ }\href {\doibase
  10.1103/RevModPhys.88.035005} {\bibfield  {journal} {\bibinfo  {journal}
  {Reviews of Modern Physics}\ }\textbf {\bibinfo {volume} {88}},\ \bibinfo
  {pages} {035005} (\bibinfo {year} {2016})},\ \bibinfo {note} {publisher:
  American Physical Society}\BibitemShut {NoStop}%
\bibitem [{\citenamefont {Altland}\ and\ \citenamefont
  {Zirnbauer}(1997)}]{AltlandZirnbauer97}%
  \BibitemOpen
  \bibfield  {author} {\bibinfo {author} {\bibfnamefont {A.}~\bibnamefont
  {Altland}}\ and\ \bibinfo {author} {\bibfnamefont {M.~R.}\ \bibnamefont
  {Zirnbauer}},\ }\href {\doibase 10.1103/PhysRevB.55.1142} {\bibfield
  {journal} {\bibinfo  {journal} {Physical Review B}\ }\textbf {\bibinfo
  {volume} {55}},\ \bibinfo {pages} {1142} (\bibinfo {year} {1997})},\ \bibinfo
  {note} {publisher: American Physical Society}\BibitemShut {NoStop}%
\bibitem [{\citenamefont {Ryu}\ \emph {et~al.}(2010)\citenamefont {Ryu},
  \citenamefont {Schnyder}, \citenamefont {Furusaki},\ and\ \citenamefont
  {Ludwig}}]{Ryu10}%
  \BibitemOpen
  \bibfield  {author} {\bibinfo {author} {\bibfnamefont {S.}~\bibnamefont
  {Ryu}}, \bibinfo {author} {\bibfnamefont {A.~P.}\ \bibnamefont {Schnyder}},
  \bibinfo {author} {\bibfnamefont {A.}~\bibnamefont {Furusaki}}, \ and\
  \bibinfo {author} {\bibfnamefont {A.~W.~W.}\ \bibnamefont {Ludwig}},\ }\href
  {\doibase 10.1088/1367-2630/12/6/065010} {\bibfield  {journal} {\bibinfo
  {journal} {New Journal of Physics}\ }\textbf {\bibinfo {volume} {12}},\
  \bibinfo {pages} {065010} (\bibinfo {year} {2010})},\ \bibinfo {note}
  {publisher: IOP Publishing}\BibitemShut {NoStop}%
\bibitem [{\citenamefont {Ludwig}(2015)}]{Ludwig15}%
  \BibitemOpen
  \bibfield  {author} {\bibinfo {author} {\bibfnamefont {A.~W.~W.}\
  \bibnamefont {Ludwig}},\ }\href {\doibase 10.1088/0031-8949/2015/T168/014001}
  {\bibfield  {journal} {\bibinfo  {journal} {Physica Scripta}\ }\textbf
  {\bibinfo {volume} {T168}},\ \bibinfo {pages} {014001} (\bibinfo {year}
  {2015})},\ \bibinfo {note} {publisher: IOP Publishing}\BibitemShut {NoStop}%
\bibitem [{\citenamefont {Qi}\ \emph {et~al.}(2010)\citenamefont {Qi},
  \citenamefont {Hughes},\ and\ \citenamefont {Zhang}}]{QiHughes10}%
  \BibitemOpen
  \bibfield  {author} {\bibinfo {author} {\bibfnamefont {X.-L.}\ \bibnamefont
  {Qi}}, \bibinfo {author} {\bibfnamefont {T.~L.}\ \bibnamefont {Hughes}}, \
  and\ \bibinfo {author} {\bibfnamefont {S.-C.}\ \bibnamefont {Zhang}},\ }\href
  {\doibase 10.1103/PhysRevB.81.134508} {\bibfield  {journal} {\bibinfo
  {journal} {Physical Review B}\ }\textbf {\bibinfo {volume} {81}},\ \bibinfo
  {pages} {134508} (\bibinfo {year} {2010})},\ \bibinfo {note} {publisher:
  American Physical Society}\BibitemShut {NoStop}%
\bibitem [{\citenamefont {Hughes}\ \emph {et~al.}(2011)\citenamefont {Hughes},
  \citenamefont {Prodan},\ and\ \citenamefont {Bernevig}}]{Hughes11}%
  \BibitemOpen
  \bibfield  {author} {\bibinfo {author} {\bibfnamefont {T.~L.}\ \bibnamefont
  {Hughes}}, \bibinfo {author} {\bibfnamefont {E.}~\bibnamefont {Prodan}}, \
  and\ \bibinfo {author} {\bibfnamefont {B.~A.}\ \bibnamefont {Bernevig}},\
  }\href {\doibase 10.1103/PhysRevB.83.245132} {\bibfield  {journal} {\bibinfo
  {journal} {Physical Review B}\ }\textbf {\bibinfo {volume} {83}},\ \bibinfo
  {pages} {245132} (\bibinfo {year} {2011})},\ \bibinfo {note} {publisher:
  American Physical Society}\BibitemShut {NoStop}%
\bibitem [{\citenamefont {Chiu}\ \emph {et~al.}(2013)\citenamefont {Chiu},
  \citenamefont {Yao},\ and\ \citenamefont {Ryu}}]{ChiuRyu13}%
  \BibitemOpen
  \bibfield  {author} {\bibinfo {author} {\bibfnamefont {C.-K.}\ \bibnamefont
  {Chiu}}, \bibinfo {author} {\bibfnamefont {H.}~\bibnamefont {Yao}}, \ and\
  \bibinfo {author} {\bibfnamefont {S.}~\bibnamefont {Ryu}},\ }\href {\doibase
  10.1103/PhysRevB.88.075142} {\bibfield  {journal} {\bibinfo  {journal}
  {Physical Review B}\ }\textbf {\bibinfo {volume} {88}},\ \bibinfo {pages}
  {075142} (\bibinfo {year} {2013})},\ \bibinfo {note} {publisher: American
  Physical Society}\BibitemShut {NoStop}%
\bibitem [{\citenamefont {Chiu}\ and\ \citenamefont
  {Schnyder}(2014)}]{ChiuSchnyder14}%
  \BibitemOpen
  \bibfield  {author} {\bibinfo {author} {\bibfnamefont {C.-K.}\ \bibnamefont
  {Chiu}}\ and\ \bibinfo {author} {\bibfnamefont {A.~P.}\ \bibnamefont
  {Schnyder}},\ }\href {\doibase 10.1103/PhysRevB.90.205136} {\bibfield
  {journal} {\bibinfo  {journal} {Physical Review B}\ }\textbf {\bibinfo
  {volume} {90}},\ \bibinfo {pages} {205136} (\bibinfo {year} {2014})},\
  \bibinfo {note} {publisher: American Physical Society}\BibitemShut {NoStop}%
\bibitem [{\citenamefont {Shiozaki}\ and\ \citenamefont
  {Sato}(2014)}]{ShiozakiSato14}%
  \BibitemOpen
  \bibfield  {author} {\bibinfo {author} {\bibfnamefont {K.}~\bibnamefont
  {Shiozaki}}\ and\ \bibinfo {author} {\bibfnamefont {M.}~\bibnamefont
  {Sato}},\ }\href {\doibase 10.1103/PhysRevB.90.165114} {\bibfield  {journal}
  {\bibinfo  {journal} {Physical Review B}\ }\textbf {\bibinfo {volume} {90}},\
  \bibinfo {pages} {165114} (\bibinfo {year} {2014})},\ \bibinfo {note}
  {publisher: American Physical Society}\BibitemShut {NoStop}%
\bibitem [{\citenamefont {Cornfeld}\ and\ \citenamefont
  {Chapman}(2019)}]{Cornfeld19}%
  \BibitemOpen
  \bibfield  {author} {\bibinfo {author} {\bibfnamefont {E.}~\bibnamefont
  {Cornfeld}}\ and\ \bibinfo {author} {\bibfnamefont {A.}~\bibnamefont
  {Chapman}},\ }\href {\doibase 10.1103/PhysRevB.99.075105} {\bibfield
  {journal} {\bibinfo  {journal} {Physical Review B}\ }\textbf {\bibinfo
  {volume} {99}},\ \bibinfo {pages} {075105} (\bibinfo {year} {2019})},\
  \bibinfo {note} {publisher: American Physical Society}\BibitemShut {NoStop}%
\bibitem [{\citenamefont {Cheng}\ \emph {et~al.}(2016)\citenamefont {Cheng},
  \citenamefont {Zaletel}, \citenamefont {Barkeshli}, \citenamefont
  {Vishwanath},\ and\ \citenamefont {Bonderson}}]{ChengZaletel16}%
  \BibitemOpen
  \bibfield  {author} {\bibinfo {author} {\bibfnamefont {M.}~\bibnamefont
  {Cheng}}, \bibinfo {author} {\bibfnamefont {M.}~\bibnamefont {Zaletel}},
  \bibinfo {author} {\bibfnamefont {M.}~\bibnamefont {Barkeshli}}, \bibinfo
  {author} {\bibfnamefont {A.}~\bibnamefont {Vishwanath}}, \ and\ \bibinfo
  {author} {\bibfnamefont {P.}~\bibnamefont {Bonderson}},\ }\href {\doibase
  10.1103/PhysRevX.6.041068} {\bibfield  {journal} {\bibinfo  {journal}
  {Physical Review X}\ }\textbf {\bibinfo {volume} {6}},\ \bibinfo {pages}
  {041068} (\bibinfo {year} {2016})},\ \bibinfo {note} {publisher: American
  Physical Society}\BibitemShut {NoStop}%
\bibitem [{\citenamefont {Lu}\ \emph {et~al.}(2020)\citenamefont {Lu},
  \citenamefont {Ran},\ and\ \citenamefont {Oshikawa}}]{Lu19}%
  \BibitemOpen
  \bibfield  {author} {\bibinfo {author} {\bibfnamefont {Y.-M.}\ \bibnamefont
  {Lu}}, \bibinfo {author} {\bibfnamefont {Y.}~\bibnamefont {Ran}}, \ and\
  \bibinfo {author} {\bibfnamefont {M.}~\bibnamefont {Oshikawa}},\ }\href
  {\doibase https://doi.org/10.1016/j.aop.2019.168060} {\bibfield  {journal}
  {\bibinfo  {journal} {Annals of Physics}\ }\textbf {\bibinfo {volume}
  {413}},\ \bibinfo {pages} {168060} (\bibinfo {year} {2020})}\BibitemShut
  {NoStop}%
\bibitem [{\citenamefont {Matsuura}\ \emph {et~al.}(2013)\citenamefont
  {Matsuura}, \citenamefont {Chang}, \citenamefont {Schnyder},\ and\
  \citenamefont {Ryu}}]{MatsuuraRyu13}%
  \BibitemOpen
  \bibfield  {author} {\bibinfo {author} {\bibfnamefont {S.}~\bibnamefont
  {Matsuura}}, \bibinfo {author} {\bibfnamefont {P.-Y.}\ \bibnamefont {Chang}},
  \bibinfo {author} {\bibfnamefont {A.~P.}\ \bibnamefont {Schnyder}}, \ and\
  \bibinfo {author} {\bibfnamefont {S.}~\bibnamefont {Ryu}},\ }\href {\doibase
  10.1088/1367-2630/15/6/065001} {\bibfield  {journal} {\bibinfo  {journal}
  {New Journal of Physics}\ }\textbf {\bibinfo {volume} {15}},\ \bibinfo
  {pages} {065001} (\bibinfo {year} {2013})},\ \bibinfo {note} {publisher: IOP
  Publishing}\BibitemShut {NoStop}%
\bibitem [{\citenamefont {Zhao}\ and\ \citenamefont {Wang}(2013)}]{ZhaoWang13}%
  \BibitemOpen
  \bibfield  {author} {\bibinfo {author} {\bibfnamefont {Y.~X.}\ \bibnamefont
  {Zhao}}\ and\ \bibinfo {author} {\bibfnamefont {Z.~D.}\ \bibnamefont
  {Wang}},\ }\href {\doibase 10.1103/PhysRevLett.110.240404} {\bibfield
  {journal} {\bibinfo  {journal} {Physical Review Letters}\ }\textbf {\bibinfo
  {volume} {110}},\ \bibinfo {pages} {240404} (\bibinfo {year} {2013})},\
  \bibinfo {note} {publisher: American Physical Society}\BibitemShut {NoStop}%
\bibitem [{\citenamefont {Simon}(1983)}]{Simon83}%
  \BibitemOpen
  \bibfield  {author} {\bibinfo {author} {\bibfnamefont {B.}~\bibnamefont
  {Simon}},\ }\href {\doibase 10.1103/PhysRevLett.51.2167} {\bibfield
  {journal} {\bibinfo  {journal} {Physical Review Letters}\ }\textbf {\bibinfo
  {volume} {51}},\ \bibinfo {pages} {2167} (\bibinfo {year} {1983})},\ \bibinfo
  {note} {publisher: American Physical Society}\BibitemShut {NoStop}%
\bibitem [{Note6()}]{Note6}%
  \BibitemOpen
  \bibinfo {note} {Note that if the Fermi surface does contain high-symmetry
  points with \(\protect \mathbf {p}_D=-\protect \mathbf {p}_D\), then
  \(\protect \mathaccentV {hat}05E{\Delta }(\protect \mathbf {p}_D)\) is an
  anti-symmetric matrix by PHS, which as we discuss in Sec. \ref {BFS} implies
  that \(\Delta '_n(\protect \mathbf {p}_D)\) that appear in the eigenvalues of
  the BdG Hamiltonian in Eq. (\ref {eq: eigenenrgies}) satisfy \(\Delta
  '_{-n}(\protect \mathbf {p}_D)=-\Delta '_n(\protect \mathbf {p}_D)\). As a
  result, there are band touchings at \(\protect \mathbf {p}_D\) and the Berry
  connection, along with the Chern number, is not well-defined unless we assume
  that the Fermi surface does not contain such high-symmetry
  points.}\BibitemShut {Stop}%
\bibitem [{\citenamefont {Fukui}\ \emph {et~al.}(2005)\citenamefont {Fukui},
  \citenamefont {Hatsugai},\ and\ \citenamefont {Suzuki}}]{Fukui2005}%
  \BibitemOpen
  \bibfield  {author} {\bibinfo {author} {\bibfnamefont {T.}~\bibnamefont
  {Fukui}}, \bibinfo {author} {\bibfnamefont {Y.}~\bibnamefont {Hatsugai}}, \
  and\ \bibinfo {author} {\bibfnamefont {H.}~\bibnamefont {Suzuki}},\ }\href
  {\doibase 10.1143/JPSJ.74.1674} {\bibfield  {journal} {\bibinfo  {journal}
  {Journal of the Physical Society of Japan}\ }\textbf {\bibinfo {volume}
  {74}},\ \bibinfo {pages} {1674} (\bibinfo {year} {2005})},\ \Eprint
  {http://arxiv.org/abs/https://doi.org/10.1143/JPSJ.74.1674}
  {https://doi.org/10.1143/JPSJ.74.1674} \BibitemShut {NoStop}%
\bibitem [{\citenamefont {Teo}\ and\ \citenamefont
  {Hughes}(2013)}]{Teo_Existence_2013}%
  \BibitemOpen
  \bibfield  {author} {\bibinfo {author} {\bibfnamefont {J.~C.~Y.}\
  \bibnamefont {Teo}}\ and\ \bibinfo {author} {\bibfnamefont {T.~L.}\
  \bibnamefont {Hughes}},\ }\href {\doibase 10.1103/PhysRevLett.111.047006}
  {\bibfield  {journal} {\bibinfo  {journal} {Phys. Rev. Lett.}\ }\textbf
  {\bibinfo {volume} {111}},\ \bibinfo {pages} {047006} (\bibinfo {year}
  {2013})}\BibitemShut {NoStop}%
\bibitem [{\citenamefont {Benalcazar}\ \emph {et~al.}(2014)\citenamefont
  {Benalcazar}, \citenamefont {Teo},\ and\ \citenamefont
  {Hughes}}]{Benalcazar_Classification_2014}%
  \BibitemOpen
  \bibfield  {author} {\bibinfo {author} {\bibfnamefont {W.~A.}\ \bibnamefont
  {Benalcazar}}, \bibinfo {author} {\bibfnamefont {J.~C.~Y.}\ \bibnamefont
  {Teo}}, \ and\ \bibinfo {author} {\bibfnamefont {T.~L.}\ \bibnamefont
  {Hughes}},\ }\href {\doibase 10.1103/PhysRevB.89.224503} {\bibfield
  {journal} {\bibinfo  {journal} {Phys. Rev. B}\ }\textbf {\bibinfo {volume}
  {89}},\ \bibinfo {pages} {224503} (\bibinfo {year} {2014})}\BibitemShut
  {NoStop}%
\bibitem [{\citenamefont {Levin}\ and\ \citenamefont
  {Gu}(2012)}]{Levin_Braiding_2012}%
  \BibitemOpen
  \bibfield  {author} {\bibinfo {author} {\bibfnamefont {M.}~\bibnamefont
  {Levin}}\ and\ \bibinfo {author} {\bibfnamefont {Z.-C.}\ \bibnamefont {Gu}},\
  }\href {\doibase 10.1103/PhysRevB.86.115109} {\bibfield  {journal} {\bibinfo
  {journal} {Phys. Rev. B}\ }\textbf {\bibinfo {volume} {86}},\ \bibinfo
  {pages} {115109} (\bibinfo {year} {2012})}\BibitemShut {NoStop}%
\bibitem [{\citenamefont {Thorngren}\ and\ \citenamefont
  {Else}(2018)}]{Thorngren_Gauging_2018}%
  \BibitemOpen
  \bibfield  {author} {\bibinfo {author} {\bibfnamefont {R.}~\bibnamefont
  {Thorngren}}\ and\ \bibinfo {author} {\bibfnamefont {D.~V.}\ \bibnamefont
  {Else}},\ }\href {\doibase 10.1103/PhysRevX.8.011040} {\bibfield  {journal}
  {\bibinfo  {journal} {Phys. Rev. X}\ }\textbf {\bibinfo {volume} {8}},\
  \bibinfo {pages} {011040} (\bibinfo {year} {2018})}\BibitemShut {NoStop}%
\bibitem [{\citenamefont {Santos}\ and\ \citenamefont
  {Hughes}(2017)}]{Santos_ParafermionicWires_2017}%
  \BibitemOpen
  \bibfield  {author} {\bibinfo {author} {\bibfnamefont {L.~H.}\ \bibnamefont
  {Santos}}\ and\ \bibinfo {author} {\bibfnamefont {T.~L.}\ \bibnamefont
  {Hughes}},\ }\href {\doibase 10.1103/PhysRevLett.118.136801} {\bibfield
  {journal} {\bibinfo  {journal} {Phys. Rev. Lett.}\ }\textbf {\bibinfo
  {volume} {118}},\ \bibinfo {pages} {136801} (\bibinfo {year}
  {2017})}\BibitemShut {NoStop}%
\bibitem [{\citenamefont {Santos}\ \emph {et~al.}(2018)\citenamefont {Santos},
  \citenamefont {Cano}, \citenamefont {Mulligan},\ and\ \citenamefont
  {Hughes}}]{Santos_Symmetry-protected_2018}%
  \BibitemOpen
  \bibfield  {author} {\bibinfo {author} {\bibfnamefont {L.~H.}\ \bibnamefont
  {Santos}}, \bibinfo {author} {\bibfnamefont {J.}~\bibnamefont {Cano}},
  \bibinfo {author} {\bibfnamefont {M.}~\bibnamefont {Mulligan}}, \ and\
  \bibinfo {author} {\bibfnamefont {T.~L.}\ \bibnamefont {Hughes}},\ }\href
  {\doibase 10.1103/PhysRevB.98.075131} {\bibfield  {journal} {\bibinfo
  {journal} {Phys. Rev. B}\ }\textbf {\bibinfo {volume} {98}},\ \bibinfo
  {pages} {075131} (\bibinfo {year} {2018})}\BibitemShut {NoStop}%
\bibitem [{\citenamefont {Santos}(2020)}]{Santos_Parafermions_2020}%
  \BibitemOpen
  \bibfield  {author} {\bibinfo {author} {\bibfnamefont {L.~H.}\ \bibnamefont
  {Santos}},\ }\href {\doibase 10.1103/PhysRevResearch.2.013232} {\bibfield
  {journal} {\bibinfo  {journal} {Phys. Rev. Research}\ }\textbf {\bibinfo
  {volume} {2}},\ \bibinfo {pages} {013232} (\bibinfo {year}
  {2020})}\BibitemShut {NoStop}%
\bibitem [{\citenamefont {Sohal}\ \emph {et~al.}(2020)\citenamefont {Sohal},
  \citenamefont {Han}, \citenamefont {Santos},\ and\ \citenamefont
  {Teo}}]{Sohal_Entanglement_2020}%
  \BibitemOpen
  \bibfield  {author} {\bibinfo {author} {\bibfnamefont {R.}~\bibnamefont
  {Sohal}}, \bibinfo {author} {\bibfnamefont {B.}~\bibnamefont {Han}}, \bibinfo
  {author} {\bibfnamefont {L.~H.}\ \bibnamefont {Santos}}, \ and\ \bibinfo
  {author} {\bibfnamefont {J.~C.~Y.}\ \bibnamefont {Teo}},\ }\href {\doibase
  10.1103/PhysRevB.102.045102} {\bibfield  {journal} {\bibinfo  {journal}
  {Phys. Rev. B}\ }\textbf {\bibinfo {volume} {102}},\ \bibinfo {pages}
  {045102} (\bibinfo {year} {2020})}\BibitemShut {NoStop}%
\bibitem [{\citenamefont {Alicea}\ and\ \citenamefont
  {Fendley}(2016)}]{alicea-fendley-2016}%
  \BibitemOpen
  \bibfield  {author} {\bibinfo {author} {\bibfnamefont {J.}~\bibnamefont
  {Alicea}}\ and\ \bibinfo {author} {\bibfnamefont {P.}~\bibnamefont
  {Fendley}},\ }\href {\doibase 10.1146/annurev-conmatphys-031115-011336}
  {\bibfield  {journal} {\bibinfo  {journal} {Annual Review of Condensed Matter
  Physics}\ }\textbf {\bibinfo {volume} {7}},\ \bibinfo {pages} {119} (\bibinfo
  {year} {2016})},\ \Eprint
  {http://arxiv.org/abs/https://doi.org/10.1146/annurev-conmatphys-031115-011336}
  {https://doi.org/10.1146/annurev-conmatphys-031115-011336} \BibitemShut
  {NoStop}%
\bibitem [{\citenamefont {Barkeshli}\ and\ \citenamefont
  {Qi}(2012)}]{BarkeshliQi-2012}%
  \BibitemOpen
  \bibfield  {author} {\bibinfo {author} {\bibfnamefont {M.}~\bibnamefont
  {Barkeshli}}\ and\ \bibinfo {author} {\bibfnamefont {X.-L.}\ \bibnamefont
  {Qi}},\ }\href {\doibase 10.1103/PhysRevX.2.031013} {\bibfield  {journal}
  {\bibinfo  {journal} {Phys. Rev. X}\ }\textbf {\bibinfo {volume} {2}},\
  \bibinfo {pages} {031013} (\bibinfo {year} {2012})}\BibitemShut {NoStop}%
\bibitem [{\citenamefont {Lindner}\ \emph {et~al.}(2012)\citenamefont
  {Lindner}, \citenamefont {Berg}, \citenamefont {Refael},\ and\ \citenamefont
  {Stern}}]{Lindner-2012}%
  \BibitemOpen
  \bibfield  {author} {\bibinfo {author} {\bibfnamefont {N.~H.}\ \bibnamefont
  {Lindner}}, \bibinfo {author} {\bibfnamefont {E.}~\bibnamefont {Berg}},
  \bibinfo {author} {\bibfnamefont {G.}~\bibnamefont {Refael}}, \ and\ \bibinfo
  {author} {\bibfnamefont {A.}~\bibnamefont {Stern}},\ }\href {\doibase
  10.1103/PhysRevX.2.041002} {\bibfield  {journal} {\bibinfo  {journal} {Phys.
  Rev. X}\ }\textbf {\bibinfo {volume} {2}},\ \bibinfo {pages} {041002}
  (\bibinfo {year} {2012})}\BibitemShut {NoStop}%
\bibitem [{\citenamefont {Clarke}\ \emph {et~al.}(2013)\citenamefont {Clarke},
  \citenamefont {Alicea},\ and\ \citenamefont {Shtengel}}]{Clarke-2013}%
  \BibitemOpen
  \bibfield  {author} {\bibinfo {author} {\bibfnamefont {D.~J.}\ \bibnamefont
  {Clarke}}, \bibinfo {author} {\bibfnamefont {J.}~\bibnamefont {Alicea}}, \
  and\ \bibinfo {author} {\bibfnamefont {K.}~\bibnamefont {Shtengel}},\ }\href
  {\doibase 10.1038/ncomms2340} {\bibfield  {journal} {\bibinfo  {journal}
  {Nature Communications}\ }\textbf {\bibinfo {volume} {4}},\ \bibinfo {pages}
  {1348 EP } (\bibinfo {year} {2013})}\BibitemShut {NoStop}%
\bibitem [{\citenamefont {Cheng}(2012)}]{Cheng-2012}%
  \BibitemOpen
  \bibfield  {author} {\bibinfo {author} {\bibfnamefont {M.}~\bibnamefont
  {Cheng}},\ }\href {\doibase 10.1103/PhysRevB.86.195126} {\bibfield  {journal}
  {\bibinfo  {journal} {Phys. Rev. B}\ }\textbf {\bibinfo {volume} {86}},\
  \bibinfo {pages} {195126} (\bibinfo {year} {2012})}\BibitemShut {NoStop}%
\bibitem [{\citenamefont {Vaezi}(2013)}]{Vaezi-2013}%
  \BibitemOpen
  \bibfield  {author} {\bibinfo {author} {\bibfnamefont {A.}~\bibnamefont
  {Vaezi}},\ }\href {\doibase 10.1103/PhysRevB.87.035132} {\bibfield  {journal}
  {\bibinfo  {journal} {Phys. Rev. B}\ }\textbf {\bibinfo {volume} {87}},\
  \bibinfo {pages} {035132} (\bibinfo {year} {2013})}\BibitemShut {NoStop}%
\bibitem [{\citenamefont {Barkeshli}\ \emph
  {et~al.}(2013{\natexlab{a}})\citenamefont {Barkeshli}, \citenamefont {Jian},\
  and\ \citenamefont {Qi}}]{BarkeshliJianQi-2013-a}%
  \BibitemOpen
  \bibfield  {author} {\bibinfo {author} {\bibfnamefont {M.}~\bibnamefont
  {Barkeshli}}, \bibinfo {author} {\bibfnamefont {C.-M.}\ \bibnamefont {Jian}},
  \ and\ \bibinfo {author} {\bibfnamefont {X.-L.}\ \bibnamefont {Qi}},\ }\href
  {\doibase 10.1103/PhysRevB.87.045130} {\bibfield  {journal} {\bibinfo
  {journal} {Phys. Rev. B}\ }\textbf {\bibinfo {volume} {87}},\ \bibinfo
  {pages} {045130} (\bibinfo {year} {2013}{\natexlab{a}})}\BibitemShut
  {NoStop}%
\bibitem [{\citenamefont {Barkeshli}\ \emph
  {et~al.}(2013{\natexlab{b}})\citenamefont {Barkeshli}, \citenamefont {Jian},\
  and\ \citenamefont {Qi}}]{BarkeshliJianQi-2013-b}%
  \BibitemOpen
  \bibfield  {author} {\bibinfo {author} {\bibfnamefont {M.}~\bibnamefont
  {Barkeshli}}, \bibinfo {author} {\bibfnamefont {C.-M.}\ \bibnamefont {Jian}},
  \ and\ \bibinfo {author} {\bibfnamefont {X.-L.}\ \bibnamefont {Qi}},\ }\href
  {\doibase 10.1103/PhysRevB.88.241103} {\bibfield  {journal} {\bibinfo
  {journal} {Phys. Rev. B}\ }\textbf {\bibinfo {volume} {88}},\ \bibinfo
  {pages} {241103} (\bibinfo {year} {2013}{\natexlab{b}})}\BibitemShut
  {NoStop}%
\bibitem [{\citenamefont {Mong}\ \emph {et~al.}(2014)\citenamefont {Mong},
  \citenamefont {Clarke}, \citenamefont {Alicea}, \citenamefont {Lindner},
  \citenamefont {Fendley}, \citenamefont {Nayak}, \citenamefont {Oreg},
  \citenamefont {Stern}, \citenamefont {Berg}, \citenamefont {Shtengel},\ and\
  \citenamefont {Fisher}}]{Mong-2014}%
  \BibitemOpen
  \bibfield  {author} {\bibinfo {author} {\bibfnamefont {R.~S.~K.}\
  \bibnamefont {Mong}}, \bibinfo {author} {\bibfnamefont {D.~J.}\ \bibnamefont
  {Clarke}}, \bibinfo {author} {\bibfnamefont {J.}~\bibnamefont {Alicea}},
  \bibinfo {author} {\bibfnamefont {N.~H.}\ \bibnamefont {Lindner}}, \bibinfo
  {author} {\bibfnamefont {P.}~\bibnamefont {Fendley}}, \bibinfo {author}
  {\bibfnamefont {C.}~\bibnamefont {Nayak}}, \bibinfo {author} {\bibfnamefont
  {Y.}~\bibnamefont {Oreg}}, \bibinfo {author} {\bibfnamefont {A.}~\bibnamefont
  {Stern}}, \bibinfo {author} {\bibfnamefont {E.}~\bibnamefont {Berg}},
  \bibinfo {author} {\bibfnamefont {K.}~\bibnamefont {Shtengel}}, \ and\
  \bibinfo {author} {\bibfnamefont {M.~P.~A.}\ \bibnamefont {Fisher}},\ }\href
  {\doibase 10.1103/PhysRevX.4.011036} {\bibfield  {journal} {\bibinfo
  {journal} {Phys. Rev. X}\ }\textbf {\bibinfo {volume} {4}},\ \bibinfo {pages}
  {011036} (\bibinfo {year} {2014})}\BibitemShut {NoStop}%
\bibitem [{\citenamefont {Khan}\ \emph {et~al.}(2014)\citenamefont {Khan},
  \citenamefont {Teo},\ and\ \citenamefont {Hughes}}]{khanteohughes-2014}%
  \BibitemOpen
  \bibfield  {author} {\bibinfo {author} {\bibfnamefont {M.~N.}\ \bibnamefont
  {Khan}}, \bibinfo {author} {\bibfnamefont {J.~C.~Y.}\ \bibnamefont {Teo}}, \
  and\ \bibinfo {author} {\bibfnamefont {T.~L.}\ \bibnamefont {Hughes}},\
  }\href {\doibase 10.1103/PhysRevB.90.235149} {\bibfield  {journal} {\bibinfo
  {journal} {Phys. Rev. B}\ }\textbf {\bibinfo {volume} {90}},\ \bibinfo
  {pages} {235149} (\bibinfo {year} {2014})}\BibitemShut {NoStop}%
\bibitem [{\citenamefont
  {Leggett}(1966)}]{Leggett_Number-PhaseFluctuations_1966}%
  \BibitemOpen
  \bibfield  {author} {\bibinfo {author} {\bibfnamefont {A.~J.}\ \bibnamefont
  {Leggett}},\ }\href {\doibase 10.1143/PTP.36.901} {\bibfield  {journal}
  {\bibinfo  {journal} {Progress of Theoretical Physics}\ }\textbf {\bibinfo
  {volume} {36}},\ \bibinfo {pages} {901} (\bibinfo {year} {1966})},\ \Eprint
  {http://arxiv.org/abs/https://academic.oup.com/ptp/article-pdf/36/5/901/5256693/36-5-901.pdf}
  {https://academic.oup.com/ptp/article-pdf/36/5/901/5256693/36-5-901.pdf}
  \BibitemShut {NoStop}%
\bibitem [{\citenamefont {Chung}\ \emph {et~al.}(2007)\citenamefont {Chung},
  \citenamefont {Bluhm},\ and\ \citenamefont {Kim}}]{ChungKim07}%
  \BibitemOpen
  \bibfield  {author} {\bibinfo {author} {\bibfnamefont {S.~B.}\ \bibnamefont
  {Chung}}, \bibinfo {author} {\bibfnamefont {H.}~\bibnamefont {Bluhm}}, \ and\
  \bibinfo {author} {\bibfnamefont {E.-A.}\ \bibnamefont {Kim}},\ }\href
  {\doibase 10.1103/PhysRevLett.99.197002} {\bibfield  {journal} {\bibinfo
  {journal} {Physical Review Letters}\ }\textbf {\bibinfo {volume} {99}},\
  \bibinfo {pages} {197002} (\bibinfo {year} {2007})},\ \bibinfo {note}
  {publisher: American Physical Society}\BibitemShut {NoStop}%
\bibitem [{\citenamefont {Agterberg}\ and\ \citenamefont
  {Tsunetsugu}(2008)}]{Agterberg08}%
  \BibitemOpen
  \bibfield  {author} {\bibinfo {author} {\bibfnamefont {D.~F.}\ \bibnamefont
  {Agterberg}}\ and\ \bibinfo {author} {\bibfnamefont {H.}~\bibnamefont
  {Tsunetsugu}},\ }\href {\doibase 10.1038/nphys999} {\bibfield  {journal}
  {\bibinfo  {journal} {Nature Physics}\ }\textbf {\bibinfo {volume} {4}},\
  \bibinfo {pages} {639} (\bibinfo {year} {2008})},\ \bibinfo {note} {number: 8
  Publisher: Nature Publishing Group}\BibitemShut {NoStop}%
\bibitem [{\citenamefont {Regan}\ \emph {et~al.}(2021)\citenamefont {Regan},
  \citenamefont {Wiman},\ and\ \citenamefont {Sauls}}]{ReganSauls21}%
  \BibitemOpen
  \bibfield  {author} {\bibinfo {author} {\bibfnamefont {R.~C.}\ \bibnamefont
  {Regan}}, \bibinfo {author} {\bibfnamefont {J.~J.}\ \bibnamefont {Wiman}}, \
  and\ \bibinfo {author} {\bibfnamefont {J.~A.}\ \bibnamefont {Sauls}},\ }\href
  {\doibase 10.1103/PhysRevB.104.024513} {\bibfield  {journal} {\bibinfo
  {journal} {Phys. Rev. B}\ }\textbf {\bibinfo {volume} {104}},\ \bibinfo
  {pages} {024513} (\bibinfo {year} {2021})}\BibitemShut {NoStop}%
\bibitem [{\citenamefont {R\"opke}\ \emph {et~al.}(1998)\citenamefont
  {R\"opke}, \citenamefont {Schnell}, \citenamefont {Schuck},\ and\
  \citenamefont {Nozi\`eres}}]{RopkeNozieres98}%
  \BibitemOpen
  \bibfield  {author} {\bibinfo {author} {\bibfnamefont {G.}~\bibnamefont
  {R\"opke}}, \bibinfo {author} {\bibfnamefont {A.}~\bibnamefont {Schnell}},
  \bibinfo {author} {\bibfnamefont {P.}~\bibnamefont {Schuck}}, \ and\ \bibinfo
  {author} {\bibfnamefont {P.}~\bibnamefont {Nozi\`eres}},\ }\href {\doibase
  10.1103/PhysRevLett.80.3177} {\bibfield  {journal} {\bibinfo  {journal}
  {Phys. Rev. Lett.}\ }\textbf {\bibinfo {volume} {80}},\ \bibinfo {pages}
  {3177} (\bibinfo {year} {1998})}\BibitemShut {NoStop}%
\bibitem [{\citenamefont {Babaev}(2004)}]{Babaev04}%
  \BibitemOpen
  \bibfield  {author} {\bibinfo {author} {\bibfnamefont {E.}~\bibnamefont
  {Babaev}},\ }\href {\doibase 10.1016/j.nuclphysb.2004.02.021} {\bibfield
  {journal} {\bibinfo  {journal} {Nuclear Physics B}\ }\textbf {\bibinfo
  {volume} {686}},\ \bibinfo {pages} {397} (\bibinfo {year}
  {2004})}\BibitemShut {NoStop}%
\bibitem [{\citenamefont {Wu}(2005)}]{Wu05}%
  \BibitemOpen
  \bibfield  {author} {\bibinfo {author} {\bibfnamefont {C.}~\bibnamefont
  {Wu}},\ }\href {\doibase 10.1103/PhysRevLett.95.266404} {\bibfield  {journal}
  {\bibinfo  {journal} {Physical Review Letters}\ }\textbf {\bibinfo {volume}
  {95}},\ \bibinfo {pages} {266404} (\bibinfo {year} {2005})},\ \bibinfo {note}
  {publisher: American Physical Society}\BibitemShut {NoStop}%
\bibitem [{\citenamefont {Berg}\ \emph {et~al.}(2009)\citenamefont {Berg},
  \citenamefont {Fradkin},\ and\ \citenamefont {Kivelson}}]{BergFradkin09}%
  \BibitemOpen
  \bibfield  {author} {\bibinfo {author} {\bibfnamefont {E.}~\bibnamefont
  {Berg}}, \bibinfo {author} {\bibfnamefont {E.}~\bibnamefont {Fradkin}}, \
  and\ \bibinfo {author} {\bibfnamefont {S.~A.}\ \bibnamefont {Kivelson}},\
  }\href {\doibase 10.1038/nphys1389} {\bibfield  {journal} {\bibinfo
  {journal} {Nature Physics}\ }\textbf {\bibinfo {volume} {5}},\ \bibinfo
  {pages} {830} (\bibinfo {year} {2009})},\ \bibinfo {note} {number: 11
  Publisher: Nature Publishing Group}\BibitemShut {NoStop}%
\bibitem [{\citenamefont {Agterberg}\ \emph {et~al.}(2011)\citenamefont
  {Agterberg}, \citenamefont {Geracie},\ and\ \citenamefont
  {Tsunetsugu}}]{Agterberg11}%
  \BibitemOpen
  \bibfield  {author} {\bibinfo {author} {\bibfnamefont {D.~F.}\ \bibnamefont
  {Agterberg}}, \bibinfo {author} {\bibfnamefont {M.}~\bibnamefont {Geracie}},
  \ and\ \bibinfo {author} {\bibfnamefont {H.}~\bibnamefont {Tsunetsugu}},\
  }\href {\doibase 10.1103/PhysRevB.84.014513} {\bibfield  {journal} {\bibinfo
  {journal} {Physical Review B}\ }\textbf {\bibinfo {volume} {84}},\ \bibinfo
  {pages} {014513} (\bibinfo {year} {2011})},\ \bibinfo {note} {publisher:
  American Physical Society}\BibitemShut {NoStop}%
\bibitem [{\citenamefont {Moon}(2012)}]{Moon12}%
  \BibitemOpen
  \bibfield  {author} {\bibinfo {author} {\bibfnamefont {E.-G.}\ \bibnamefont
  {Moon}},\ }\href {\doibase 10.1103/PhysRevB.85.245123} {\bibfield  {journal}
  {\bibinfo  {journal} {Physical Review B}\ }\textbf {\bibinfo {volume} {85}},\
  \bibinfo {pages} {245123} (\bibinfo {year} {2012})},\ \bibinfo {note}
  {publisher: American Physical Society}\BibitemShut {NoStop}%
\bibitem [{\citenamefont {Lee}(2014)}]{Lee14}%
  \BibitemOpen
  \bibfield  {author} {\bibinfo {author} {\bibfnamefont {P.~A.}\ \bibnamefont
  {Lee}},\ }\href {\doibase 10.1103/PhysRevX.4.031017} {\bibfield  {journal}
  {\bibinfo  {journal} {Physical Review X}\ }\textbf {\bibinfo {volume} {4}},\
  \bibinfo {pages} {031017} (\bibinfo {year} {2014})},\ \bibinfo {note}
  {publisher: American Physical Society}\BibitemShut {NoStop}%
\bibitem [{\citenamefont {Fradkin}\ \emph {et~al.}(2015)\citenamefont
  {Fradkin}, \citenamefont {Kivelson},\ and\ \citenamefont
  {Tranquada}}]{FradkinKivelson15}%
  \BibitemOpen
  \bibfield  {author} {\bibinfo {author} {\bibfnamefont {E.}~\bibnamefont
  {Fradkin}}, \bibinfo {author} {\bibfnamefont {S.~A.}\ \bibnamefont
  {Kivelson}}, \ and\ \bibinfo {author} {\bibfnamefont {J.~M.}\ \bibnamefont
  {Tranquada}},\ }\href {\doibase 10.1103/RevModPhys.87.457} {\bibfield
  {journal} {\bibinfo  {journal} {Reviews of Modern Physics}\ }\textbf
  {\bibinfo {volume} {87}},\ \bibinfo {pages} {457} (\bibinfo {year} {2015})},\
  \bibinfo {note} {publisher: American Physical Society}\BibitemShut {NoStop}%
\bibitem [{\citenamefont {Jiang}\ \emph {et~al.}(2017)\citenamefont {Jiang},
  \citenamefont {Li}, \citenamefont {Kivelson},\ and\ \citenamefont
  {Yao}}]{JiangKivelson17}%
  \BibitemOpen
  \bibfield  {author} {\bibinfo {author} {\bibfnamefont {Y.-F.}\ \bibnamefont
  {Jiang}}, \bibinfo {author} {\bibfnamefont {Z.-X.}\ \bibnamefont {Li}},
  \bibinfo {author} {\bibfnamefont {S.~A.}\ \bibnamefont {Kivelson}}, \ and\
  \bibinfo {author} {\bibfnamefont {H.}~\bibnamefont {Yao}},\ }\href {\doibase
  10.1103/PhysRevB.95.241103} {\bibfield  {journal} {\bibinfo  {journal}
  {Physical Review B}\ }\textbf {\bibinfo {volume} {95}},\ \bibinfo {pages}
  {241103} (\bibinfo {year} {2017})},\ \bibinfo {note} {publisher: American
  Physical Society}\BibitemShut {NoStop}%
\bibitem [{\citenamefont {Fernandes}\ \emph {et~al.}(2019)\citenamefont
  {Fernandes}, \citenamefont {Orth},\ and\ \citenamefont
  {Schmalian}}]{FernandesSchmalian19}%
  \BibitemOpen
  \bibfield  {author} {\bibinfo {author} {\bibfnamefont {R.~M.}\ \bibnamefont
  {Fernandes}}, \bibinfo {author} {\bibfnamefont {P.~P.}\ \bibnamefont {Orth}},
  \ and\ \bibinfo {author} {\bibfnamefont {J.}~\bibnamefont {Schmalian}},\
  }\href {\doibase 10.1146/annurev-conmatphys-031218-013200} {\bibfield
  {journal} {\bibinfo  {journal} {Annual Review of Condensed Matter Physics}\
  }\textbf {\bibinfo {volume} {10}},\ \bibinfo {pages} {133} (\bibinfo {year}
  {2019})},\ \bibinfo {note} {publisher: Annual Reviews}\BibitemShut {NoStop}%
\bibitem [{\citenamefont {Fernandes}\ and\ \citenamefont
  {Fu}(2021)}]{FernandesFu21}%
  \BibitemOpen
  \bibfield  {author} {\bibinfo {author} {\bibfnamefont {R.~M.}\ \bibnamefont
  {Fernandes}}\ and\ \bibinfo {author} {\bibfnamefont {L.}~\bibnamefont {Fu}},\
  }\href {\doibase 10.1103/PhysRevLett.127.047001} {\bibfield  {journal}
  {\bibinfo  {journal} {Physical Review Letters}\ }\textbf {\bibinfo {volume}
  {127}},\ \bibinfo {pages} {047001} (\bibinfo {year} {2021})},\ \bibinfo
  {note} {publisher: American Physical Society}\BibitemShut {NoStop}%
\bibitem [{\citenamefont {Babaev}(2002)}]{Babaev02}%
  \BibitemOpen
  \bibfield  {author} {\bibinfo {author} {\bibfnamefont {E.}~\bibnamefont
  {Babaev}},\ }\href {\doibase 10.1103/PhysRevLett.89.067001} {\bibfield
  {journal} {\bibinfo  {journal} {Physical Review Letters}\ }\textbf {\bibinfo
  {volume} {89}},\ \bibinfo {pages} {067001} (\bibinfo {year} {2002})},\
  \bibinfo {note} {publisher: American Physical Society}\BibitemShut {NoStop}%
\bibitem [{\citenamefont {Smiseth}\ \emph {et~al.}(2005)\citenamefont
  {Smiseth}, \citenamefont {Sm\o{}rgrav}, \citenamefont {Babaev},\ and\
  \citenamefont {Sudb\o{}}}]{SmithsethBabaev05}%
  \BibitemOpen
  \bibfield  {author} {\bibinfo {author} {\bibfnamefont {J.}~\bibnamefont
  {Smiseth}}, \bibinfo {author} {\bibfnamefont {E.}~\bibnamefont
  {Sm\o{}rgrav}}, \bibinfo {author} {\bibfnamefont {E.}~\bibnamefont {Babaev}},
  \ and\ \bibinfo {author} {\bibfnamefont {A.}~\bibnamefont {Sudb\o{}}},\
  }\href {\doibase 10.1103/PhysRevB.71.214509} {\bibfield  {journal} {\bibinfo
  {journal} {Phys. Rev. B}\ }\textbf {\bibinfo {volume} {71}},\ \bibinfo
  {pages} {214509} (\bibinfo {year} {2005})}\BibitemShut {NoStop}%
\bibitem [{\citenamefont {Kang}\ and\ \citenamefont
  {Vafek}(2018)}]{KangVafek18}%
  \BibitemOpen
  \bibfield  {author} {\bibinfo {author} {\bibfnamefont {J.}~\bibnamefont
  {Kang}}\ and\ \bibinfo {author} {\bibfnamefont {O.}~\bibnamefont {Vafek}},\
  }\href {\doibase 10.1103/PhysRevX.8.031088} {\bibfield  {journal} {\bibinfo
  {journal} {Physical Review X}\ }\textbf {\bibinfo {volume} {8}},\ \bibinfo
  {pages} {031088} (\bibinfo {year} {2018})},\ \bibinfo {note} {publisher:
  American Physical Society}\BibitemShut {NoStop}%
\bibitem [{\citenamefont {Koshino}\ \emph {et~al.}(2018)\citenamefont
  {Koshino}, \citenamefont {Yuan}, \citenamefont {Koretsune}, \citenamefont
  {Ochi}, \citenamefont {Kuroki},\ and\ \citenamefont {Fu}}]{KoshinoFu18}%
  \BibitemOpen
  \bibfield  {author} {\bibinfo {author} {\bibfnamefont {M.}~\bibnamefont
  {Koshino}}, \bibinfo {author} {\bibfnamefont {N.~F.}\ \bibnamefont {Yuan}},
  \bibinfo {author} {\bibfnamefont {T.}~\bibnamefont {Koretsune}}, \bibinfo
  {author} {\bibfnamefont {M.}~\bibnamefont {Ochi}}, \bibinfo {author}
  {\bibfnamefont {K.}~\bibnamefont {Kuroki}}, \ and\ \bibinfo {author}
  {\bibfnamefont {L.}~\bibnamefont {Fu}},\ }\href {\doibase
  10.1103/PhysRevX.8.031087} {\bibfield  {journal} {\bibinfo  {journal}
  {Physical Review X}\ }\textbf {\bibinfo {volume} {8}},\ \bibinfo {pages}
  {031087} (\bibinfo {year} {2018})},\ \bibinfo {note} {publisher: American
  Physical Society}\BibitemShut {NoStop}%
\bibitem [{\citenamefont {Yuan}\ and\ \citenamefont
  {Fu}(2018{\natexlab{b}})}]{YuanFu18_2}%
  \BibitemOpen
  \bibfield  {author} {\bibinfo {author} {\bibfnamefont {N.~F.~Q.}\
  \bibnamefont {Yuan}}\ and\ \bibinfo {author} {\bibfnamefont {L.}~\bibnamefont
  {Fu}},\ }\href {\doibase 10.1103/PhysRevB.98.045103} {\bibfield  {journal}
  {\bibinfo  {journal} {Physical Review B}\ }\textbf {\bibinfo {volume} {98}},\
  \bibinfo {pages} {045103} (\bibinfo {year} {2018}{\natexlab{b}})},\ \bibinfo
  {note} {publisher: American Physical Society}\BibitemShut {NoStop}%
\bibitem [{\citenamefont {Andrews}\ and\ \citenamefont
  {Soluyanov}(2020)}]{AndrewsSoluyanov20}%
  \BibitemOpen
  \bibfield  {author} {\bibinfo {author} {\bibfnamefont {B.}~\bibnamefont
  {Andrews}}\ and\ \bibinfo {author} {\bibfnamefont {A.}~\bibnamefont
  {Soluyanov}},\ }\href {\doibase 10.1103/PhysRevB.101.235312} {\bibfield
  {journal} {\bibinfo  {journal} {Physical Review B}\ }\textbf {\bibinfo
  {volume} {101}},\ \bibinfo {pages} {235312} (\bibinfo {year} {2020})},\
  \bibinfo {note} {publisher: American Physical Society}\BibitemShut {NoStop}%
\bibitem [{\citenamefont {Chandrasekhar}(1962)}]{chandrasekhar1962note}%
  \BibitemOpen
  \bibfield  {author} {\bibinfo {author} {\bibfnamefont {B.}~\bibnamefont
  {Chandrasekhar}},\ }\href@noop {} {\bibfield  {journal} {\bibinfo  {journal}
  {Applied Physics Letters}\ }\textbf {\bibinfo {volume} {1}},\ \bibinfo
  {pages} {7} (\bibinfo {year} {1962})}\BibitemShut {NoStop}%
\bibitem [{\citenamefont {Clogston}(1962)}]{clogston1962upper}%
  \BibitemOpen
  \bibfield  {author} {\bibinfo {author} {\bibfnamefont {A.~M.}\ \bibnamefont
  {Clogston}},\ }\href@noop {} {\bibfield  {journal} {\bibinfo  {journal}
  {Physical Review Letters}\ }\textbf {\bibinfo {volume} {9}},\ \bibinfo
  {pages} {266} (\bibinfo {year} {1962})}\BibitemShut {NoStop}%
\bibitem [{\citenamefont {Cao}\ \emph {et~al.}(2018)\citenamefont {Cao},
  \citenamefont {Fatemi}, \citenamefont {Fang}, \citenamefont {Watanabe},
  \citenamefont {Taniguchi}, \citenamefont {Kaxiras},\ and\ \citenamefont
  {Jarillo-Herrero}}]{cao2018unconventional}%
  \BibitemOpen
  \bibfield  {author} {\bibinfo {author} {\bibfnamefont {Y.}~\bibnamefont
  {Cao}}, \bibinfo {author} {\bibfnamefont {V.}~\bibnamefont {Fatemi}},
  \bibinfo {author} {\bibfnamefont {S.}~\bibnamefont {Fang}}, \bibinfo {author}
  {\bibfnamefont {K.}~\bibnamefont {Watanabe}}, \bibinfo {author}
  {\bibfnamefont {T.}~\bibnamefont {Taniguchi}}, \bibinfo {author}
  {\bibfnamefont {E.}~\bibnamefont {Kaxiras}}, \ and\ \bibinfo {author}
  {\bibfnamefont {P.}~\bibnamefont {Jarillo-Herrero}},\ }\href@noop {}
  {\bibfield  {journal} {\bibinfo  {journal} {Nature}\ }\textbf {\bibinfo
  {volume} {556}},\ \bibinfo {pages} {43} (\bibinfo {year} {2018})}\BibitemShut
  {NoStop}%
\end{thebibliography}%

\end{document}